%% file: N1VMS-arXiv-v1.tex
\documentclass[11pt, final, a4paper]{article}

\PassOptionsToPackage{smalltableaux}{ytableau}

\usepackage[nojhep,cm]{MRnotes}
\usepackage[
  a4paper,
  top=27mm,
  bottom=30mm,
  left=27mm,
  right=27mm,
  headheight=14pt,
  headsep=7mm,
  footskip=10mm
]{geometry}
\usepackage{pgfplots}
\pgfplotsset{compat=1.18}
\usetikzlibrary{
  arrows,
  patterns.meta,
  pgfplots.fillbetween,
  positioning,
  shapes.geometric,
  shapes.misc
}
\usepackage{pdfpages}
\usepackage{tabularx}
\usepackage{textgreek}
\usepackage{verbatim}
\ytableausetup{centertableaux}

\hypersetup{
  pdfencoding=unicode,
  colorlinks=true,
  urlcolor=Maroon,
  linkcolor=RoyalBlue,
  citecolor=Maroon,
  pdftitle={Super Liouville theory on the torus: bootstrap and the super Virasoro minimal string},
  pdfauthor={Scott Collier, Beatrix M\"uhlmann, Mukund Rangamani, Ioannis Tsiares, Jianming Zheng},
  pdfdisplaydoctitle=true,
  pdfstartview=FitH,
  linktocpage=true
}

\definecolor{vert}{rgb}{0.1367 0.543 0.1367}

\renewcommand{\d}{\mathrm{d}}
\allowdisplaybreaks

\input{svms-macros.tex}

\newcommand{\drafttitle}{
  Super Liouville theory on the torus:\\
  bootstrap and the super Virasoro minimal string \\[2ex]
  \hrule
}

\newcommand{\draftauthors}{
  Scott Collier$^{1,2}$, Beatrix M\"uhlmann$^{3}$, Mukund Rangamani$^{4}$,\\[1ex]
  Ioannis Tsiares$^{5}$, Jianming Zheng$^{4}$
}
\newcommand{\draftaffiliation}{
$^1$ Department of Physics, Syracuse University, Syracuse, NY, 13244, USA\\ [1ex]
$^2$ Institute for Quantum \& Information Sciences, Syracuse University, Syracuse, NY, 13244, USA\\ [1ex]
  $^3$ School of Natural Sciences, Institute for Advanced Study, Princeton, NJ
  08540, USA\\[1ex]
  $^4$ Center for Quantum Mathematics \& Physics (QMAP), Department of Physics
  \& Astronomy, \\
  University of California, Davis, CA 95616, USA\\ [1ex]
  $^5$ Laboratoire de Physique, \'Ecole Normale Sup\'erieure,
Universit\'e PSL, CNRS, Sorbonne Universit\'e, Universit\'e Paris Cit\'e, 24 rue Lhomond, F-75005 Paris, France
}

\newcommand{\draftemails}{scolli32@syr.edu, beatrix@ias.edu, mukund@physics.ucdavis.edu, ioannis.tsiares@phys.ens.fr, jmzheng@ucdavis.edu}

\newcommand{\draftabstract}{
  Motivated by applications to low-dimensional worldsheet string constructions, we analyze $\mathcal{N} = 1$ super Liouville theory on the torus. Specifically, we re-examine the recursive representations of the torus one-point super Virasoro blocks that have been constructed in the literature and correct several of their ingredients. In the process we identify a new block in the odd spin structure, associated with a privileged superdescendant whose torus one-point function transforms like that of a Virasoro primary. We then demonstrate that the structure constants for both spacelike and timelike super Liouville theory satisfy modular crossing. Furthermore, we use a bootstrap argument to construct modular crossing kernels for all four spin structure sectors, valid for central charges $c \in \mathbb{C}\setminus(-\infty,\tfrac{3}{2}]$, and verify that they correctly implement the crossing transformations of the blocks. As a concrete application of the technology, we analyze the super Virasoro minimal string. There are four such theories, \zA{\pm} and \zB{\pm}, which are distinguished by the GSO projection together with the choice of worldsheet supercharge. 
  The resulting NS-sector sphere three-point and torus one-point amplitudes are shown to agree with topological recursion on the spectral curves of the dual matrix integrals, which were recently derived from 3d supergravity. 
  Notably, the one-point function of the privileged superdescendant plays an essential role in the contribution of the odd spin structure to the torus amplitude via a careful treatment of the picture changing operator. We also identify an exceptional zero-momentum Ramond insertion in the \zA{-} theory that we interpret geometrically as a Ramond puncture, and match its mixed sphere three-point amplitude to the corresponding intersection theory prediction.
}

\newcommand{\makedrafttitle}{%
  \thispagestyle{empty}%
  \begin{center}
  {\LARGE\bfseries\sffamily \drafttitle\par}
    \vspace{1.2cm}
    {\large\bfseries \draftauthors\par}
    \vspace{0.45cm}
    {\small \draftaffiliation\par}
    \vspace{0.3cm}
    {\small\texttt{\draftemails}\par}
  \end{center}
  \vspace{0.8cm}
  \begin{abstract}
    \draftabstract
  \end{abstract}
  \vspace{0.8cm}
}

\begin{document}

\makedrafttitle
\newpage
\hrule \hrule
\tableofcontents
\bigskip
\hrule \hrule

\section{Introduction}\label{sec:intro}

Exactly solvable models of string theory and quantum gravity are rare gems. In recent years a systematic source of such models has emerged in two dimensions, where there is a growing landscape of worldsheet string theories that have been solved by dual double-scaled matrix integrals. These dual descriptions provide controlled laboratories in which structural questions about string theory and quantum gravity can be studied exactly. Notably, Jackiw-Teitelboim dilaton gravity was shown to be equivalent to a double-scaled matrix integral in~\cite{Saad:2019lba}, and this duality was subsequently extended to JT supergravity and the Altland-Zirnbauer classification of random matrix ensembles in~\cite{Stanford:2019vob}. These dualities have since been embedded in two-dimensional string theory. The Virasoro minimal string (VMS)~\cite{Collier:2023cyw}, defined by coupling spacelike and timelike Liouville CFTs on the worldsheet, was shown to be dual to a double-scaled matrix integral whose leading density of eigenvalues is the universal Cardy density of states of two-dimensional conformal field theory~\cite{Cardy:1986ie}. This duality reduces to the JT gravity/RMT duality in a semiclassical limit. The complex Liouville string~\cite{Collier:2024kmo, Collier:2024kwt, Collier:2024lys}, built instead from two complex-conjugate copies of spacelike Liouville CFT, is similarly dual to a two-matrix integral whose string amplitudes moreover compute certain integrated cosmological correlators in dS$_3$ quantum gravity~\cite{Collier:2025lux}. In these solvable examples the string amplitudes are analytic functions of the central charge and of the external vertex operator data, and are all exactly computable to all orders in the genus expansion through topological recursion on the spectral curve of the dual matrix integral (or equivalently, through intersection theory on the moduli space of Riemann surfaces).

A key aspect of the mechanism underlying the solvability of this class of models is Virasoro representation theory. In each case the worldsheet theory is built out of Liouville CFT, which is in a precise sense the simplest CFT with pure Virasoro symmetry (under mild assumptions, it is the unique such theory whose spectrum contains only scalar Virasoro primaries~\cite{Ribault:2014hia,Collier:2017shs}). Indeed, Liouville CFT has been solved exactly, and its data is expressible in terms of universal objects in Virasoro representation theory. In a convenient operator normalization, its spectral density is fixed by the modular transform of the identity character, while the DOZZ structure constants~\cite{Dorn:1994xn, Zamolodchikov:1995aa, Teschner:1995yf} can be expressed in terms of the fusion kernel that branches the sphere four-point identity block in the dual channel~\cite{Ponsot:1999uf,Ponsot:2000mt,Eberhardt:2023mrq}. Thus, the ingredients are universal quantities governing the statistics of heavy primaries in a generic 2d CFT: the former is the Cardy formula mentioned above, and the latter is the universal formula $C_0$ that controls the variance of structure constants of heavy operators~\cite{Collier:2019weq}. The other ingredients needed to assemble string amplitudes are the Virasoro conformal blocks, which are purely kinematic objects fixed by Virasoro symmetry that can be efficiently computed by Zamolodchikov-type recursion relations~\cite{Zamolodchikov:1984eqp,Zamolodchikov:1987avt,Hadasz:2009db,Cho:2017oxl}. Together, this data allows one to subject the proposed dualities to stringent numerical tests. 

It is natural to ask to what extent this structure carries over to the case of $\mathcal{N}=1$ supersymmetry, particularly in the simplest setting of the super Virasoro minimal string ($\sVMS$) formulated at the level of the worldsheet SCFT in~\cite{Muhlmann:2025ngz,Rangamani:2025wfa}. Likewise, the supersymmetric version of the complex Liouville string has been analyzed from the worldsheet at the level of low-point genus-zero Neveu-Schwarz amplitudes in~\cite{Du:2025lya}.\footnote{See~\cref{sec:discuss} for more comments on the super complex Liouville string.} More specifically, the $\sVMS$ was recently analyzed from the perspective of $\mathcal{N}=1$ AdS$_3$ supergravity in~\cite{Eberhardt:2026hfh}. That analysis clarified the $\mathcal{N}=1$ landscape, showing that there are in fact \emph{four} super Virasoro minimal string theories, distinguished by a choice of GSO projection together with an ambiguity in the definition of the total supercharge (preliminary evidence for this intricate structure was the observation in~\cite{Muhlmann:2025ngz,Rangamani:2025wfa} of a relative sign ambiguity in the solution of structure constants of the two super Liouville theories). It led to an explicit proposal for the spectral curves of the dual matrix integrals in each of the four sectors, along with concrete predictions for the string amplitudes in terms of intersection numbers on the moduli space of the worldsheet surfaces. These models have also been studied directly from the matrix model perspective~\cite{Johnson:2024fkm,Johnson:2025vyz,Johnson:2026ugu}, in particular via string equations that supply non-perturbative completions.

However, the worldsheet conformal field theory is considerably less developed than its bosonic counterpart. While the structure constants of the $\mathcal{N}=1$ spacelike Liouville CFT were determined long ago in~\cite{Rashkov:1996np, Poghossian:1996agj, Belavin:2007gz}, those of the $\mathcal{N}=1$ timelike Liouville CFT were only recently bootstrapped in~\cite{Muhlmann:2025ngz, Rangamani:2025wfa}. Similarly, while elliptic recursive representations for the sphere four-point $\mathcal{N}=1$ super-Virasoro blocks were derived in~\cite{Hadasz:2006qb,Belavin:2007gz,Hadasz:2008dt,Suchanek:2010kq}, the situation for the torus one-point blocks is comparatively incomplete. Recursive representations for the $\mathcal{N}=1$ torus one-point blocks were proposed in~\cite{Hadasz:2012im}. However, we will see that they  require some corrections and must be supplemented with an additional structure in order to compute the full string amplitude, as we explain below. Moreover, the modular crossing kernels for the $\mathcal{N}=1$ torus one-point blocks have only been written down in the Neveu-Schwarz sector~\cite{Apresyan:2023vxl}. As a result, modular covariance of the torus one-point function in $\mathcal{N}=1$ super Liouville CFT (spacelike or timelike) has not yet been fully demonstrated.

The main goal of this paper is to close these gaps. As a result we will be able to perform a complete worldsheet computation of the torus one-point amplitude in all four $\sVMS$ theories, finding perfect agreement with the matrix model predictions of~\cite{Eberhardt:2026hfh}. Along the way, we will see that incorporating $\mathcal{N}=1$ supersymmetry is a (perhaps surprisingly) non-trivial generalization, and treating it carefully requires us to confront head-on some technical subtleties of string perturbation theory as currently formulated in the RNS formalism. We now summarize our main results.

\paragraph{Recursive representation of torus one-point super-Virasoro blocks:}
Torus observables in $\mathcal{N}=1$ SCFT are conventionally organized into four sectors, corresponding to the choice of Neveu-Schwarz (NS) or Ramond (R) boundary conditions along the spatial cycle, together with the possible insertion of $(-1)^F$ along the temporal cycle. We label the four sectors $\set{\mathrm{NS}, \widetilde{\mathrm{NS}},\mathrm{R},\widetilde{\mathrm{R}}}$ by $a\in\set{3,4,2,1}$, the mnemonic following from the fact that the trace over the vacuum module produces the corresponding theta function $\vartheta_a$. Under the modular S-transformation that exchanges the cycles of the torus, the $\widetilde{\mathrm{NS}}$ and $\mathrm{R}$ sectors are exchanged while the $\mathrm{NS}$ and $\widetilde{\mathrm{R}}$ sectors are mapped to themselves. The torus one-point functions of the superconformal primary $V_P$ and its superdescendant $W_P = \widetilde{G}_{-\frac{1}{2}}G_{-\frac{1}{2}}V_P$ may be decomposed into super Virasoro conformal blocks in each of the four sectors. Building on~\cite{Hadasz:2012im}, we construct Zamolodchikov-type recursive representations for these blocks in all sectors that render them efficiently computable, fixing in the process several issues with the expressions already available in the literature. 

The odd spin structure ($\widetilde{\mathrm{R}}$) is special: the one-point block of the superdescendant $W_P$ vanishes identically due to fermion zero modes. We instead identify a distinguished level three-half\footnote{We label the NS sector descendant operator by the total mode number of the creation operator, which can be half-integral. } superdescendant $\widetilde{G}_{-\frac{3}{2}} G_{-\frac{3}{2}}V_P$ whose torus one-point function transforms like that of a Virasoro primary and is itself only non-vanishing in the odd spin structure. Using superconformal Ward identities we show that it admits a standard super Virasoro conformal block decomposition and propose a recursive representation of the corresponding block. The torus one-point function of this distinguished superdescendant operator in spacelike and timelike super Liouville theories will turn out to play an important role in the assembly of the corresponding string amplitude in $\sVMS$; we will see that it arises from a careful treatment of the picture-changing operator in the odd spin structure sector.\footnote{Such level three-half superdescendant operators are similarly distinguished in other $\mathcal{N}=1$ superconformal field theories and will generically have non-vanishing torus expectation values in the odd spin structure. Our super Liouville expressions can be adapted to other theories with a suitable replacement of structure constants and spectral densities.}

\paragraph{Modular covariance of super-Liouville CFT and modular crossing kernels:}
With the blocks under control, we provide strong numerical evidence that the structure constants of both spacelike and timelike $\mathcal{N}=1$ super Liouville CFT define modular covariant torus one-point functions in all spin structure sectors. In particular, this provides a new consistency check of the timelike super Liouville structure constants that were bootstrapped in~\cite{Muhlmann:2025ngz,Rangamani:2025wfa}. Prior to this work, only the sphere 4-point crossing of the NS structure constants was numerically checked in~\cite{Rangamani:2025wfa}.

In order to prove modular covariance \emph{analytically} in both the NS and R sectors, one needs to construct the corresponding modular crossing kernels that express the super-Virasoro blocks in a complete basis of blocks in the modular dual channel. In this paper we construct all the modular crossing kernels that implement the modular S-transform in the regime of central charges relevant for the spacelike theory, $c\in\mathbb{C}\setminus(-\infty,\tfrac{3}{2}]$. 

The presence of the two independent one-point functions (of $V_P$ and its superdescendant), together with the mixing of the spin structures and the covariant superdescendant block in the odd spin structure, implies that the full suite consists of \emph{eight} kernels. Two of the eight kernels (corresponding to the $\mathrm{NS}$ sector) were previously obtained in~\cite{Apresyan:2023vxl} by solving Moore-Seiberg type consistency conditions. The remaining six are new. We determine all eight kernels by positing a set of bootstrap conditions and solving them. We verify that our procedure reproduces the known NS-sector results and subject the new kernels to numerical tests of the defining crossing relations. We expect that these kernels will find applications similar in scope to those of their bosonic counterparts: in boundary conditions for super Liouville CFT, in the universal asymptotics of $\mathcal{N}=1$ SCFT data, and in the quantization of super-Teichm\"uller space and the formulation of $\mathcal{N}=1$ super-Virasoro TQFT (which has already started to be developed in~\cite{Eberhardt:2026hfh}).

\paragraph{Application: the super Virasoro minimal string:}
As an immediate application of this technology, we study string amplitudes in the super Virasoro minimal string, defined by the worldsheet SCFT
\begin{equation}\label{eq:svmsdef}
\begin{array}{c}
\text{spacelike $\mathcal{N}=1$}\\
\text{super Liouville}\\
c = \frac{3}{2} + 3\,Q^2
\end{array}
\ \oplus \
\begin{array}{c}
\text{timelike $\mathcal{N}=1$}\\
\text{super Liouville}\\
\ch = \frac{3}{2} - 3\,\Qh^2
\end{array}
\ \oplus \
\begin{array}{c}
\text{$\mathfrak{b}\mathfrak{c}$-$\beta\gamma$ ghosts}\\
c_\mathrm{gh} = -15
\end{array}
\end{equation}
with $Q^2 - \Qh^2 = 4$ enforcing the vanishing of the total central charge. As mentioned above, coupling the matter SCFT to worldsheet supergravity involves \emph{two} binary choices: the GSO projection, and a relative sign in the definition of the total supercurrent. As a result there are \emph{four} distinct consistent super Virasoro minimal string theories, which we denote by Type \zA{\pm} and \zB{\pm}, following~\cite{Eberhardt:2026hfh}. The \zA{-} and \zB{-} theories reduce to JT supergravity  analyzed in~\cite{Stanford:2019vob} in a semiclassical limit, while the \zA{+} and \zB{+} theories more closely parallel bosonic VMS.\footnote{The 0A theories are defined by a GSO projection that removes physical Ramond-sector vertex operators at generic momenta. However, this convention turns out to allow an exceptional zero-momentum Ramond insertion in the \zA{-} theory, as discussed around (\ref{eq:Rstar}). Our theory labels agree with those of \cite{Eberhardt:2026hfh} for the plus theories, but have the \zA{-} and \zB{-} identifications reversed.  
The rationale is explained in the text, and hinges on the GSO projections and the distinction between generic Ramond-sector vertex operators and Ramond punctures. 
The translation between various conventions is summarized below (here we are referring to the GSO conventions; recall that $\mathcal{N}=1$ JT supergravity is further related to our minus theories by a semiclassical $b\to 0$ limit):
\[
\mathrm{0A}^-_{\mathrm{here}} = \mathrm{0A}_{\mathrm{JT},\text{\tiny{\cite{Stanford:2019vob}}} } 
= \mathrm{0B}^-_{\text{\tiny{\cite{Eberhardt:2026hfh}}}} \,, \qquad 
\mathrm{0B}^-_{\mathrm{here}} = \mathrm{0B}_{\mathrm{JT},\text{\tiny{\cite{Stanford:2019vob}}} } 
= \mathrm{0A}^-_{\text{\tiny{\cite{Eberhardt:2026hfh}}}} \,. 
\]
\label{fn:taxonomy}
}

The direct worldsheet computation of the torus one-point amplitude is the ideal arena in which to employ our technology; in each of the four theories, it involves integrating the super-Virasoro blocks against the super-Liouville structure constants in all four sectors. In particular the odd spin structure must be treated with care: superconformal ghost zero modes require a delicate treatment of the picture changing operator~\cite{Yin:2026str,Sen:2026xfy}, and ultimately necessitate the computation of the torus one-point function of the special level three-half superdescendant operator mentioned above. Carefully assembling these ingredients and evaluating the moduli space integral numerically, we find to a high degree of numerical precision that the resulting amplitudes are strikingly simple functions of the external operator data:  
\begin{equation}\label{eq:torus1ptintro}
\wsamp{1,(1|0)}{\mathrm{0A}^+} = \frac{4\,h_\mathrm{ext} - 1}{48}\,, \qquad
\wsamp{1,(1|0)}{\mathrm{0B}^+} = \frac{4\,h_\mathrm{ext} - 1}{24}\,, \qquad
\wsamp{1,(1|0)}{\mathrm{0A}^-} = -\frac{1}{16}\,, \qquad
\wsamp{1,(1|0)}{\mathrm{0B}^-} = 0\,,
\end{equation}
where $h_\mathrm{ext} = \frac{1}{8}Q^2 + \frac{1}{2}P_\mathrm{ext}^2$ is the weight of the superconformal primary labeling the external vertex operator. The subscript in the amplitudes, $g, (n|m)$, indicates the genus ($g$) followed by the number of NS ($n$) and R ($m$) vertex operator insertions.  These results are in perfect agreement with topological recursion on the spectral curves following from the intersection theory predictions of~\cite{Eberhardt:2026hfh}. The \zA{+} and \zB{+} amplitudes are proportional to the corresponding bosonic quantum volumes; the \zB{-} amplitude vanishes to within numerical precision, consistently with the intersection-theory prediction that all stable NS amplitudes vanish; and perhaps most interestingly, the \zA{-} amplitude, which is computed by topological recursion on a spectral curve with a hard edge~\cite{Eberhardt:2026hfh}, is a momentum-independent constant. 

We also identify an exceptional zero-momentum Ramond vertex operator in the \zA{-} theory, despite the absence of physical Ramond states at generic momentum. The sphere amplitude involving two of these insertions with one NS-sector vertex operator is non-vanishing and in our normalization conventions equals unity. This agrees with an intersection theory prediction \cite{Eberhardt:2026hfh} following from the fact that the corresponding supermoduli space is a point. We hence interpret it as providing a worldsheet candidate for the Ramond ``punctures'' of \cite{Stanford:2019vob}.\footnote{This exceptional Ramond state could be regarded as the worldsheet avatar in $\sVMS$ of spacetime Ramond fluxes that exist in the $\hat{c}=1$ Type 0A theory~\cite{Douglas:2003up,Takayanagi:2003sm,Klebanov:2003wg}.}

\paragraph{Outline:} This paper is organized as follows. We begin with a brief overview of the super Liouville theories in~\cref{sec:superL} which allows us to set our notation and conventions. We then specialize to a discussion of the superconformal theory on the torus in~\cref{sec:torussL}, where we present the torus superconformal blocks, taking the opportunity to address the issues we encountered with their recursive representations. We also present numerical checks of modular covariance of the one-point functions for both the spacelike and timelike theories. 
In~\cref{sec:modularkernel} we turn to a related question: constructing the modular crossing kernels for all spin structure sectors in the regime of central charges relevant for the spacelike theory. 
Given that these kernels implement the modular transformation of the blocks, which we verify numerically, modular covariance of the spacelike super Liouville torus one-point functions is reduced to a set of functional identities among the kernels.
Finally, in~\cref{sec:svms} we turn to the $\sVMS$, taking the opportunity to define the worldsheet theory in a manner consistent with that suggested by the 3d supergravity analysis~\cite{Eberhardt:2026hfh}. The worldsheet torus one-point amplitudes, which we compute directly from the worldsheet using the technology we develop, provide an ideal testing ground for our torus superconformal blocks, and satisfyingly reproduce the matrix model predictions of~\cite{Eberhardt:2026hfh}. We conclude with a brief discussion in~\cref{sec:discuss}. Two short appendices~\labelcref{sec:conventions,sec:wsbblocks} contain useful compilations of conventions and explicit torus amplitude formulae for quick reference.

\section{Preliminaries: the super Liouville theories}\label{sec:superL}

We will be interested in $\mathcal{N} = (1,1)$ SCFTs in $1+1$ dimensions. There are two distinct superalgebras, both of which have Virasoro generators $L_n$ with $n\in \mathbb{Z}$. The distinction lies in the fermionic generators $G_r$;  $r \in \mathbb{Z} + \frac{1}{2}$ for the Neveu-Schwarz (NS) superalgebra and $r \in \mathbb{Z}$ for the Ramond (R) superalgebra. These algebras are realized by the supersymmetric Liouville CFT, whose classical description comprises a bosonic field $\phi$ and its left and right moving fermionic superpartners $\psi$ and $\widetilde{\psi}$. The quantum superconformal field theory exists for any complex central charge.  We will distinguish the generic case where the central charge $c \in \mathbb{C} \backslash (-\infty,\frac{3}{2}]$ from the one where the central charge lies along the ray $c\in (-\infty, \frac{3}{2}]$. The former is  the~\emph{spacelike super Liouville theory}, while the latter is the~\emph{timelike super Liouville theory}. As in the bosonic case, we emphasize that the spacelike and timelike super Liouville CFTs are distinct solutions of the bootstrap equations and are not directly related by analytic continuation.

To begin with we collate some basic bootstrap data for the theories in question. For the spacelike theories this was worked out originally in~\cite{Rashkov:1996np,Poghossian:1996agj}, and further explored in~\cite{Fukuda:2002bv}. The timelike theory was bootstrapped recently in~\cite{Muhlmann:2025ngz,Rangamani:2025wfa}.

\paragraph{Spacelike super Liouville:} We parameterize the theory by $b$ and let
\begin{equation}\label{eq:QbcsL}
Q = b + \frac{1}{b} \,,\qquad
c = \frac{3}{2} + 3\, Q^2\,.
\end{equation}
The spectrum of the theory is continuous, and is parameterized by a Liouville momentum $P \in \mathbb{R}_{\geq 0}$. The primaries are either in the NS sector, where they fit into a full supermultiplet, or in the Ramond sector where they are related by a supercharge zero mode action.

The NS supermultiplet is built atop a superconformal primary operator $V_P$ and is captured by the superfield\footnote{We adopt here a convention that was implicitly used in~\cite{Rangamani:2025wfa}. The action of supercharges is taken to be in a (slightly unconventional) right-left order, i.e., $\widetilde{G} \, G$. This affects the level-$1$ bosonic NS sector descendant. The choice is made to keep structure constants of the spacelike theory unmodified, while ensuring that torus modular crossing is respected.} 
\begin{equation}\label{eq:Snsop}
\SF{S}_P = V_P +  \theta\, \Lambda_P +  \thetab\, \Lambdat_P + \theta\,\thetab\, W_P \,.
\end{equation}
Here $\Lambda_P,\Lambdat_P$ and $W_P$ are superdescendants
\begin{equation}
    \Lambda_P = G_{-\frac12}V_P,\quad \Lambdat_P = \widetilde{G}_{-\frac12}V_P,\quad W_P = \widetilde{G}_{-\frac12} G_{-\frac12}V_P\, .
\end{equation}
The superconformal primary $V_P$ has weights
\begin{equation}\label{eq:sLhp}
\hP= \hPt = \frac{1}{2}\, \pqty{\frac{Q^2}{4} + P^2} \,,
\end{equation}
with the rest of the supermultiplet picking up the additional $\frac{1}{2}$ unit weight from the supercharge raising operators. In the Ramond sector we work with the primary operators $\RsL{\pm}{P}$.
The operators are eigenstates of $(-1)^{F_s}$, with $F_s$ being the fermion number of the spacelike Liouville SCFT (with the superscript indicating their parity), and have conformal weight
\begin{equation}
\hP + \frac{1}{16}, \quad \hPt + \frac{1}{16}\,.
\end{equation}
For reference, we also record here both the chiral decomposition used for these operators in~\cite{Fukuda:2002bv}, and the semiclassical picture in terms of the spin ($\sigma$) and disorder ($\mu$) operators dressed by a Liouville exponential 
\begin{equation}\label{eq:Srop}
\begin{aligned}
\RsL{+}{P}
 & =
\frac{e^{-i\,\frac{\pi}{4}}}{\sqrt{2}}\, \pqty{\Theta^{++}_P +\Theta^{--}_P}
\sim \sigma\, e^{\pqty{\frac{Q}{2} \pm i\,P }\, \phi} \,,
\\
\RsL{-}{P}
 & =
\frac{1}{\sqrt{2}}\,\pqty{\Theta^{+-}_P + \Theta^{-+}_P }
\sim  \mu\, e^{\pqty{\frac{Q}{2} \pm i\,P }\, \phi} \,.
\end{aligned}
\end{equation}
It is also useful to record the conventions we want for the supercharge action on the Ramond vertex operators. Following the conventions in~\cite{Muhlmann:2025ngz,Rangamani:2025wfa}  we have
\begin{equation}
\begin{aligned}
G_0\, R^{\epsilon}_P
 & =
\frac{P}{\sqrt{2}}\, e^{- i\, \frac{\pi}{4}\,\epsilon}\, R^{-\epsilon}_P \,,
\\
\widetilde{G}_0\, R^{\epsilon}_P
 & =
-\frac{P}{\sqrt{2}}\, e^{i\frac{\pi}{4}\,\epsilon}\, R^{-\epsilon}_P \,.
\end{aligned}
\label{eq:Rzeromodeaction}
\end{equation}

The vertex operators are normalized such that the two-point functions of the NS superprimary and the Ramond primaries are (NB: $\epsilon, \epsilon' \in \{+,-\}$)
\begin{equation}\label{eq:nsr2ptnorm}
\begin{aligned}
\expval{V_{P_1}(0)\, V_{P_2}(1)}
 & =
\frac{1}{\rhoNS(P_1)}\,
\bqty{
\delta(P_1  - P_2)  +
\delta(P_1  + P_2)}\,,
\\
\expval{\RsL{\epsilon}{P_1}(0)\, \RsL{\epsilon'}{P_2}(1)}
 & =
\frac{1}{2\,\rhoR(P_1)}
\bqty{\delta(P_1 -P_2) + \epsilon\, \delta(P_1 + P_2)}\,
\delta_{\epsilon \epsilon'}\,.
\end{aligned}
\end{equation}
The spectral densities are in turn
\begin{equation}\label{eq:rhons}
\rhoNS(P) = 4\, \sinh(\pi\,b\, P)\, \sinh(\pi\,b^{-1}\, P)\,,
\qquad
\rhoR(P) = 2\sqrt{2}\, \cosh(\pi\,b\, P)\, \cosh(\pi\,b^{-1}\, P)\,.
\end{equation}
The operators $V_P$, $W_P$ and $\RsL{+}{P}$ are even under $P \to -P$ but $\RsL{-}{P}$ is odd. 

There are four basic structure constants characterizing  the theory. Two correspond to three-point functions of the NS primaries, in terms of which any nonvanishing NS sector 3-point function can be obtained by using the superconformal Ward identities. Two others are mixed NS-R-R correlators, involving the superconformal primary $V_P$. Again the other mixed NS-R-R correlator can be obtained using the action of the supercharges. A concise summary can be found in~\cite[Appendix C]{Rangamani:2025wfa}. Thus, the basic data of interest is captured by four correlation functions
\begin{equation}\label{eq:CVWdef}
\begin{aligned}
\expval{V_{P_1}(0)\, V_{P_2}(1)\, V_{P_3}(\infty)}
 & =
\CV{b}(P_1,P_2,P_3)\,,
\\
\expval{W_{P_1}(0)\, V_{P_2}(1)\, V_{P_3}(\infty)}
 & =
\CW{b}(P_1, P_2,P_3)\,,
\\
\expval{V_{P_1}(0)\, \RsL{+}{P_2}(1)\, \RsL{+}{P_3}(\infty)}
 & =
\frac{1}{2} \,
\pqty{\Ce{b}(P_1;P_2,P_3) + \Co{b}(P_1;P_2,P_3) } \,,
\\
\expval{V_{P_1}(0)\, \RsL{-}{P_2}(1)\, \RsL{-}{P_3}(\infty)}
 & =
\frac{1}{2} \,
\pqty{\Ce{b}(P_1;P_2,P_3) - \Co{b}(P_1;P_2,P_3) }\,.
\end{aligned}
\end{equation}
For the mixed correlators, we have demarcated the momentum argument of the Ramond vertex operators. However, we do not distinguish the location of the superdescendant $W_P$ in $\CW{b}$ --- the result will turn out to be symmetric in the three arguments.

The explicit form of the structure constants with a reflection symmetric normalization is
\begin{equation}\label{eq:sLCVandW}
\begin{aligned}
\CV{b}(P_1,P_2,P_3)
 & =
\frac{\dGNS{b}{2\,Q}}{2\, \dGNS{b}{Q}^3}\,
\frac{\dGNS{b}{\frac{Q}{2} \pm i\,P_1 \pm i\, P_2 \pm i\, P_3}}{\Prod{k=1}{3}\,
\dGNS{b}{Q \pm 2\,i\, P_k}} \,,
\\
\CW{b}(P_1,P_2,P_3)
 & =
i\,\frac{\dGNS{b}{2\,Q}}{\dGNS{b}{Q}^3}\,
\frac{\dGR{b}{\frac{Q}{2} \pm i\,P_1 \pm i\,P_2 \pm i\,P_3}}{\Prod{k=1}{3}\,
\dGNS{b}{Q \pm 2\,i\, P_k}}\,.
\end{aligned}
\end{equation}
Here we have adopted the standard notation for the product with sign choices, viz.,
$f(a \pm y+ c) \equiv \prod_{\eta = \pm}\, f(a + \eta\, y + c)$.
In the Ramond sector,  non-vanishing three-point functions  are determined by
the functions $\Ce{b}$ and $\Co{b}$ which are\footnote{We note in passing a factor of $\sqrt{2}$ difference in normalization compared to the conventions used in~\cite{Rangamani:2025wfa}; it is consistent with the conventions of~\cite{Muhlmann:2025ngz}.}
\begin{equation}\label{eq:sLCeo}
\begin{aligned}
\Ce{b}(P_1;P_2,P_3)
 & =
\frac{\dGNS{b}{2\,Q}}{\sqrt{2}\, \dGNS{b}{Q}^3}\,
\frac{\dGR{b}{\frac{Q}{2} \pm  i\,(P_1+P_2+P_3)}\, \dGR{b}{\frac{Q}{2} \pm
    i\,(P_1 - P_2-P_3)}}{\dGNS{b}{Q \pm 2\, i\,P_1}} \\
 & \qquad \qquad \times
\frac{\dGNS{b}{\frac{Q}{2} \pm i \, (P_1-P_2 + P_3)}\,
\dGNS{b}{\frac{Q}{2} \pm i\,(P_1 + P_2 - P_3)}}{\dGR{b}{Q\pm 2\,i\, P_2}\,
\dGR{b}{Q\pm 2\, i\,P_3}}\,, \\
\Co{b}(P_1;P_2,P_3)
 & =
\frac{\dGNS{b}{2\,Q}}{\sqrt{2}\, \dGNS{b}{Q}^3}\,\frac{
    \dGNS{b}{\frac{Q}{2} \pm
    i\,(P_1+P_2+P_3)}\, \dGNS{b}{\frac{Q}{2} \pm
    i\, (P_1 - P_2- P_3)}}{\dGNS{b}{Q \pm 2\, i\,P_1}} \\
 & \qquad \qquad  \times
\frac{\dGR{b}{\frac{Q}{2} \pm  i\,(P_1-P_2 + P_3)}\,
\dGR{b}{\frac{Q}{2} \pm i\,(P_1 + P_2 - P_3)}}{\dGR{b}{Q\pm 2\,i\,P_2}\,
\dGR{b}{Q \pm2\, i\,P_3}}\,.
\end{aligned}
\end{equation}

We have given the structure constants in terms of NS and R double-Gamma functions.
These are in turn defined as combinations of the bare Barnes
double-Gamma functions
\begin{equation}
\begin{aligned}
\dGNS{b}{z}
 & =
\dGb{\frac{z}{2}}\, \dGb{\frac{z+Q}{2}}\,, 
\\
\dGR{b}{z}
 & =
\dGb{\frac{z+b}{2}}\, \dGb{\frac{z+b^{-1}}{2}} \,.
\end{aligned}
\label{eq:GammaNSR}
\end{equation}
The reader can find useful properties of these functions summarized in~\cite{Muhlmann:2025ngz,Rangamani:2025wfa}. A nice compendium of results regarding the Barnes double-Gamma functions themselves can be found in~\cite{Eberhardt:2023mrq}.

\paragraph{Timelike super Liouville:} The timelike theory with $\ch \leq \frac{3}{2}$ is characterized by a similar set of primary operators. The key distinction is in the range of the parameters. The theory is defined by a parameter $\bh\in \mathbb{R}_+$, with
\begin{equation}
\Qh = \bh^{-1}- \bh\,, \qquad \ch = \frac{3}{2}- 3\,\Qh^2 \ \leq \ \frac{3}{2}\,.
\end{equation}
A priori the primaries are labeled by momenta $\Ph \in \mathbb{R}$. However, when we decompose correlators using conformal blocks the contour of integration over the momenta of internal primaries must be shifted off the real $\Ph$ axis~\cite{Ribault:2015sxa}.
The NS supermultiplet is
\begin{equation}\label{eq:Tnsop}
\SF{T}_{\Ph}
= \Vh_{\Ph} +  \theta\, \Lambdah_{\Ph} +  \thetab\, \Lambdaht_{\Ph}
+ \theta \,\thetab\, \Wh_{\Ph} \,,
\end{equation}
and the Ramond operators (eigenstates of $(-1)^{F_t}$, where $F_t$ is the fermion number of the timelike Liouville SCFT) are $\RtL{\pm}{\Ph}$. Notice that we are again setting 
$\Wh_\Ph = \widetilde{G}_{-\frac{1}{2}} G_{-\frac{1}{2}}\, \Vh_{\Ph}$. 
The conformal weights are parameterized in terms of  $\Ph$ as
\begin{equation}\label{eq:tLhp}
\hhP= \hhPt = \frac{1}{2}\, \pqty{-\frac{\Qh^2}{4} + \Ph^2} \,.
\end{equation}
These are the weights of the superprimary $\Vh_{\Ph}$. The  remaining operators in the
NS supermultiplet have weights shifted up by $\frac{1}{2}$ by the action of the NS supercharges. The Ramond operators are scalars with an additional $\frac{1}{16}$
weight contribution from the Ramond ground state energy. As in the spacelike theory, the NS sector operators and $\RtL{+}{\Ph}$ are even under $\Ph \to -\Ph$ but $\RtL{-}{\Ph}$ is odd.

The structure constants of this theory were determined in~\cite{Muhlmann:2025ngz,Rangamani:2025wfa}. They are given in terms of those of the spacelike theory with
suitable identification of parameters. Let the timelike structure constants be defined
in terms of the correlation functions as
\begin{equation}\label{eq:CtLVeoWdef}
\begin{aligned}
\expval{\Vh_{\Ph_1}(0)\, \Vh_{\Ph_2}(1)\, \Vh_{\Ph_3}(\infty)}
   & =
\CVh{\bh}(\Ph_1,\Ph_2,\Ph_3)\,,
\\
\expval{\Wh_{\Ph_1}(0)\, \Vh_{\Ph_2}(1)\, \Vh_{\Ph_3}(\infty)}
   & =
\CWh{\bh}(\Ph_1,\Ph_2,\Ph_3)\,,
\\
\expval{\Vh_{\Ph_1}(0)\, \RtL{+}{\Ph_2}(1)\, \RtL{+}{\Ph_3}(\infty)}
   & =
\frac{1}{2} \,
\pqty{\Ceh{\bh}(\Ph_1;\Ph_2,\Ph_3) + \Coh{\bh}(\Ph_1;\Ph_2,\Ph_3) }\,,
\\
\expval{\Vh_{\Ph_1}(0)\, \RtL{-}{\Ph_2}(1)\, \RtL{-}{\Ph_3}(\infty)}
   & =
\frac{1}{2} \,
\pqty{\Ceh{\bh}(\Ph_1;\Ph_2,\Ph_3) - \Coh{\bh}(\Ph_1;\Ph_2,\Ph_3) }\,.
\end{aligned}
\end{equation}
The functions appearing on the r.h.s.~can be deduced from their spacelike counterparts
through the relations\footnote{In~\cite{Rangamani:2025wfa} it was noted that the shift relations determine $\CWh{\bh}(\Ph_1,\Ph_2,\Ph_3)$ and $\Coh{\bh}(\Ph_1;\Ph_2,\Ph_3)$ up to a sign (denoted $\eta_{_\mathrm{W}}$  and $\eta_{_\mathrm{R}}$ in the aforementioned reference). A similar ambiguity is present in the spacelike theory. For consistency with our right-left supercharge convention we have taken $\eta_{\mathrm{W}} = -1$ and included the sign explicitly in~\eqref{eq:tLCs}. 
This choice is important to get correct both for modular invariance, and for the worldsheet string construction where this sign choice correlates with  choices for the left moving worldsheet supercharge.\label{fn:CWoddsigns}}
\begin{equation}\label{eq:tLCs}
\begin{aligned}
\CVh{\bh}(\Ph_1,\Ph_2,\Ph_3)
   & =
\frac{2i}{\CW{\bh}(i\,\Ph_1,i\,\Ph_2, i\,\Ph_3)}  \,,
\\
\CWh{\bh}(\Ph_1,\Ph_2,\Ph_3)
   & =
-\frac{2i}{\CV{\bh}(i\,\Ph_1,i\,\Ph_2, i\,\Ph_3)} \,,
\\
\Ceh{\bh}(\Ph_1;\Ph_2,\Ph_3)
   & =
\frac{1}{\Co{\bh}(i\,\Ph_1 ; i\,\Ph_2, i\,\Ph_3)} \,,
\\
\Coh{\bh}(\Ph_1;\Ph_2,\Ph_3)
   & =
\frac{1}{\Ce{\bh}(i\,\Ph_1;i\,\Ph_2, i\,\Ph_3)}\,.
\end{aligned}
\end{equation}
Finally, note that the spectral densities of the timelike theory are given by
\begin{equation}\label{eq:tLrho}
\varrho_{\mathrm{NS}}^{(\mathsf{b})}(\mathsf{P}) = \frac{(i\mathsf{P})^2}{2\rho_{\mathrm{NS}}^{(\mathsf{b})}(i\mathsf{P})}, \qquad \varrho_{\mathrm{R}}^{(\mathsf{b})}(\mathsf{P}) = \frac{1}{2\rho_{\mathrm{R}}^{(\mathsf{b})}(i\mathsf{P})}\, .
\end{equation}

The form of the structure constants was determined by solving the shift relations arising from the null state decoupling equation of a degenerate four-point function. These relations do not, however, fix the overall normalization of the structure constants. In fact, owing to this, it was noticed in~\cite{Muhlmann:2025ngz,Rangamani:2025wfa} that one can introduce a relative sign in the definition of $\CWh{\bh}$ and $\Coh{\bh}$ as noted in~\cref{fn:CWoddsigns}. As noted there we made a choice for $\CWh{\bh}$ that is perhaps unexpected. One might have expected that from a  CFT perspective there is, per se, no intrinsic significance to the sign. This is untrue; the choice of sign is constrained by modular covariance of torus one-point amplitudes. Our conventions are chosen to keep the structure constants as close as possible to the previous literature; hence $W_P$ and $\Wh_{\Ph}$ differ from standard definitions by an overall sign. When we couple super Liouville theories (and perhaps other matter) to two-dimensional $\mathcal{N} = (1,1)$ supergravity to construct a worldsheet string theory, there will be an additional sign choice to be made in the definition of the total matter supercharge. 

\paragraph{Genus-$0$ bootstrap:}  Before we proceed to our main line of development, viz.,
analyzing the correlation function on the torus, we should explain what is known for
sphere correlators. Recursive representations of the conformal blocks for the $\mathcal{N} =1$ superconformal algebra
have been determined in a series of papers~\cite{Hadasz:2007nt,Belavin:2007gz,Suchanek:2010kq}. The last of these checked 4-point crossing explicitly numerically
for the spacelike super Liouville theory. For the NS sector blocks a clear summary and applications to the Type 0B non-critical string were discussed in~\cite{Balthazar:2022atu}.
Crossing symmetry has also been numerically checked for the timelike theory~\cite{Rangamani:2025wfa}, albeit only for the NS sector  structure constants. There is preliminary evidence that with the stated conventions crossing holds for all factorization channels for sphere 4-point functions with both NS and Ramond operators in both spacelike and timelike theories~\cite{Collier:2026wip}.

\section{Super Liouville theory on the torus}\label{sec:torussL}

One of our primary interests is to examine $\mathcal{N} =1$ SCFTs on the torus.
To this end we will explore the one-point functions of various operators. In order to do so,
we have to recognize that observables of a SCFT depend on the choice of boundary
conditions along the two cycles of the torus. A torus correlator has a Hilbert space trace representation once we pick a choice of space and time directions.
Having done so, we can compute the trace of the operator with either NS or R boundary conditions along the spatial circle of the torus, and additionally choose to weigh the temporal direction with $(-1)^{F_\mathrm{s}}$. 
So we have four traces, which are traditionally characterized as the
NS, $\widetilde{\mathrm{NS}}$, R, and $\widetilde{\mathrm{R}}$ sectors, respectively.

We shall find it convenient to label these sectors by $a \in \set{3,4,2,1}$
in the order specified, i.e., $\set{\mathrm{NS}, \widetilde{\mathrm{NS}}, \mathrm{R}, \widetilde{\mathrm{R}}}$, respectively. The label choice has a simple mnemonic ---
the trace with the identity operator leads to the superconformal character,
which is a theta function of index $a$, $\vartheta_a$ (we use the notation of~\cite{Polchinski:1998rq}). Of the four sectors, the three corresponding to the even spin structure ($a\in \{2,3,4\}$) mix under modular transformations, while the fourth $a=1$, the odd spin structure, does not mix with the others. We shall see some interesting consequences of this in our analysis, as we will discover a new torus block corresponding to the one-point function of a special superdescendant that itself transforms covariantly under modular transformations.

Fortunately for us, torus superconformal blocks relevant for genus-$1$ correlators have
already been analyzed in~\cite{Hadasz:2012im}, who decomposed the one-point functions.
We will explain the salient features of their results, adapting them to our notation.
Along the way we take the opportunity to  fix some minor issues with their expressions for the recursive representations of the blocks. As we describe in the course of our analysis, modular covariance of the torus 1-point function provides a strong constraint. In fact, the issues we describe below surfaced in the course of numerically checking modular crossing. Subsequently, in~\cref{sec:modularkernel} we will also construct modular $S$-kernels for the blocks.

\subsection{Torus one-point functions}\label{sec:torus1pt}

Among the components of the NS supermultiplet, $\Lambda_P$ and $\widetilde{\Lambda}_P$ have vanishing torus one-point functions.
Furthermore, Ramond operators must appear in pairs. Hence, insofar as conformal primaries are concerned we need only consider the torus one-point functions of $V_P$ and $W_P$.\footnote{In the following, we will phrase the discussion in the language of the spacelike theory for concreteness; the story for the timelike case is entirely analogous.}
Let us denote the one-point functions of the superprimary by the following notation:\footnote{Below, recall that $(-1)^{F_s}$ keeps track of the fermion number in the spacelike Liouville CFT. The correlators in the timelike theory are similarly defined with $(-1)^{F_t}$.}
\begin{equation}
\begin{aligned}
\expval{V_P}_3
   & =
\Tr_{\mathrm{NS}} \pqty{V_P(0)\, q^{L_0 - \frac{c}{24}}\,
    \overline{q}^{\widetilde{L}_0 - \frac{c}{24}}} \,,
\\
\expval{V_P}_4
   & =
\Tr_{\mathrm{NS}}\pqty{ (-1)^{F_s} \, V_P(0)\, q^{L_0 - \frac{c}{24}}\,  \overline{q}^{\widetilde{L}_0 - \frac{c}{24}}} \,,
\\
\expval{V_P}_2
   & =
\Tr_{\mathrm{R}} \pqty{V_P(0)\, q^{L_0 - \frac{c}{24}}\,  \overline{q}^{\widetilde{L}_0 - \frac{c}{24}}}\,,
\\
\expval{V_P}_1
   & =
\Tr_{\mathrm{R}}\pqty{ (-1)^{F_s}\, V_P(0)\, q^{L_0 - \frac{c}{24}}\,  \overline{q}^{\widetilde{L}_0 - \frac{c}{24}}} \,.
\end{aligned}
\end{equation}
The one-point functions of the superdescendant $\expval{W_P}_a$ are similarly defined.
We set $\tau$ to be the modular/complex structure parameter of the torus,
and let $q = e^{2\pi i\, \tau}$ be the corresponding elliptic nome.

\paragraph{A distinguished superdescendant:}  Somewhat curiously, we will also have reason to consider the following level three-half 
descendant operator's expectation value:
\begin{equation}\label{eq:GVdef}
\expval{DV_P}_1 \equiv
\Tr_{\mathrm{R}}\pqty{ (-1)^{F_s}\, \widetilde{G}_{-\frac{3}{2}} G_{-\frac{3}{2}}
    V_P(0)\, q^{L_0 - \frac{c}{24}}\,  \overline{q}^{\widetilde{L}_0 - \frac{c}{24}}} \,.
\end{equation}
We have retained our convention for the right-left order of supercharge action above. 
A priori from a SCFT perspective it is unclear why we distinguish this particular superdescendant correlator. First, let us note that although $\widetilde{G}_{-\frac{3}{2}}G_{-\frac{3}{2}}V$ is not itself a Virasoro primary, its torus one-point function acts like that of a primary. The reason for this is the following. The superconformal descendant operator
\begin{equation}\label{eq:level-3/2-primary}
\mathcal{O}_V\equiv \left(\widetilde{G}_{-\frac{3}{2}}-\frac{1}{\hPt+\frac{1}{2}}\,\widetilde{L}_{-1}\, \widetilde{G}_{-\frac{1}{2}}\right)
\left(G_{-\frac{3}{2}}-\frac{1}{\hP+\frac{1}{2}}\, L_{-1}\, G_{-\frac{1}{2}}\right)
V_P
\end{equation}
\emph{is} a Virasoro primary with conformal weights $(\hP+\frac{3}{2},\hPt+\frac{3}{2})$. However,  the contribution of the $\widetilde{L}_{-1}$ and $L_{-1}$ terms to the torus one-point function in any spin sector vanishes due to translation invariance of the torus one-point function. Combining with the anti-holomorphic sector, we thus have
\begin{equation}\label{eq:GVcorra}
\expval{\mathcal{O}_V}_a = 
\expval{\widetilde{G}_{-\frac{3}{2}} G_{-\frac{3}{2}}V_P}_a\, .
\end{equation}
Thus, the torus one-point function of $\widetilde{G}_{-\frac{3}{2}} G_{-\frac{3}{2}}V_P$ behaves like that of a primary.

There is an additional constraint to consider. The torus one-point functions of this superdescendant operator vanish in all but the $a=1$ sector, which is only nonzero due to the fermion zero mode on the torus in the odd spin structure. 
This may be seen from a superconformal Ward identity. Writing $U \equiv \widetilde{G}_{-\frac{3}{2}}V_P$, consider the correlator $\expval{G(z)U(0)}_a$ in an even spin structure. Since $V_P$ is an NS superconformal primary, the operator product of the supercurrent with $U$ is
\begin{equation}
    G(z)U(0) = \frac{1}{z}G_{-\frac{1}{2}}U(0) + G_{-\frac{3}{2}}U(0) + \mathcal{O}(z)\,.
\end{equation}
The correlator therefore has at most a simple pole at the origin. It moreover has the same periodicity properties as the Szeg\H{o} kernel, which is given by
\begin{equation}
S_a(z \mid \tau) = \frac{\vartheta_a(z\mid \tau)\, \partial_z\vartheta_1(0\mid \tau)}{\vartheta_a(0\mid \tau)\, \vartheta_1(z\mid \tau)}\,, \qquad a \in \set{2,3,4}\,.
\end{equation}
Subtracting the appropriate multiple of the Szeg\H{o} kernel from the correlator thus leaves a non-singular holomorphic contribution. Such a nontrivial remainder would correspond to a fermion zeromode, which is absent in the even spin structures. Thus the correlator is proportional to the Szeg\H{o} kernel itself. Since this kernel is odd in $z$, its Laurent expansion contains no constant term, and thus $\expval{\widetilde{G}_{-\frac{3}{2}} G_{-\frac{3}{2}}V_P}_a = 0$ for $a=2,3,4$.

The story for the odd spin structure is different,  since now we have fermion zero modes.
Indeed, we will see that, in the $a=1$ sector, the superdescendant correlator $\langle DV_P\rangle_1$ admits a standard superconformal block decomposition. Moreover, the superconformal blocks themselves are computed by a structurally identical recursion relation to those of $V$ and $W$, and are branched into the dual channel with a modular crossing kernel that is a close cousin of the more standard modular kernels. Most importantly, we will also see that this operator arises naturally in the computation of the torus one-point amplitude in the $\mathcal{N}=1$ super Virasoro minimal string, which was our original motivation for investigating it.

\paragraph{The conformal block decomposition:}
The main effort of~\cite{Hadasz:2012im} was obtaining the decomposition of
these expectation values in terms of the three-point structure constants,
and the elliptic torus superconformal blocks.
We can write the one-point functions separately in each of the four sectors
for the operators of interest.

For the NS and $\widetilde{\mathrm{NS}}$ sectors (hence $a=3$, $4$)
the correlators are given by
\begin{equation}
\begin{aligned}
\expval{V_{P_\mathrm{ext}}}_a
   & =
\int_0^\infty \, dP \, \rhoNS(P)\, \CV{b}(P_{\mathrm{ext}},P,P)\,
\norm{\mathcal{F}^{\mathrm{V}}_a(P_\mathrm{ext};P,P \mid q) }^2\,,
\\
\expval{W_{P_\mathrm{ext}}}_a
   & =
\int_0^\infty \, dP \, \rhoNS(P)\, \CW{b}(P_\mathrm{ext},P,P)\,
\norm{\mathcal{F}^{\mathrm{W}}_a(P_\mathrm{ext};P,P \mid q)}^2 \,.
\end{aligned}
\label{eq:1ptNS}
\end{equation}
Likewise, in the R and $\widetilde{\mathrm{R}}$ sectors ($a=2$ and $a=1$, respectively) the expectation values are
\begin{equation}
\begin{aligned}
\expval{V_{P_\mathrm{ext}}}_2
   & =
2 \, \int_0^\infty dP \, \rhoR(P) \,\Ce{b}(P_\mathrm{ext};P,P)\,
\norm{\mathcal{F}_2^{\mathrm{V}}(P_\mathrm{ext};P,P \mid q)}^2 \,,
\\
\expval{V_{P_\mathrm{ext}}}_1
   & =
2 \, \int_0^\infty dP \, \rhoR(P) \,\Co{b}(P_\mathrm{ext};P,P)
\norm{\mathcal{F}_1^{\mathrm{V}}(P_\mathrm{ext};P,P \mid q)}^2 \,,
\\
\expval{W_{P_\mathrm{ext}}}_2
   & =
2 \, \int_0^\infty dP  \, \rhoR(P) \, \Co{b}(P_\mathrm{ext};P,P)\,
\norm{\mathcal{F}_2^{\mathrm{W}}(P_\mathrm{ext};P,P \mid q)}^2\,,
\\
\expval{W_{P_\mathrm{ext}}}_1
   & =
0\,.
\end{aligned}
\label{eq:1ptR}
\end{equation}
As promised, the presence of fermionic zero modes in the $\widetilde{\mathrm{R}}$ sector causes the superdescendant to have vanishing expectation value.
The expectation value of the superconformal primary is, however, non-vanishing.
Consequently, there is an asymmetry between the even and odd structure constants.
The symmetry is restored when we include the correlator of a conformal descendant
introduced above in~\eqref{eq:GVdef}. We will demonstrate below that in super Liouville CFT\footnote{ 
We emphasize that although this is the conformal block decomposition of the special superdescendant correlator in Liouville SCFT, a well-defined conformal block decomposition for the analogous superdescendant exists in any CFT with $\mathcal{N}=1$ supersymmetry. Indeed,  the recursive representation of the corresponding conformal blocks and the associated crossing kernels are theory-independent. The resulting structure constants can be derived from those of the primaries using superconformal Ward identities in the same way that they are determined below.} 
\begin{equation}\label{eq:GGV1}
\expval{\widetilde{G}_{-\frac{3}{2}} G_{-\frac{3}{2}} V_{P_\mathrm{ext}}}_1
= \, \int_0^\infty dP\, \rhoR(P)\, \Ce{b}(P_{\mathrm{ext}}; P,P)
\norm{\mathcal{F}_1^{\mathrm{DV}}(P_{\mathrm{ext}}; P,P \mid q)}^2 \,.
\end{equation}

The momentum of the operator whose expectation value we are computing is made explicit
in the above formulae as $P_{\mathrm{ext}}$. The internal momentum of the operator
that runs in the loop is $P$.  In writing these expressions we have been somewhat succinct with our notation for combining the holomorphic and antiholomorphic blocks.
The shorthand $\norm{\bullet}^2$ is to be read as follows:  combine the holomorphic and antiholomorphic blocks with elliptic nomes $q$ and $\overline{q}$, respectively, without complex-conjugating the Liouville parameter and momenta. To wit,
\begin{equation}
\norm{\mathcal{F}(P_\mathrm{ext}; P,P \mid q)}^2
\equiv
\mathcal{F}_{b}(P_\mathrm{ext}; P , P  \mid q) \,
\mathcal{F}_{b}(P_\mathrm{ext}; P, P |\overline{q}) \,,
\label{eq:normdef}
\end{equation}
for any of the blocks defined above.\footnote{Notice that the blocks $\mathcal{F}^{\mathrm{DV}}_1$ and $\mathcal{F}^{\mathrm{W}}_2$ contain explicit phases and factors of $P$, arising from a $G_0$ action in the superconformal Ward identity. These conspire between left- and right-movers to produce an overall factor $iP^2$, as we verify below.}

Our next task is to explain how to determine the blocks $\mathcal{F}$ by elliptic recursion, and to rationalize the descendant correlator with the odd spin structure.
We will proceed to elucidate the blocks themselves first, and then tackle how the descendant correlator can be deduced. Having done so we will explain some basic consistency checks, and the constraints imposed by modular covariance, which we finally verify explicitly by numerical computation.

\subsection{Elliptic torus superconformal blocks}\label{sec:torusblocks}

The blocks involved are $\mathcal{F}_a^{\mathrm{V,W}}$ which can be described in
terms of the corresponding elliptic blocks which can be recursively determined~\cite{Hadasz:2012im}. We will parameterize the blocks in our notation and record a translation
to the conventions of the original reference in~\cref{sec:conventions}.

The holomorphic blocks in the NS sector are (NB: $\mathrm{I} \in \set{\mathrm{V}, \mathrm{W}}$),
\begin{equation}
\begin{aligned}
\mathcal{F}^{\mathrm{I}}_3(P_\mathrm{ext};P,P \mid q)
   & =
q^{\frac{P^2}{2}} \, \sqrt{\frac{\vartheta_3(q)}{\eta(q)^3}} \,
\mathcal{H}^{\mathrm{I}}_3(P_\mathrm{ext},P \mid q) \,,
\\
\mathcal{F}^{\mathrm{I}}_4(P_\mathrm{ext};P,P \mid q)
   & =
q^{\frac{P^2}{2}} \, \sqrt{\frac{\vartheta_4(q)}{\eta(q)^3}} \,
\mathcal{H}^{\mathrm{I}}_4(P_\mathrm{ext},P \mid q) \,.
\end{aligned}
\label{eq:FblockNS}
\end{equation}
The elliptic block data is obtained by recursion, and the difference between
$\mathcal{H}_3$ and $\mathcal{H}_4$ is that the recursion coefficients are
weighted in the latter case by the fermion number. In the R sector, there are four
blocks according to~\cite{Hadasz:2012im}. However,
one of these vanishes, which is the origin of the asymmetry mentioned above.
We will introduce a new block, which corresponds to the descendant operator in the odd spin structure.
Including it, the blocks are given by
\begin{equation}
\begin{aligned}
\mathcal{F}^{\mathrm{V}}_2(P_\mathrm{ext};P,P \mid q)
   & =
\frac{1}{\sqrt{2}} \,q^{\frac{P^2}{2}} \, \sqrt{\frac{\vartheta_2(q)}{\eta(q)^3}} \,
\mathcal{H}_\mathrm{even}^{\mathrm{V}}(P_\mathrm{ext},P \mid q) \,,
\\
\mathcal{F}^{\mathrm{V}}_1(P_\mathrm{ext};P,P \mid q)
   & =
q^{\frac{P^2}{2}} \,
\mathcal{H}_{\mathrm{odd}}^{\mathrm{V}}(P_\mathrm{ext},P \mid q)\,,
\\
\mathcal{F}^{\mathrm{DV}}_1(P_{\mathrm{ext}};P,P \mid q)
   & =
e^{i\frac{\pi}{4}}\, P\,  q^{\frac{P^2}{2}} \,
\mathcal{H}_{\mathrm{even}}^{\mathrm{W}}(P_\mathrm{ext},P \mid q)\,,
\\
\mathcal{F}^{\mathrm{W}}_2(P_\mathrm{ext};P,P \mid q)
   & =
e^{i\frac{\pi}{4}}\, P\, q^{\frac{P^2}{2}} \, \sqrt{\frac{\vartheta_2(q)}{\eta(q)^3}} \,
\mathcal{H}_{\mathrm{odd}}^{\mathrm{W}}(P_\mathrm{ext},P \mid q) \,,
\\
\mathcal{F}^{\mathrm{W}}_1(P_\mathrm{ext};P,P \mid q)
   & =
0 \,.
\end{aligned}
\label{eq:FblockR}
\end{equation}
In presenting the blocks, we have adopted a notation that should make clear the association between the blocks and the correlators. As noted above, a translation of our convention to that used in~\cite{Hadasz:2012im} is in~\cref{sec:conventions}. We refer the reader to~\cref{sec:wsbblocks} for explicit decompositions of all the relevant torus correlators in both the spacelike and timelike theories.

\paragraph{The elliptic recursion relations:}  The elliptic blocks are given as a Fourier
series in the elliptic nome $q$, with the series coefficients determined recursively from
a starting seed value. The $\mathcal{H}_a^{\mathrm{I}}$ blocks summarized above can be uniformly given
a series expansion with coefficients $H_a^{\mathrm{I}}$. The powers of $q$ appearing are half-integer
quantized in the NS sector and integral in the R sector. Explicitly, we have
\begin{equation}
\begin{aligned}
\mathcal{H}^{\mathrm{I}}_3(P_\mathrm{ext},P \mid q)
   & =
\sum_{m\in \frac{1}{2}\,\mathbb{Z}_{\geq0}}\,
q^m\, H_m^\mathrm{I}(P_\mathrm{ext},P)\,,
\\
\mathcal{H}^\mathrm{I}_4(P_\mathrm{ext},P \mid q)
   & =
\sum_{m\in \frac{1}{2}\,\mathbb{Z}_{\geq0}}\,
    (-1)^{2m}\,q^m\, H_m^\mathrm{I}(P_\mathrm{ext},P)\,,
\\
\mathcal{H}^{\mathrm{I}}_{\mathrm{even}}(P_\mathrm{ext},P \mid q)
   & =
\sum_{m\in \mathbb{Z}_{\geq0}}\,
q^m\, H^{\mathrm{I, even}}_m(P_\mathrm{ext},P)\,,
\\
\mathcal{H}^{\mathrm{I}}_{\mathrm{odd}}(P_\mathrm{ext},P \mid q)
   & =
\sum_{m\in \mathbb{Z}_{\geq0}}\,
q^m\, H^{\mathrm{I, odd}}_m(P_\mathrm{ext},P)\,.
\end{aligned}
\label{eq:Hblockqexp}
\end{equation}
Parameterized thus, one can give a simple universal expression for the recursion relation that determines the coefficients of the elliptic blocks. We have
\begin{equation}
\begin{aligned}
H_m^A(P_\mathrm{ext},P)
   & =
\delta_{m,0} + \sum_{1 \leq r\,s \leq 2\,m} \,\frac{\Res^A_{r,s}(P_\mathrm{ext})}{h_P - h_{rs}} \, H^A_{m - \frac{r\,s}{2}}(P_\mathrm{ext},P_{r,-s})\,,
\end{aligned}
\label{eq:Hcoeffrecurse}
\end{equation}
where $A$ is a collective index $A \in \set{\mathrm{I},\set{\emptyset, \mathrm{even}, \mathrm{odd}}}$ and the quantization of $m$ is fixed by the block in question. We will give the expression for the residues in each case below. The poles in the conformal weight recursion occur at degenerate representations of the superconformal algebra. Note that the sum has to include the end point values as indicated above explicitly.

\paragraph{Degenerate representations:} To write down the formulae
for the residues, we need some kinematic data. Firstly, degenerate
representations of the $\mathcal{N} =1$ superconformal algebra occur at momenta
\begin{equation}
P_{r,s}  = \frac{i}{2} \pqty{r\,b + s\, b^{-1}} \,,
\qquad  r, s \in \mathbb{Z}_{\geq 1}\,.
\label{eq:Pdegen}
\end{equation}
The degenerate operators appearing at even parity, $r+s \in 2\,\mathbb{Z}_{\geq 1}$, are in
the NS sector, while those at odd parity, $r+s \in 2\, \mathbb{Z}_{\geq 1} +1$, are in the R sector.  In addition, we also have use for the momenta
\begin{equation}
P_{r,-s} = \frac{i}{2}\, (-1)^s \, \pqty{r\,b - s\, b^{-1}} \,.
\end{equation}
The conformal weights for the two sets of momenta are readily ascertained to be
\begin{equation}
\begin{aligned}
h_{r,s}
   & =
\frac{Q^2}{8} + \frac{1}{2}\, P_{r,s}^2 =\frac{1-r^2}{8}\,b^2 +
\frac{1-s^2}{8}\, b^{-2} + \frac{1-r\,s}{4}
\,, \\
h_{r,-s}
   & =
h_{r,s} + \frac{rs}{2} \,.
\end{aligned}
\label{eq:hdegen}
\end{equation}

\paragraph{The Gram matrix inverse coefficients:}
Next, we require the coefficients appearing in the Gram matrix inverse. These were originally conjectured in~\cite{Belavin:2006pv} after explicitly examining low-lying states. We will use the parameterization
\begin{equation}
A_{r,s} =
2^{rs-2} \prod_{m=1-r}^r \, \prod_{n=1-s}^s\,
\frac{1}{m\,b+ n\,  b^{-1}}\,,
\qquad m+n \in 2\,\mathbb{Z}\,, \quad (m,n) \neq (0,0) \;\text{or}\; (r,s)\,
\label{eq:Arsdef}
\end{equation}
which is closely related to the location of the zeros of the Gram matrix. The latter appear at $Z_{r,s}$ defined as
\begin{equation}
Z_{r,s} = 2^{1- rs} \prod_{k,\,\ell \in [r,s]} \, (k\, b + \ell\, b^{-1})
\end{equation}
with the index range of the product constrained by sector as\footnote{Here the notation $(a:b)_2$ is meant to denote the set of integers from $a$ to $b$ proceeding in steps of 2.}
\begin{equation}
\begin{aligned}
[m,n]_{_\mathrm{NS}}
   & =
\{(1-m:m-1)_2 \mid (1-n:n-1)_2\} \cup \{(2-m:m)_2 \mid (2-n:n)_2\} \backslash \{0,0\} \,,
\\
[m,n]_{_\mathrm{R}}
   & =
\{(1-m:m-1)_2 \mid (2-n:n)_2\} \cup \{(2-m:m)_2 \mid (1-n:n-1)_2\} \backslash \{0,0\} \,.
\end{aligned}
\end{equation}
The two sets of data satisfy a simple relation
\begin{equation}\label{eq:BZ-identity}
Z_{r,s} \cdot A_{r,s} = 
\begin{dcases}
\frac{r\, b + s \,b^{-1}}{2} = -i\, P_{r,s}\,, 
&\qquad r + s \ \text{even (NS)}\,,
\\
\frac{1}{2} \,, 
& \qquad r+s \ \text{odd (R)} \,.
\end{dcases}
\end{equation}

\paragraph{Fusion monomials:} The final piece of kinematic data is the fusion monomials.
These are graded by parity, and for succinctness we define the primitives in terms of which
the residues can be conveniently written
\begin{equation}
\begin{aligned}
X_{r,s}^\mathrm{e} (P)
   & =
\mathop{\Prod{p=1-r}{r-1}\; \Prod{q=1-s}{s-1}}_{\mathrm{steps\; of \;2}}\,
\frac{2\, i\, P - p\, b - q\, b^{-1}}{2\,\sqrt{2}}
\,, \qquad p+q+r+s \equiv 2 \; \pmod{4} \,, \\
X_{r,s}^\mathrm{o} (P)
   & =
\mathop{\Prod{p=1-r}{r-1}\; \Prod{q=1-s}{s-1}}_{\mathrm{steps\; of \;2}}\,
\frac{2\, i\, P - p\, b - q\, b^{-1}}{2\,\sqrt{2}}
\,, \qquad p+q+r+s \equiv 0 \; \pmod{4} \,.
\end{aligned}
\label{eq:Xfusion}
\end{equation}

\paragraph{The residues at the degenerate poles:} We now have all the information
to present the expression for the residues for the weight recursion coefficients $H_{m}$
defined in~\eqref{eq:Hcoeffrecurse}.  
The residues in the NS sector are given by 
\begin{equation}\label{eq:HNSresidues}
\begin{aligned}
\Res^\mathrm{V}_{r,s}(P_\mathrm{ext})
   & =
A_{r,s}\,
X_{r,s}^\mathrm{e}(P_{r,s} + P_\mathrm{ext})\,
X_{r,s}^\mathrm{e}(P_{r,s} - P_\mathrm{ext})
\\
   &
\quad \times  \bqty{
    \underbrace{
    X_{r,s}^\mathrm{e}(P_{r,- s} + P_\mathrm{ext})\,
    X_{r,s}^\mathrm{e}(P_{r,-s} - P_\mathrm{ext})}
    _{rs \,\in\, 2\, \mathbb{N}, \; r+s \,\in\, 2\,\mathbb{N}}
    +
    \underbrace{(-1)^{rs}\,
    X_{r,s}^\mathrm{o}(P_{r,-s} + P_\mathrm{ext})\,
    X_{r,s}^\mathrm{o}(P_{r,-s} - P_\mathrm{ext})}
    _{rs \,\in\, 2\,\mathbb{N}+1,\;  r+s \,\in \,2\, \mathbb{N}}
    }\,,
\\
\Res^\mathrm{W}_{r,s}(P_\mathrm{ext})
   & =
A_{r,s}\,
X_{r,s}^\mathrm{o}(P_{r,s} + P_\mathrm{ext})\,
X_{r,s}^\mathrm{o}(P_{r,s} - P_\mathrm{ext})
\\
   &
\quad \times  \bqty{
    \underbrace{
    X_{r,s}^\mathrm{o}(P_{r,- s} + P_\mathrm{ext})\,
    X_{r,s}^\mathrm{o}(P_{r,-s} - P_\mathrm{ext})}
    _{rs \,\in\, 2\, \mathbb{N}, \; r+s \,\in\, 2 \mathbb{N}}
    +
    \underbrace{
    X_{r,s}^\mathrm{e}(P_{r,-s} + P_\mathrm{ext})\,
    X_{r,s}^\mathrm{e}(P_{r,-s} - P_\mathrm{ext})}
    _{rs \,\in\, 2 \mathbb{N}+1,\; r+s \,\in\, 2\,\mathbb{N}}
    }\,,
\end{aligned}
\end{equation}
and the R sector residues are
\begin{equation}\label{eq:HRresidues}
\begin{aligned}
\Res^{\mathrm{I, even}}_{r,s}(P_\mathrm{ext})
   & =
A_{r,s}\, \mathfrak{f}_{r,s}^{\mathrm{I}}\,\underbrace{
    X_{r,s}^\mathrm{e}(P_\mathrm{ext} + P_{r,-s})\,
    X_{r,s}^\mathrm{o}(P_\mathrm{ext} - P_{r,-s}) \,
    X_{r,s}^\mathrm{e}(P_\mathrm{ext} + P_{r,s})\,
    X_{r,s}^\mathrm{o}(P_\mathrm{ext} - P_{r,s})}
    _{r+s \,\in\, 2\, \mathbb{N}+1}\,,
\\
\Res^{\mathrm{I, odd}}_{r,s}(P_\mathrm{ext})
   & =
A_{r,s}\, \mathfrak{f}^{\mathrm{I}}_{r,s}\, \underbrace{
    X_{r,s}^\mathrm{e}(P_\mathrm{ext} - P_{r,-s})\,
    X_{r,s}^\mathrm{o}(P_\mathrm{ext} + P_{r,-s}) \,
    X_{r,s}^\mathrm{e}(P_\mathrm{ext} - P_{r,s})\,
    X_{r,s}^\mathrm{o}(P_\mathrm{ext} + P_{r,s})}
    _{r+s \,\in\, 2\,\mathbb{N}+1}\,.
\end{aligned}
\end{equation}
In the Ramond blocks the Gram matrix coefficient $A_{r,s}$ is dressed by a
factor $\mathfrak{f}^{\mathrm{I}}_{r,s}$ which differs for the two operators. Explicitly,
\begin{equation}
\mathfrak{f}^{\mathrm{I}}_{r,s} =
\begin{dcases}
(r\, b + s\, b^{-1})\,,       & \qquad \mathrm{I} = \mathrm{V} \,, \\
(-1)^s\, (r\, b - s\, b^{-1}) & \qquad \mathrm{I} = \mathrm{W}\,.
\end{dcases}
\end{equation}
These are just the rescaled degenerate momenta, whose presence in the residues has a ready explanation, as we elucidate below.  These formulae complete the specification of the recursive method to determine the superconformal blocks.

The expressions above differ in some minor details from those reported in~\cite{Hadasz:2012im}. To be clear, we have not derived the blocks from first principles as they
did, but rather used the expressions provided therein, and attempted to check modular covariance. In the process we discovered a few issues, which we record here for the benefit of the reader.

\paragraph{1. Ramond block asymptotics:}  The expression for the Ramond block $\mathcal{F}_1^{\mathrm{V}}$  reported in~\eqref{eq:FblockR} has a factor of $q^{P^2/2}$, which is absent in the aforementioned reference, cf.~\cite[Eq (4.18)]{Hadasz:2012im}. The factor was argued to be unity from the asymptotics for the block in the odd spin structure sector. There is no
obvious character available, since the trace in the $\widetilde{\mathrm{R}}$ sector vanishes owing to fermion zero modes. However, one expects that the weight of the operator still contributes. In fact, in the absence of the factor of $q^{P^2/2}$ the block is not well-behaved at the cusp, viz., as $\Im(\tau) \to \infty$. Relatedly, this factor turns out to be necessary for the modular covariance of $\expval{V_{P_{\mathrm{ext}}}}$.

\paragraph{2. Momentum factors in the Ramond sector blocks:} Notice that the Gram matrix coefficients are multiplied by a factor proportional to the degenerate momenta $P_{r,\pm s}$
in the residue formula. These factors can be motivated directly from the action of the Ramond zero mode, $G_0$, which involves  a momentum factor, cf.~\eqref{eq:Rzeromodeaction}.

For concreteness, let us derive the momentum prefactor for the blocks involving the $W$ operator. Looking back at the defining equation~\eqref{eq:FblockR},  we notice that
$\mathcal{F}_2^{\mathrm{W}} \propto P\, \mathcal{H}_{\mathrm{odd}}^{\mathrm{W}}$. Now we expand both sides in a $q$-series. For the r.h.s.~we can use~\eqref{eq:Hblockqexp} and expand the character contribution in the standard fashion, say
\begin{equation}\label{eq:theta2exp}
\sqrt{\frac{\vartheta_2(q)}{\eta(q)^3}} = \sqrt{2}\, \sum_{n=0}^\infty\,
p_2(n)\, q^n\,.
\end{equation}
For the l.h.s.~let $F_{m}^{2,\mathrm{W}}$ be the corresponding Fourier coefficients. Then from the definition we infer that the recursion relation implies
\begin{equation}
F_m^{2,\mathrm{W}} = \sqrt{2}\, e^{i\frac{\pi}{4}}\, P\,
\sum_{m_1 + m_2 = m}\,
p_2(m_1)\, H_{m_2}^{\mathrm{W,odd}}.
\end{equation}
Using the thus obtained expression for $F_m^{2,\mathrm{W}}$ in the recursion formula for $\mathcal{F}_2^{\mathrm{W}}$ we ascertain the recursion for the coefficients $H_m^{\mathrm{W, odd}}$. The prefactor $P$ in the above expression
ends up implying that each recursion coefficient is determined in terms of the
preceding ones up to multiplication by a factor of $P_{r,-s}/P$. Taking this into account we arrive at our expressions for the residues.

For the $V$ blocks the factor $P_{r,s}$ arises similarly. However, in this case there is a slightly more elementary way to motivate it directly from the expressions for the Gram matrix zeros obtained in~\cite{Belavin:2007gz}. Notice that the indices of the products $m,n$ have to  sum to an even integer. Since in the Ramond sector the degenerate operators arise at odd $r+s \in 2 \mathbb{Z} +1$, the endpoint contribution need not be excluded from the product. This immediately realizes the necessary factor. Since we expressed the result for $A_{r,s}$ in a sector agnostic way, we have to reinstate it in the residues. The expressions in~\cite{Hadasz:2012im} for the residue appear to miss this, and so their Ramond blocks are missing the coefficient $\mathfrak{f}^{\mathrm{I}}_{r,s}$, which we claim is required for consistency with the Ward identities. We will also see that its inclusion is consistent with modular covariance of the torus one-point functions in both spacelike and timelike super Liouville CFT.

\paragraph{3. Fusion monomial argument ordering:}
We have made explicit the order of the arguments in the fusion monomials. The residues require the ordering $P_{r,\pm s} \pm P_\mathrm{ext}$ (degenerate momentum first), which matches the convention of~\cite{Hadasz:2012im}. With this ordering the signs of the recursion coefficients differ from the explicit expressions given in the aforementioned reference; they have the effect of swapping $H_3^{\mathrm{I}}$ and $H_4^{\mathrm{I}}$ in the NS sector. One can check this explicitly:
\begin{equation}
\begin{aligned}
X_{1,1}^\mathrm{e}(P_{1,1} + P) \, X_{1,1}^\mathrm{e}(P_{1,1} -P)
   & =
h_P \,,
\\
X_{1,1}^\mathrm{e}(P_{1,-1} + P) \, X_{1,1}^\mathrm{e}(P_{1,-1} -P)
   & =
h_P - \frac{1}{2}\,,
\\
X_{1,1}^\mathrm{o}(P_{1,1} + P) \, X_{1,1}^\mathrm{o}(P_{1,1} -P)
   & =
X_{1,1}^\mathrm{o}(P_{1,-1} + P) \, X_{1,1}^\mathrm{o}(P_{1,-1} -P) = 1 \,.
\end{aligned}
\end{equation}
Then using the expressions for the Gram matrix coefficients one deduces that
\begin{equation}
H^{\mathrm{V}}_{\frac{1}{2}}
= - \frac{h_{P_\mathrm{ext}}}{2\,h_P} \,,
\qquad
H^{\mathrm{W}}_{\frac{1}{2}}
=  \frac{2\,h_{P_\mathrm{ext}}-1}{4\,h_P} \,.
\end{equation}
The sign in the first equation comes from
the $(-1)^{rs}$ which is activated in the $\mathrm{V}$-block, but there is no such term in the $\mathrm{W}$-block.
The $\mathrm{V}$-block leading coefficient may also be verified against a direct computation using the superconformal algebra.

We may also exploit the above computation to gain insight into the blocks at special external momenta: $P_{\mathrm{ext}} = P_{1,1}$, corresponding to the identity operator $h_{\mathrm{ext}} =0$, and $P_{\mathrm{ext}} = P_{1,-1}$, corresponding to a weight $h_{\mathrm{ext}} =\frac{1}{2}$ operator.  Observe that
\begin{equation}
H_m^{\mathrm{V/W}}
\propto
H_{\frac{1}{2}}^{\mathrm{V/W}}, \quad m>0 \,.
\end{equation}
Hence, for $h_{P_\mathrm{ext}} = 0$, the $\mathcal{H}^\mathrm{V}$  simplifies to unity, while $\mathcal{H}^\mathrm{W}$  reduces to unity when $h_\mathrm{ext} = \frac{1}{2}$, viz.,
\begin{equation}\label{eq:simpleblocks1}
\begin{aligned}
\mathcal{H}^\mathrm{V}_3(P_{1,1},P \mid q)
   & =
\mathcal{H}^\mathrm{V}_4(P_{1,1},P \mid q)
=  1\,,
\\
\mathcal{H}^\mathrm{W}_3(P_{1,-1},P \mid q)
   & =
\mathcal{H}^\mathrm{W}_4(P_{1,-1},P \mid q)
=  1 \,.
\end{aligned}
\end{equation}

We can investigate the Ramond blocks in a similar vein. The blocks again ought to simplify when the external operator has momentum $P_{1,1}$ or $P_{1,-1}$. Based on the structure constants, we expect the former should pertain for the even block and the latter for the odd block. To see this we note that  $X^{\mathrm{o}}_{r,s}(P_{\mathrm{ext}} -P_{r,s} )$ vanishes for $P_{\mathrm{ext}} = P_{1,1}$, while $X^{\mathrm{e}}_{r,s}(P_{\mathrm{ext}} -P_{r,-s})$ vanishes for $P_{\mathrm{ext}} = P_{1,-1}$.  This suffices to guarantee
\begin{equation}\label{eq:simpleblocks2}
\mathcal{H}^{\mathrm{I}}_{\mathrm{even}}(P_{1,1} ,P  \mid q)
=
\mathcal{H}^{\mathrm{I}}_\mathrm{odd}(P_{1,-1}, P \mid q)
= 1 \,.
\end{equation}
In fact, we found it helpful to use this to fix the  combination of signs that can appear in the residues:
$\Res^{\mathrm{I,even}}_{r,s} (P_{\mathrm{ext}}) \propto X^{\mathrm{o}} (P_{\mathrm{ext}} - P_{r,s})$ and
$\Res^{\mathrm{I,odd}}_{r,s}(P_{\mathrm{ext}}) \propto X^{\mathrm{e}} (P_{\mathrm{ext}}-P_{r,-s})$, respectively. Furthermore, demanding that the even block does not vanish for $P_{\mathrm{ext}} = P_{1,-1}$ and the odd one does not vanish for $P_{\mathrm{ext}} = P_{1,1}$ turns out to completely fix the combination of fusion monomials, providing a simple consistency check.

\subsection{Ward identities and the odd spin structure}\label{sec:odd}

Our next step is to justify the expressions for the odd spin structure expectation value of descendant operators. By examining the sphere three-point function using superconformal Ward
identities, we shall fix a general form of the result. Allowing all but the external operator of interest to run in the loop would immediately then give us the required answer.

\paragraph{The superconformal Ward identities:}
The Ward identity of interest involves the consideration of single valued functions on the sphere. Consider, for instance, as in~\cite{Suchanek:2009ths} the following dressed correlator  
\begin{equation}
w^n\, \sqrt{w\, (w-z)}\, \expval{V_{P_1}(\infty)\,
    G(w,\bar{w}) R^{\epsilon}_{P_2}(z, \zb)\, R^{\epsilon'}_{P_3}(0)}.
\end{equation}
We want to consider the limit $w \to z$ and exploit the OPE to get the descendant of the Ramond primary. However, the supercharge action on the Ramond operator can be moved over to the other two operators by standard contour manipulations, thus giving us relations to the descendant of interest.

Extracting the behavior for individual modes of the supercharge we find at fixed mode number $n$ the relation
\begin{equation}
\begin{aligned}
\sum_{p=0}^\infty \, \Bigg[
    \binom{\frac{1}{2}}{p}\, (-z)^p\,
    \expval{
    G_{p-n-\frac{1}{2}}\, V_{P_1}(\infty)\, R^{\epsilon}_{P_2}(z)\, R^{\epsilon'}_{P_3}(0)}
    -
    \binom{n+\frac{1}{2}}{p}\, z^{n+\frac{1}{2} - p}\,
    \expval{
    V_{P_1}(\infty)\, G_p\, R^{\epsilon}_{P_2}(z)\, R^{\epsilon'}_{P_3}(0)}
    \\
    -
    i\,\epsilon\, \binom{\frac{1}{2}}{p}\, (-1)^p\,
    z^{\frac{1}{2} - p}\,
    \expval{
    V_{P_1}(\infty)\, R^{\epsilon}_{P_2}(z)\, G_{n+p} R^{\epsilon'}_{P_3}(0)}
    \Bigg] = 0\,.
\end{aligned}
\end{equation}
A similar manipulation in the antiholomorphic sector starting from
\begin{equation}
\bar{w}^n\, \sqrt{\bar{w}\, (\bar{w}-\zb)}\,
\expval{V_{P_1}(\infty)\,
    \widetilde{G}(w,\bar{w}) R^{\epsilon}_{P_2}(z, \zb)\, R^{\epsilon'}_{P_3}(0)}\,,
\end{equation}
leads to
\begin{equation}
\begin{aligned}
\sum_{p=0}^\infty \, \Bigg[
    \binom{\frac{1}{2}}{p}\, (-\zb)^p\,
    \expval{
    \widetilde{G}_{p-n-\frac{1}{2}}\, V_{P_1}(\infty)\,
    R^{\epsilon}_{P_2}(z)\, R^{\epsilon'}_{P_3}(0)}
    -
    \binom{n+\frac{1}{2}}{p}\, \zb^{n+\frac{1}{2} - p}\,
    \expval{
    V_{P_1}(\infty)\, \widetilde{G}_p\, R^{\epsilon}_{P_2}(z)\, R^{\epsilon'}_{P_3}(0)}
    \\
    + \, i\,\epsilon\, \binom{\frac{1}{2}}{p}\, (-1)^p\,
    \zb^{\frac{1}{2} - p}\,
    \expval{
    V_{P_1}(\infty)\, R^{\epsilon}_{P_2}(z)\, \widetilde{G}_{n+p} R^{\epsilon'}_{P_3}(0)}
    \Bigg] = 0\,.
\end{aligned}
\end{equation}

Our interest is in the $n=0$ and $n=1$ identities.
For $n=0$ this gives first the basic identity quoted in the reference:
\begin{equation}
\begin{aligned}
\expval{G_{-\frac{1}{2}} V_{P_1} (\infty)\, R^{\epsilon}_{P_2}(1)\, R^{\epsilon'}_{P_3}(0)}
   & =
\expval{V_{P_1}(\infty) \, G_0 R^{\epsilon}_{P_2}(1)\, R^{\epsilon'}_{P_3}(0)}
+ i \epsilon \, \expval{V_{P_1}(\infty) \, R^{\epsilon}_{P_2}(1)\,
    G_0R^{\epsilon'}_{P_3}(0)} \,,
\\
\expval{\widetilde{G}_{-\frac{1}{2}} V_{P_1} (\infty)\,
    R^{\epsilon}_{P_2}(1)\, R^{\epsilon'}_{P_3}(0)}
   & =
\expval{V_{P_1}(\infty) \, \widetilde{G}_0 R^{\epsilon}_{P_2}(1)\, R^{\epsilon'}_{P_3}(0)}
- i \epsilon \, \expval{V_{P_1}(\infty) \, R^{\epsilon}_{P_2}(1)\,  \widetilde{G}_0R^{\epsilon'}_{P_3}(0)} \,.
\end{aligned}
\end{equation}
Similarly, at the next level $n=1$ we find for example in the holomorphic sector
\begin{equation}
\begin{aligned}
\expval{G_{-\frac{3}{2}} V_{P_1}(\infty) \, R^{\epsilon}_{P_2}(1)\, R^{\epsilon'}_{P_3}(0)}
   & =
\frac{1}{2}\, \expval{G_{-\frac{1}{2}} V_{P_1}(\infty) \, R^{\epsilon}_{P_2}(1)\, R^{\epsilon'}_{P_3}(0)} +
\expval{V_{P_1}(\infty) \, G_0 R^{\epsilon}_{P_2}(1)\, R^{\epsilon'}_{P_3}(0)} \,.
\end{aligned}
\end{equation}
Combining the two expressions we can deduce the correlator involving $G_{-\frac{3}{2}} V_P$.
The final result for the holomorphic and antiholomorphic parts is
\begin{equation}
\begin{aligned}
\expval{G_{-\frac{3}{2}} V_{P_1}(\infty) \, R^{\epsilon}_{P_2}(1)\, R^{\epsilon'}_{P_3}(0)}
   & =
\frac{3}{2}\, \expval{V_{P_1}(\infty) \, G_0 R^{\epsilon}_{P_2}(1)\, R^{\epsilon'}_{P_3}(0)}
+ \frac{i\,\epsilon}{2} \, \expval{V_{P_1}(\infty) \, R^{\epsilon}_{P_2}(1)\,
    G_0R^{\epsilon'}_{P_3}(0)} \,,
\\
\expval{\widetilde{G}_{-\frac{3}{2}} V_{P_1}(\infty) \, R^{\epsilon}_{P_2}(1)\,
    R^{\epsilon'}_{P_3}(0)}
   & =
\frac{3}{2}\, \expval{V_{P_1}(\infty) \, \widetilde{G}_0 R^{\epsilon}_{P_2}(1)\,
    R^{\epsilon'}_{P_3}(0)}
- \frac{i\,\epsilon}{2} \, \expval{V_{P_1}(\infty) \, R^{\epsilon}_{P_2}(1)\,
    \widetilde{G}_0R^{\epsilon'}_{P_3}(0)} \,.
\end{aligned}
\end{equation}

\paragraph{The $W$ correlator:} Since $W = -\,G_{-1/2}\,\widetilde{G}_{-1/2}\,V$ is a
superdescendant, we can first derive its expectation value from that of $V$.
Using the $n=0$ Ward identities directly, we end up with
\begin{equation}\label{eq:GGhalfV}
\begin{aligned}
\expval{
    G_{-\frac{1}{2}} \widetilde{G}_{-\frac{1}{2}}
    V_{P_1}\, R^{\epsilon}_{P_2}\, R^{\epsilon'}_{P_3}}
   & =
\expval{V_{P_1} \, G_0\widetilde{G}_0\, R^\epsilon_{P_2} \, R^{\epsilon'}_{P_3}}
+ \expval{V_{P_1} \, R^\epsilon_{P_2} \, G_0\widetilde{G}_0\, R^{\epsilon'}_{P_3}}
\\
   & \quad\
- i\,\epsilon\,
\expval{V_{P_1} \, \widetilde{G}_0 R^\epsilon_{P_2} \, G_0 R^{\epsilon'}_{P_3}}
- i\,\epsilon\,
\expval{V_{P_1} \, G_0 R^\epsilon_{P_2} \, \widetilde{G}_0 R^{\epsilon'}_{P_3}}\,.
\end{aligned}
\end{equation}
The signs above follow from applying $\widetilde{G}_{-1/2}$ first, then $G_{-1/2}$.
There is a parity flip when $G_0$ crosses $\widetilde{G}_0 R^\epsilon_{P_2}$.
Now we can substitute the supercharge action~\eqref{eq:Rzeromodeaction}  to fix
$G_0\,\widetilde{G}_0\, R^\epsilon_P = -\frac{P^2}{2}\,e^{i\frac{\pi}{2}\,\epsilon}\, R^\epsilon_P$.

We now have determined the sphere 3-point function. The torus one-point function can be
obtained by gluing together the two Ramond operators. This involves not only setting
$P_2 = P_3 = P$ and integrating over the spectrum of allowed $P \in \mathbb{R}_{\geq 0}$, but
also summing over the internal fermion number eigenstates. For fixed fermion number polarization
we find upon setting the internal momenta equal
\begin{equation}\label{eq:GGhalfVeval}
\begin{aligned}
\expval{
    \widetilde{G}_{-\frac{1}{2}} G_{-\frac{1}{2}}
    V_{P_1}\, R^{\epsilon}_{P}\, R^{\epsilon'}_{P}}
   & =
\frac{P^2}{2}\,
\pqty{e^{i\frac{\pi}{2}\,\epsilon} + e^{i\frac{\pi}{2}\,\epsilon'}}
\expval{V_{P_1}\, R^\epsilon_P \, R^{\epsilon'}_P}
\\
   & \quad
- i\,\epsilon\, \frac{P^2}{2}\,
\pqty{e^{i\frac{\pi}{4}\,(\epsilon-\epsilon')}
    + e^{-i\frac{\pi}{4}(\epsilon-\epsilon')}}
\expval{V_{P_1}\, R^{-\epsilon}_P \, R^{-\epsilon'}_P}\,.
\end{aligned}
\end{equation}

The final step is to sum over fermion polarizations. The correlators on the r.h.s.~are non-vanishing only for $\epsilon = \epsilon'$. However, we can choose to combine the contributions for the two choices, $\epsilon = \epsilon' = 1$ and $\epsilon = \epsilon' = -1$ without or with a relative sign. These give two independent answers, corresponding to the two Ramond sector traces R and $\widetilde{\mathrm{R}}$. Denoting the two choices by $(++) \pm (--)$ with the brackets denoting the choice for the pair $\epsilon$ and $\epsilon'$ one finds
\begin{equation}\label{eq:GGhalfpolsum}
\begin{aligned}
(++) + (--)
   & \sim  2i\,P^2\,\expval{VR^+R^+}
   - 2i\,P^2\,\expval{VR^-R^-}
= 2i\,P^2\, \Co{b}(P_1;P,P)  \,,
\\
(++) - (--)
   & \sim 0\,P^2\,\expval{VR^+R^+}
+ 0\,P^2\,\expval{VR^-R^-}
= 0 \,.
\end{aligned}
\end{equation}
Adding up the two polarizations with no sign picks up the structure constant $C_\mathrm{odd}$
--- exactly the coefficient in $\expval{W}_2$, cf.~\eqref{eq:1ptR}. Moreover, the overall sign is also fixed to be $+ 2i\, P^2$. 
As seen in~\cref{eq:FblockR}, the factor of $P^2$ is included in the definition of the corresponding left- and right-moving conformal blocks. As a simple consistency check, one can compute the Hilbert space traces in the R and $\widetilde{\mathrm{R}}$ sectors using the chiral decomposition of the vertex operators~\cite{Hadasz:2008dt,Suchanek:2010kq} and arrive at this result. The sum with a relative sign
vanishes identically and is consistent with $\mathcal{F}_1^W = 0$
for the  $\widetilde{\mathrm{R}}$ sector $W$ correlator. This benchmarking will be useful below,
as it makes clear that we should sum the two choices for the R sector trace, but subtract them for the $\widetilde{\mathrm{R}}$ sector trace.

\paragraph{The descendant correlator:} The new correlation function we want to discuss is that of the descendant operator $-G_{-\frac{3}{2}} \widetilde{G}_{-\frac{3}{2}}  V_{P}$. Combining the holomorphic and antiholomorphic Ward identities we end up with
\begin{equation}
\begin{aligned}
\expval{
    G_{-\frac{3}{2}} \widetilde{G}_{-\frac{3}{2}}
    V_{P_1}\, R^{\epsilon}_{P_2}\, R^{\epsilon'}_{P_3}}
   & =
\frac{9}{4}\,
\expval{V_{P_1}(\infty) \, G_0\widetilde{G}_0 \,R^{\epsilon}_{P_2}(1)\,
    R^{\epsilon'}_{P_3}(0)}
+ \frac{1}{4}\,
\expval{V_{P_1}(\infty) \, R^{\epsilon}_{P_2}(1)\,G_0\widetilde{G}_0 \,
    R^{\epsilon'}_{P_3}(0)}
\\
   &
- \frac{3i\,\epsilon}{4} \, \expval{V_{P_1}(\infty) \,\widetilde{G}_0
    R^{\epsilon}_{P_2}(1)\,  G_0 R^{\epsilon'}_{P_3}(0)}
- \frac{3i\,\epsilon}{4} \, \expval{V_{P_1}(\infty) \, G_0 R^{\epsilon}_{P_2}(1)\,
    \widetilde{G}_0R^{\epsilon'}_{P_3}(0)}\,.
\end{aligned}
\label{eq:Rtildecorr}
\end{equation}
There is an additional sign in the second line because we have to move $G_0$ through $\widetilde{G}_0 R^{\epsilon}$, which has the opposite fermion parity to $R^{\epsilon}$.

Recall from our discussion above that the trace in the $\widetilde{\mathrm{R}}$ sector involves setting the internal momenta equal, and summing over the fermion polarizations $\epsilon = \epsilon'$ with a relative sign between the $(++)$ and $(--)$ terms. From the r.h.s.~of~\eqref{eq:Rtildecorr} we obtain
the contribution
\begin{equation}\label{eq:GGV1step1}
(++) - (--) \sim - \frac{i}{2} \,  P^2\, \Ce{b}(P_1; P,P) \, .
\end{equation}
It can be checked that the combination without a relative sign, viz., $(++) + (--)$, which gives the R sector trace will be proportional to $-2i\, P^2\, \Co{b}(P_1;P,P)$.

To pass from the polarization sum to the expectation value $\expval{DV_P}_1$ we should reverse the order of supercharge action in the l.h.s.~and fix the normalization. Furthermore, the two point function in the Ramond sector is normalized as $\frac{1}{2\rhoR}$ and writing the block with the measure for the internal states given by $\rhoR(P)\, dP$ thus supplies a factor of two. Therefore, 
\begin{equation}
\expval{DV_{P_1}}_1 = i\,\int_0^\infty\, dP\, \rhoR(P)\,  P^2\, \Ce{b}(P_1; P,P)\, 
(q\,\bar{q})^{\frac{1}{2}\, P^2} \bqty{1 + \cdots} \,,
\end{equation}
where the omitted terms assemble into the product of elliptic blocks $\norm{\mathcal H_{\mathrm{even}}^{\mathrm W}}^2$. 
The same factor of two from the inverse Ramond two-point
normalization also enters $\expval{W_P}_2$. In that case,
it is already supplied by the leading term of the
theta-function prefactor in the product of chiral blocks, cf.~\eqref{eq:theta2exp}, that explicitly accounts for the two Ramond ground states. The odd-spin descendant block has no such prefactor, so this normalization factor must be included directly. Together with the supercharge ordering, it produces the factor $iP^2 C_{\mathrm{even}}$ appearing above.

Having determined the constraint from the Ward identity and the polarization sums, we are left with completing the sum over intermediate states (integrating over the internal loop). For this we reinstate the Szeg\H{o} kernel in the respective sectors for the fermion propagator. This will end up causing the contribution to the R sector correlator (the polarization sum leading to $\Co{b}$ above) to vanish, and leave behind a non-vanishing answer for the $\widetilde{\mathrm{R}}$ sector. Carefully following this chain of logic would result in a first principles derivation of the block $\mathcal{F}_1^{\mathrm{DV}}$. However, at this point, we can simply postulate that it is given by the even elliptic block, modified to account for the additional $P$ factors appearing with the structure constant. 
We thus propose the block given in (\ref{eq:FblockR}). The modular-covariance and crossing-kernel tests below provide nontrivial checks of this identification.
 
\subsection{Modular covariance}\label{sec:modc}

A clear test of the torus superconformal blocks is that they lead to modular covariant expectation values of conformal primaries.  For a
scalar primary operator $\mathcal{O}$ inserted on the torus, the one-point functions must satisfy the following $S$-modular transformation identity
\begin{equation}
\expval{\mathcal{O}(0\mid-\tfrac{1}{\tau})}_a  =
\abs{\tau}^{2h}\, \expval{\mathcal{O}(0\mid\tau)}_{\mathfrak{s}(a)}\,.
\label{eq:genmodcov}
\end{equation}
Here we introduced a shorthand to indicate the mixing between the different spin structures under modular $S$-transformation. The function $\mathfrak{s}(a)$ is explicitly given by
\begin{equation}
\mathfrak{s}(3) = 3\,, \qquad
\mathfrak{s}(1) = 1\,, \qquad
\mathfrak{s}(2) = 4\,, \qquad
\mathfrak{s}(4) =2\,.
\label{eq:sfndef}
\end{equation}
This just says that the NS and $\widetilde{\mathrm{R}}$ map back to themselves, while $\widetilde{\mathrm{NS}}$ and R are exchanged by the $S$-transformation.

We will test the modular covariance of $\expval{V}_a$ and $\expval{W}_a$ to verify the blocks $\mathcal{H}(P_{\mathrm{ext}},P  \mid q)$ directly. For illustration, consider the expectation value of $V$ in the NS sector. Setting $\widetilde{q} = e^{-\frac{2\pi i}{\tau}}$, we have
\begin{equation}
\begin{aligned}
\expval{V_{P_\mathrm{ext}}(0\mid -\tfrac{1}{\tau})}_3
   & =
\abs{\frac{\vartheta_3(-\frac{1}{\tau})}{\eta(-\frac{1}{\tau})^3}}\,
\int\, dP\, \rhoNS(P)\, \CV{b}(P_\mathrm{ext},P,P) \;
\norm{\,\widetilde{q}^{\frac{P^2}{2}}\, \mathcal{H}_3^{\mathrm{V}}(P_\mathrm{ext},P \mid \widetilde{q})}^2
\\
   & = \frac{1}{\abs{\tau}} \,
\abs{\frac{\vartheta_3(\tau)}{\eta(\tau)^3}}
\int\, dP\, \rhoNS(P)\, \CV{b}(P_\mathrm{ext},P,P) \;
\norm{\,\widetilde{q}^{\frac{P^2}{2}}\, \mathcal{H}_3^{\mathrm{V}}(P_\mathrm{ext},P \mid \widetilde{q})}^2
\\
\expval{V_{P_\mathrm{ext}}(0 \mid\tau)}_3
   & =
\abs{\frac{\vartheta_3(\tau)}{\eta(\tau)^3}}
\int\, dP\, \rhoNS(P)\, \CV{b}(P_\mathrm{ext},P,P) \;
\norm{q^{\frac{P^2}{2}}\, \mathcal{H}_3^{\mathrm{V}}(P_\mathrm{ext},P \mid q)}^2\,.
\end{aligned}
\end{equation}
We have explicitly used the modular transformation properties of the theta and Dedekind eta functions in deriving the second equality. Similar manipulations in the other sectors lead to the following identities for the $V$ expectation value:
\begin{equation}
\begin{aligned}
   &
\int\, dP\, \rhoNS(P)\, \CV{b}(P_\mathrm{ext},P,P) \;
\norm{\,\widetilde{q}^{\frac{P^2}{2}}\, \mathcal{H}_3^{\mathrm{V}}(P_\mathrm{ext},P \mid \widetilde{q})}^2
\\
   & \quad
=
\abs{\tau}^{2\,h_{\mathrm{ext}} + 1}\,
\int\, dP'\, \rhoNS(P')\, \CV{b}(P_\mathrm{ext},P',P')
\norm{q^{\frac{P'^2}{2}}\, \mathcal{H}^{\mathrm{V}}_3(P_\mathrm{ext},P' \mid q)}^2 \,,
\\
   &
\int\, dP\, \rhoNS(P)\, \CV{b}(P_\mathrm{ext},P,P) \;
\norm{\,\widetilde{q}^{\frac{P^2}{2}}\, \mathcal{H}_4^{\mathrm{V}}(P_\mathrm{ext},P \mid \widetilde{q})}^2
\\
   & \quad
=
\abs{\tau}^{2\,h_{\mathrm{ext}} + 1}\,
\int\, dP'\, \rhoR(P')\, \Ce{b}(P_\mathrm{ext};P',P')
\norm{q^{\frac{P'^2}{2}}\, \mathcal{H}^{\mathrm{V}}_\mathrm{even}(P_\mathrm{ext},P' \mid q)}^2 \,,
\\
   &
\int\, dP\, \rhoR(P)\, \Co{b}(P_\mathrm{ext};P,P) \;
\norm{\,\widetilde{q}^{\frac{P^2}{2}}\, \mathcal{H}^{\mathrm{V}}_{\mathrm{odd}}(P_\mathrm{ext},P \mid \widetilde{q})}^2
\\
   & \quad
=
\abs{\tau}^{2\,h_{\mathrm{ext}}}\,
\int\, dP'\, \rhoR(P')\, \Co{b}(P_\mathrm{ext};P',P')
\norm{q^{\frac{P'^2}{2}}\, \mathcal{H}^{\mathrm{V}}_{\mathrm{odd}}(P_\mathrm{ext},P' \mid q)}^2 \,,
\end{aligned}
\label{eq:Vmodchecks}
\end{equation}
The correlators in the $\widetilde{\mathrm{NS}}$ and R sectors get exchanged as noted above (the corresponding theta functions transform as necessary).
Notice that the last equation has one fewer power of $\abs{\tau}$ due to the absence of characters in the $\expval{V_{P_{\mathrm{ext}}}}_1$ correlator. For the $W$ operator we similarly deduce the following identities
\begin{equation}
\begin{aligned}
   &
\int\, dP\, \rhoNS(P)\, \CW{b}(P_\mathrm{ext},P,P) \;
\norm{\,\widetilde{q}^{\frac{P^2}{2}}\, \mathcal{H}_3^{\mathrm{W}}(P_\mathrm{ext},P \mid \widetilde{q})}^2
\\
   & \quad
=
\abs{\tau}^{2\,h_{\mathrm{ext}} + 2}\,
\int\, dP'\, \rhoNS(P')\, \CW{b}(P_\mathrm{ext},P',P')
\norm{q^{\frac{P'^2}{2}}\, \mathcal{H}^{\mathrm{W}}_3(P_\mathrm{ext},P' \mid q)}^2 \,,
\\
   &
\int\, dP\, \rhoNS(P)\, \CW{b}(P_\mathrm{ext},P,P) \;
\norm{\,\widetilde{q}^{\frac{P^2}{2}}\, \mathcal{H}_4^{\mathrm{W}}(P_\mathrm{ext},P \mid \widetilde{q})}^2
\\
   & \quad
=
2i\,\abs{\tau}^{2\,h_{\mathrm{ext}} + 2}\,
\int\, dP'\, \rhoR(P')\, P'^2\, \Co{b}(P_\mathrm{ext};P',P') \;
\norm{q^{\frac{P'^2}{2}}\, \mathcal{H}^{\mathrm{W}}_\mathrm{odd}(P_\mathrm{ext},P' \mid q)}^2 \,.
\end{aligned}
\label{eq:Wmodchecks}
\end{equation}
The reader might find the explicit expressions for the one-point functions in~\cref{sec:wsbblocks} helpful in interpreting these formulae.

There are some simple sanity checks we can do before embarking on numerical verification. When $P_\mathrm{ext} = P_{1,-1}$ we saw that the blocks for $W$ simplify, cf.~\eqref{eq:simpleblocks1} and~\eqref{eq:simpleblocks2}. In addition, one can check
\begin{equation}
\begin{aligned}
\CW{b}(P_{\mathrm{ext}}=P_{1,-1},P,P)\,\rhoNS(P) 
&= 
\frac{ibP^2}{\pi(1+b^2)},\\ 
\Co{b}(P_{\mathrm{ext}}=P_{1,-1};P,P)\, \rhoR(P) 
&= 
\frac{b}{2\pi(1+b^2)}\, .
\end{aligned}
\end{equation}
Hence, the modular covariance of $\expval{W}_3$ for this special case demands
\begin{equation}\label{eq:gaussian1}
\int_0^\infty\, dP\, P^2\, \abs{\widetilde{q}^{\frac{P^2}{2}}}^2 
= \abs{\tau}^{3}\, \int_0^\infty\, dP'\, P'^2\, \abs{q^{\frac{P'^2}{2}}}^2
\;\; \Longrightarrow \;\;
\frac{1}{\sqrt{\widetilde{\tau}_2^3}} = \frac{\abs{\tau}^3}{\sqrt{\tau_2^3}}\,.
\end{equation}
The last equality is, of course, a consequence of the $S$-transform.  The selfsame identity suffices to establish the transformation between $\expval{W}_4$ and $\expval{W}_2$.  In this case, the factors of $2i$ work out correctly  --- for instance, $\CW{b}$ has a factor of $i$ and there is a factor of $\frac{1}{2}$ from the limiting value of the structure constant
$\Co{b}$ in~\eqref{eq:sLCeo}.

We can now fill in another gap in our discussion and prove that there should be a zero mode factor  $q^{\frac{P^2}{2}}$ in $\mathcal{F}_1^{\mathrm{V}}$. The argument is analogous to the above. Again restricting to $P_{\mathrm{ext}} = P_{1,-1}$ we examine the final equation in~\eqref{eq:Vmodchecks}. The r.h.s.~has a factor of $\abs{\tau}$, since $h_{\mathrm{ext}} = \frac{1}{2}$. The integral largely collapses because $\mathcal{H}^{\mathrm{V}}_{\mathrm{odd}}(P_{1,-1},P \mid q) = 1$ from~\eqref{eq:simpleblocks2}. Therefore, we require
\begin{equation}\label{eq:gaussian2}
\int_0^\infty\, dP\, \abs{\widetilde{q}^{\frac{P^2}{2}}}^2 = \abs{\tau}\, \int_0^\infty\, dP'\, \abs{q^{\frac{P'^2}{2}}}^2
\;\; \Longrightarrow \;\;
\frac{1}{\sqrt{\widetilde{\tau}_2}} = \frac{\abs{\tau}}{\sqrt{\tau_2}}\,,
\end{equation}
which is, of course, the same condition encountered from~\eqref{eq:gaussian1}. Should the $q^{\frac{P^2}{2}}$ be absent, modular covariance would have failed.

\paragraph{The $\expval{\widetilde{G}_{-\frac{3}{2}} G_{-\frac{3}{2}} V}$ correlator:}
Although $\widetilde G_{-\frac32}G_{-\frac32}V$ is a descendant,
its torus one-point function coincides with that of the
Virasoro primary $\mathcal O_V$ defined in
\eqref{eq:level-3/2-primary}. It thus transforms with conformal weights $(h_{\mathrm{ext}}+\frac32,h_{\mathrm{ext}}+\frac32)$. In terms of the proposed conformal block decomposition, we thus have
\begin{equation}
\begin{aligned}
   &
\int\, dP\, \rhoR(P)\, P^2\, \Ce{b}(P_\mathrm{ext};P,P) \;
\norm{\,\widetilde{q}^{\frac{P^2}{2}}\, \mathcal{H}_{\mathrm{even}}^\mathrm{W}(P_\mathrm{ext},P \mid \widetilde{q})}^2
\\
   & \quad
=
\abs{\tau}^{2\, h_{\mathrm{ext}} + 3}\,
\int\, dP'\, \rhoR(P')\, P'^2\, \Ce{b}(P_\mathrm{ext};P',P')
\norm{q^{\frac{P'^2}{2}}\, \mathcal{H}_{\mathrm{even}}^{\mathrm{W}}(P_\mathrm{ext},P' \mid q)}^2 \,.
\end{aligned}
\label{eq:GVmodchecks}
\end{equation}
We find strong numerical evidence (see below) that this identity holds. Below we also show how it follows from corresponding crossing kernel identities. Later we will explain how this correlator enters the contribution of the odd spin structure to the string amplitudes.

\begin{figure}[htbp!]
    \centering
    {\includegraphics[width=0.95\textwidth]{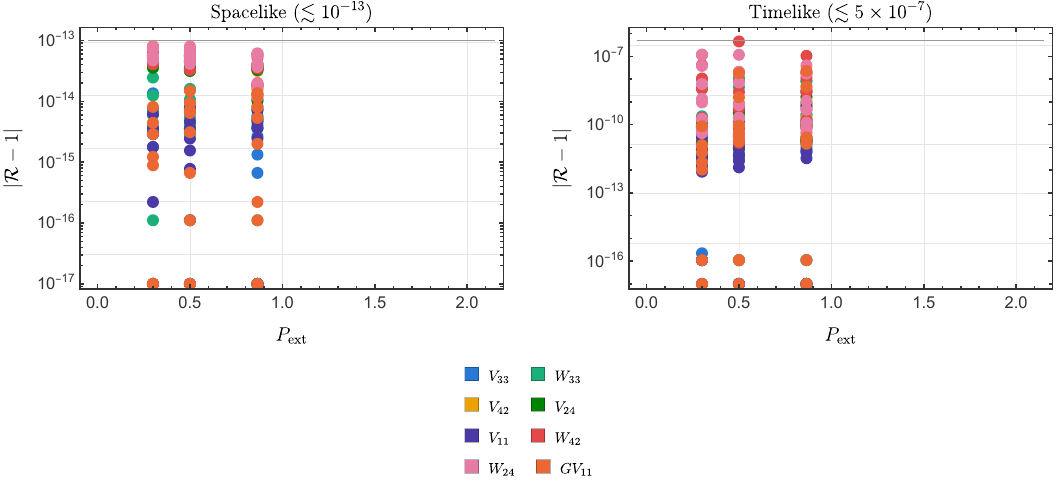}}
    \caption{Checks of modular covariance of the eight torus one-point functions for the spacelike and timelike super Liouville theories for the parameters scanned over the sets in~\eqref{eq:parscan-mod-cov}.}
    \label{fig:mod-cov-check}
\end{figure}

\paragraph{Numerical check of modular covariance:}
To verify that the identities~\eqref{eq:Vmodchecks},~\eqref{eq:Wmodchecks},
and~\eqref{eq:GVmodchecks} hold we implemented the recursion relation numerically.
Our explicit checks were carried out for the following parameters:
\begin{equation}\label{eq:parscan-mod-cov}
b\in \set{\frac{3}{4}, \frac{4}{5}, \frac{3}{\pi} } \,,\qquad
P_{\mathrm{ext}} \in \set{\frac{3}{10}, \frac{1}{2}, \frac{e}{\pi}}\,.
\end{equation}
For each tuple $(b,P_{\mathrm{ext}})$ sourced from the above sets, we sampled covariance
at three different values of $\tau$: the self-dual point $\tau = i$,
a point on the imaginary axis $\tau = 1.1 \,i$ and $\tau = \frac{1+4\,i}{5}$.
In addition, for the $\widetilde{\mathrm{NS}}$ and $\mathrm{R}$ sectors, we computed both sets of correlators
for the above values of $\tau$ and for the $S$-dual value, so that one can compare the ratios in both directions. 
Define for any of the operators (denoted $\mathcal{O}$)
\begin{equation}
\mathcal{R}_{ab} =  
\abs{\tau}^{2h_{\mathcal{O}}}\, 
\frac{\expval{\mathcal{O}(0 | q)}_a}{\expval{\mathcal{O}(0| \widetilde{q})}_b}\,,
\qquad
(a,b) \in \set{(3,3), (4,2), (2,4), (1,1)}\,.
\label{eq:ratiomodc}
\end{equation}

The recursion was carried out to order $8$, with $40$-digit numerical precision.
We were able to check that all the ratios defined in~\eqref{eq:ratiomodc} are near unity, with
\begin{equation}
\abs{\mathcal{R}_{ab} -1} \lesssim 10^{-13} \,,
\end{equation}
in the spacelike theory. 
The left panel of~\cref{fig:mod-cov-check} illustrates this explicitly.

\paragraph{Modular covariance of timelike super Liouville:} Our numerical implementation can also be used to test the modular covariance of the timelike Liouville theory structure constants. In this we have to be careful to ensure that the internal momentum is not on the real axis, but its contour is shifted up into the complex $\Ph$ plane by an amount $i\,\epsilon$.   We found $\epsilon = 0.08$ to suffice for our purposes. The other change is that one should use the timelike spectral densities defined in~\eqref{eq:tLrho}. 

There are two subtleties to be wary of in working with the timelike theory. The right chiral block is not complex conjugate of the left one since $\Ph \notin \mathbb{R}$. This was the rationale for our notation~\eqref{eq:normdef}. In addition, the Ward identities discussed in~\cref{sec:odd} make amply clear that factor of $\Ph^2$ in $\expval{\Wh}_2$
and $\expval{\mathsf{DV}}_1$ arise from action of the supercharge zero modes. These factors should likewise not be analytically continued, and should be evaluated on the shifted real line. The explicit expression in~\eqref{eq:genus1-time} should help make these statements clear. 

As seen in the right panel of~\cref{fig:mod-cov-check}, we were able to numerically verify that the analogous ratios defined for the timelike theory are small, although larger than the spacelike case ($\abs{\mathcal{R}_{ab} -1} \lesssim 5\times10^{-7}$), for the same set of parameters reported above for the spacelike case.  

\section{Modular \texorpdfstring{$S$}{S}-kernels for the torus}\label{sec:modularkernel}

We now turn to the modular $S$-kernel for the $\mathcal{N} = 1$ superconformal field theory.
The goal here is to present the full set of crossing kernels for the S-transform $S: \tau \to -\frac{1}{\tau}$ of the (chiral) conformal blocks on the once-punctured torus in all spin structure sectors, for central charges $c\in \mathbb{C}\backslash(-\infty,\frac{3}{2}]$ (i.e.,  those relevant for the spacelike case).\footnote{Constructing the crossing kernels in the ``timelike'' regime of central charges (i.e., $c\in (-\infty,\tfrac{3}{2}]$) is an intricate problem even in the more familiar case of bosonic CFT. Recently,~\cite{Roussillon:2025tmv} constructed the bosonic modular and sphere-four-point kernels for the case of rational central charges in the timelike regime $\ch\in\mathbb{Q}_{(-\infty,1]}$. 
In the $\mathcal{N}=1$ case this is still an open question; we have nothing to report here on that front.} The main issue we have to deal with is the fact that a subset of spin structures mix under this operation, and the presence of two distinct primaries $V_P$ and $W_P$ that have non-trivial one-point functions.  We will
demonstrate that there are eight crossing kernels including one that corresponds to the transformation of the level three-half operator $\widetilde{G}_{-\frac{3}{2}} G_{-\frac{3}{2}}  V_P$.  Hitherto, two of these kernels were constructed and analyzed in~\cite{Apresyan:2023vxl}, which we recover from our analysis.

Before plunging into the details of the super Liouville case, it is helpful to review the crossing transformations in the non-supersymmetric Liouville theory.  An excellent recent account of these can be found in~\cite{Eberhardt:2023mrq}. For the present, we will focus only on the $S$-transformation, and define the basic object of interest. As we have already discussed, the torus one-point function is built by composing the tree-level (sphere) structure constants, convolved with the torus blocks, which we denote in the bosonic Liouville theory as $\mathcal{F}_{\mathrm{bos}}(P_0,P \mid q)$ (cf.~\eqref{eq:1ptNS} in the super Liouville case). The correlator itself transforms as a modular form of definite weight. Folding in the fact that the blocks are themselves chiral, it follows that under an $S$-transform the chiral torus blocks should transform accordingly. One requires (NB: recall $\widetilde{q} = e^{- \frac{2\pi i}{\tau}}$)
\begin{equation}
\mathcal{F}_{\mathrm{bos}}(P_0, P_s \mid q)
= (-i\tau)^{-h_0} \int_{\mathbb{R}} \, dP_t\;
\mathbb{S}_{P_s P_t}[P_0] \; \mathcal{F}_{\mathrm{bos}}(P_0,P_t \mid \widetilde{q}) \,.
\label{eq:FSbosonic}
\end{equation}
We use $P_s$ and $P_t$ to indicate the internal momenta that run in the loop.
The explicit expression for the bosonic kernel $\mathbb{S}_{P_s P_t}[P_0]$ was originally derived in~\cite{Teschner:2003at} and, as might be expected, it is built out of the same class of special functions that appear in the structure constants. Having a concrete expression for this object has proven very fruitful, and it has since found numerous applications in mathematical physics: from the construction of the bulk-to-boundary structure constants in boundary Liouville theory~\cite{Hosomichi:2001xc, Numasawa:2022cni, Ang:2023fky} and the AGT correspondence~\cite{Alday:2009aq,Drukker:2009id, Hosomichi:2010vh}, to universal bootstrap aspects of irrational 2d CFTs~\cite{Collier:2019weq} and the quantization of Teichm\"uller space in direct relation with the quantization of \AdS{3} gravity via Virasoro TQFT~\cite{Collier:2023fwi}. We anticipate that our present construction of the $\mathcal{N}=1$ modular kernels will leave an analogous imprint on these topics and their supersymmetric generalizations.

We will first present our results for the 8 $S$-kernels in~\cref{sec:modSsummary}, and then explain how we obtained them. Rather than solve for the kernels using the defining properties (also known as `Moore-Seiberg relations') and the corresponding shift relations, as was done in~\cite{Apresyan:2023vxl} for the NS sector observables, we posit a set of bootstrap conditions, which we elucidate in~\cref{sec:modSbootstrap}. The remainder of the discussion is devoted to exhibiting explicit solutions for the eight kernels, and providing evidence that they satisfy the required properties. To anchor our discussion we will first verify that our procedure reproduces the known results in the NS sector, and then proceed to outline the other sectors.

\subsection{A torus \texorpdfstring{$S$}{S}-kernel summary}\label{sec:modSsummary}

To characterize the modular $S$-kernel for the supersymmetric case, we will use
$P_0$, $P_s$ and $P_t$ to denote the Liouville momenta of the superconformal primary $V$. This, therefore, means that $h_0 = \frac{Q^2}{8} + \frac{P_0^2}{2}$ will refer to the superconformal primary weight. We will compensate for the weights of other operators explicitly in the formulae below.

The modular $S$ kernel in $\mathcal{N}=1$ Liouville theory has $8$ incarnations. In the spacelike theory we denote them as follows:
\begin{equation}
\renewcommand{\arraystretch}{1.5}
\setlength{\tabcolsep}{11pt}
\begin{tabular}{|c||c|c|c|c|}
\hline
Channel & $3\leftrightarrow 3$ & $4\rightarrow 2$ & $2\rightarrow 4$ & $1\leftrightarrow 1$ \\
\hline\hline
$\langle V \rangle$
& $\Sm{V}{33}{P_s P_t}[P_0]$
& $\Sm{V}{42}{P_s P_t}[P_0]$
& $\Sm{V}{24}{P_s P_t}[P_0]$
& $\Sm{V}{11}{P_s P_t}[P_0]$ \\
\hline
$\langle W \rangle$
& $\Sm{W}{33}{P_s P_t}[P_0]$
& $\Sm{W}{42}{P_s P_t}[P_0]$
& $\Sm{W}{24}{P_s P_t}[P_0]$
& $\Sm{DV}{11}{P_s P_t}[P_0]$ \\
\hline
\end{tabular}
\label{eq:N1Snotation}
\end{equation}
The kernels $\Sm{V}{33}{}$ and $\Sm{W}{33}{}$ corresponding to modular transformations of the blocks in sector 3 were constructed in~\cite{Apresyan:2023vxl};
we will revisit them below. The remaining five kernels denoted as $\Sm{I}{ij}{}$ for $i,j=2,4,1$ and $\mathrm{I} \in \set{\mathrm{V}, \mathrm{W}}$ have not previously appeared. The kernel $\Sm{DV}{11}{}$ a priori appears distinct; it implements the modular transformation of blocks built upon the descendant external operator $\widetilde{G}_{-3/2}G_{-3/2}V_P$.
As described in~\cref{sec:torussL}  its torus one-point function is required to be modular covariant, necessitating therefore an additional crossing kernel.

\paragraph{Structural properties of $\mathbb{S}$:} Each of the eight kernels we obtain has a universal product structure representation, which allows one to express them as
\begin{equation}
\Sm{I}{\bullet}{P_sP_t}[P_0]
\;=\;
\mathcal{C}_K\,\mathcal{P}_K(P_0,P_s,P_t)\,
\int_{i\mathbb{R}}\!\frac{du}{i}\;
e^{-2\pi u P_s}\, \mathcal{I}_K(u,P_0,P_s,P_t) \,.
\end{equation}
Here $\mathcal C_K$ is a constant, and the \emph{prefactor} $\mathcal P_K$ and
\emph{integrand} $\mathcal I_K$ both have fixed arguments, as will become clear below.
We also use $\bullet$ as a placeholder for specifying the detailed argument, as shall be made clear in the following.

The definition of the kernel w.r.t.~the branching of the $\mathcal{N}=1$ chiral torus 1-point blocks $\mathcal{F}$ (cf. (\ref{eq:FblockNS}), (\ref{eq:FblockR})) takes the form analogous to~\eqref{eq:FSbosonic} and reads
\begin{equation}\label{eq:SFtransform}
(-i\tau)^{h_0+\delta} \
\mathcal{F}^{\mathrm{I}}_a(P_0, P_s \mid q)
= \int_{\mathbb{R}} \, dP_t \
\Sm{I}{a \, \mathfrak{s}(a)}{P_sP_t}[P_0] \
\mathcal{F}^{\mathrm{I}}_{\mathfrak{s}(a)}(P_0, P_t \mid \widetilde{q}) \,.
\end{equation}
Here $\mathfrak{s}(a)$ is the channel map defined in~\eqref{eq:sfndef} and $\delta \in \set{\frac{1}{2}, \frac{3}{2}}$ is an offset when considering operators other than the superconformal primary $V$.\footnote{We recall our notation: $\mathrm{I} \in \set{\mathrm{V}, \mathrm{W}, \mathrm{DV}}$, $\delta = 0$ for $\mathrm{I} = \mathrm{V}$, $\delta = \frac{1}{2}$ for $\mathrm{I} = \mathrm{W}$, and $\delta = \frac{3}{2}$ for the level three-half superdescendant.} We now unpack the ingredients entering the formula.
To write these in a compact form, we will not only invoke the spectral densities~\eqref{eq:rhons} and the double-Gamma functions~\eqref{eq:GammaNSR}, but will also require
\begin{equation}
\begin{aligned}
\SNS{b}{z}
   & = \frac{\dGNS{b}{z}}{\dGNS{b}{Q-z}} = \Sb{\frac{z}{2}}\, \Sb{\frac{z+Q}{2}} \,, \\
\SR{b}{z}
   & = \frac{\dGR{b}{z}}{\dGR{b}{Q-z}} = \Sb{\frac{z+b}{2}}\, \Sb{\frac{z+b^{-1}}{2}} \,.
\end{aligned}
\label{eq:SNSR}
\end{equation}

\paragraph{The prefactor:} The prefactor is built out of the double-Gamma functions, and takes the following universal form
\begin{equation}\label{eq:genpref}
\mathcal{P}_K =
\frac{
    \rho_{\mathrm{D}}(P_t)\;
    \dGbX{E}{Q\pm2i\,P_s}\;
    \dGbX{F}{\frac{Q}{2}-i\,P_0 \pm 2i\,P_t}
}{
    \SbX{A}{\frac{Q}{2}+i\,P_0}\;
    \dGbX{B}{Q \pm 2i\,P_t}\;
    \dGbX{C}{\frac{Q}{2} -i \, P_0 \pm 2i\, P_s}
}.
\end{equation}
The six labels $(\mathrm{A},\mathrm{B},\mathrm{C}\,;\,\mathrm{D},\mathrm{E},\mathrm{F})
\in\{\mathrm{NS},\mathrm R\}^6$ specify each kernel in question. One can explicitly
characterize these in the following manner:
\begin{equation}
\renewcommand{\arraystretch}{1.25}
\begin{tabular}{l| | c  c  cccccc}
Kernel & covariant factor & $\mathcal C_K$ &
A & B & C & D & E & F\\
\hline
\hline
$\Sm{V}{33}{}$    &  $(-i\tau)^{h_0}$  &  $\tfrac14$              & NS & NS & NS & NS & NS & NS\\[1ex]
$\Sm{W}{33}{}$    &   $(-i\tau)^{h_0+1/2}$   &  $\tfrac14$              & R  & NS & R  & NS & NS & R \\[1ex]
$\Sm{V}{42}{}$      &  $(-i\tau)^{h_0}$  &  $\tfrac12$              & NS & R  & NS & R  & NS & R \\[1ex]
$\Sm{W}{42}{}$    &  $(-i\tau)^{h_0+1/2}$ &  $\tfrac{e^{i\pi/4}}{2\sqrt 2}$ & R  & R  & R  & R  & NS & NS\\[1ex]
$\Sm{V}{11}{}$     &  $(-i\tau)^{h_0}$   &  $\tfrac{1}{2\sqrt 2}$  & R  & R  & NS & R  & R  & NS\\[1ex]
$\Sm{DV}{11}{}$    & $(-i\tau)^{h_0+3/2}$  &  $\tfrac{1}{2i\sqrt 2}$ & NS & R  & R  & R  & R  & R
\end{tabular}
\end{equation}
Note that the parity of the prefactor $\mathcal{P}_K$ w.r.t.~$P_s$ and $P_t$ is always trivial, as can be seen from~\eqref{eq:genpref}.

\paragraph{The integrand:} To write the integrand compactly, introduce the following two combinations of the momenta
\begin{equation}
\begin{aligned}
\mathfrak{p}_{+} & = \frac{Q}{4}+\frac{i}{2}(2\,P_t+P_0),     & \quad
\mathfrak{p}_{-} & = \frac{Q}{4}+\frac{i}{2}(-2\,P_t+P_0), \\
\mathfrak{r}_{+} & = \frac{3\,Q}{4}+\frac{i}{2}(2\,P_t-P_0),  & \quad
\mathfrak{r}_{-} & = \frac{3\,Q}{4}+\frac{i}{2}(-2\,P_t-P_0).
\end{aligned}
\end{equation}
Then each kernel's integrand can be expressed as a sum of two terms of the form
\begin{equation}
\mathcal I_K \;=\;
\frac{\SbX{A_1}{u+ \mathfrak{p}_{+}}\, \SbX{A_2}{u+ \mathfrak{p}_{-}}}{
    \SbX{B_1}{u+\mathfrak{r}_{+}}\,\SbX{B_2}{u+\mathfrak{r}_{-}}}
\;+\;\sigma_K\;
\frac{\SbX{A'_1}{u+\mathfrak{p}_{+}}\,\SbX{A'_2}{u+ \mathfrak{p}_{-}}}{
    \SbX{B'_1}{u+\mathfrak{r}_{+}}\,
    \SbX{B'_2}{u+\mathfrak{r}_{-}}}\,.
\end{equation}
Here $\sigma_K \in \set{+1, -1}$ is a relative sign, and the labels $(\mathrm{A}_1, \mathrm{A}_2 ; \mathrm{B}_1, \mathrm{B}_2)$ are reserved for the first term, and their primed counterparts $(\mathrm{A}'_1, \mathrm{A}'_2 ; \mathrm{B}'_1, \mathrm{B}'_2)$ are reserved for the second term. Explicitly, we have
\begin{equation}
\renewcommand{\arraystretch}{1.25}
\begin{tabular}{l| |c  cccc c cccc | cc}
Kernel & $\sigma_K$ &
$\mathrm{A}_1$ & $\mathrm{A}_2$ & $\mathrm{B}_1$ & $\mathrm{B}_2$ & &
$\mathrm{A}'_1$ & $\mathrm{A}'_2$ & $\mathrm{B}'_1$ & $\mathrm{B}'_2$ 
& \multicolumn{2}{c}{Parity in $(P_s,P_t)$}\\
\hline
\hline
$\Sm{V}{33}{}$  & $+$ & NS & NS & NS & NS & & R  & R  & R  & R & even & even \\
$\Sm{W}{33}{}$  & $+$ & NS & NS & R  & R  & & R  & R  & NS & NS & even & even  \\
$\Sm{V}{42}{}$  & $+$ & NS & R  & NS & R  & & R  & NS & R  & NS & even & even \\
$\Sm{W}{42}{}$ & $-$ & NS & R  & R  & NS & & R  & NS & NS & R & even & odd \\
$\Sm{V}{11}{}$         & $+$ & NS & R  & R  & NS & & R  & NS & NS & R & even & even \\
$\Sm{DV}{11}{}$         & $-$ & NS & R  & NS & R  & & R  & NS & R  & NS & odd &  odd \\
\end{tabular}
\end{equation}
In the last column we record the parity of the \textit{full} integral $\int_{i\mathbb{R}}\!\frac{du}{i}\;e^{-2\pi u P_s}\,\mathcal{I}_K(u,P_0,P_s,P_t)$ w.r.t. reflections of $P_s$ and $P_t$.
\par We note finally that the convergence condition of the integrals
\begin{equation*}
    \int_{i\mathbb{R}}\!\frac{du}{i}\;
e^{-2\pi u P_s}\, \mathcal{I}_K(u,P_0,P_s,P_t)
\end{equation*}
associated to the six kernels described in the table above is \emph{universally} 
\begin{equation}
    \abs{\Im P_s}<\frac{Q}{4}+\frac{\Im P_0}{2}.
\end{equation}
This is easily seen by the asymptotics of $\SNS{b}{z}$ and $\SR{b}{z}$ as $\abs{z}\rightarrow \infty$ for fixed $b$, which are in turn dictated by the asymptotics of $S_b(z)$.\footnote{The latter has the following behavior:
\begin{equation*}
    \log S_{b}(z)=-\frac{i\pi}{2}z^2+\frac{i\pi}{2}Qz-\frac{i\pi}{12}(Q^2+1)+O(z^{-1}), \quad \abs{z}\rightarrow \infty \, , \, \Im z\geq0.
\end{equation*}
   For $\abs{z}\rightarrow \infty$ in the lower half-plane, one uses the property $S_b(x)=1/S_b(Q-x)$ and the previous asymptotics.}

\subsection{The \texorpdfstring{$S$}{S}-kernel bootstrap}\label{sec:modSbootstrap}

We now outline a bootstrap process that determines the form of the six unknown kernels starting from the knowledge of $\Sm{I}{33}{}$ from~\cite{Apresyan:2023vxl}.
In particular, we will bootstrap explicitly the kernels
$\Sm{I}{42}{P_s P_t}[P_0]$, $\Sm{V}{11}{P_s P_t}[P_0]$, and $\Sm{DV}{11}{P_sP_t}[P_0]$.
The kernels $\Sm{I}{24}{P_s P_t}[P_0]$ will be determined easily from crossing and idempotency.

The bootstrap conditions which constitute  the basis of our construction are the following:
\begin{enumerate}
\item The kernels branch the corresponding modular-transformed blocks over the Liouville spectrum ($P\in\mathbb{R}_+$ in NS and R sectors) with the appropriate covariant factor $(-i\tau)^{h_0 + \delta}$ in each case.
\item The kernels are invariant under $b\leftrightarrow b^{-1}$.
\item They are meromorphic functions of the momenta $P_s$, $P_t$, and $P_0$ and reflect the singularities of the corresponding blocks.
\item All eight kernels admit a qualitatively similar (product structure) representation in some domain of the momenta, namely that of a meromorphic prefactor times a meromorphic integral.
\item They respect the parity properties of the corresponding blocks under reflections $P_i\rightarrow -P_i$.
\item They satisfy idempotency in the appropriate form, namely the condition $S^2=1$. 
\item Whenever applicable, we get a consistent $P_0\rightarrow iQ/2$ limit to the modular kernel that branches torus characters.
\end{enumerate}
Armed with these conditions, we will essentially be able to write down the hitherto unknown kernels uniquely.

\paragraph{The kernels in the NS sector:}
Before we proceed to implement our bootstrap for the hitherto unknown $S$ modular kernels, let us take stock of the results from~\cite{Apresyan:2023vxl}.
First, the kernel $\Sm{V}{33}{}$ was determined to be (cf.~their Eq.~(67))
\begin{equation}\label{eq:sv33}
\begin{aligned}
\Sm{V}{33}{P_sP_t}[P_0]
   & =
\frac{\rhoNS(P_t)}{4\,\SNS{b}{\frac{Q}{2}+iP_0}}\frac{
    \dGNS{b}{Q\pm 2i\, P_s}\, \dGNS{b}{\frac{Q}{2}-i\,P_0 \pm 2i\, P_t}}{
    \dGNS{b}{Q\pm 2i \,P_t}\, \dGNS{b}{\frac{Q}{2}-i\,P_0\pm 2i\, P_s}} \\
   & \times
\int_{i\mathbb{R}}\frac{du}{i} \,
e^{-2\pi \,u\, P_s}
\bqty{
    \frac{
    \SNS{b}{u+ \frac{Q}{4} \pm i\,P_t + \frac{i}{2}\,P_0 }}{
    \SNS{b}{u + \frac{3\,Q}{4} \pm i\,P_t - \frac{i}{2}\, P_0}}+
    \frac{
    \SR{b}{u + \frac{Q}{4} \pm i\,P_t + \frac{i}{2}\, P_0}}{
    \SR{b}{u + \frac{3\,Q}{4} \pm i\,P_t - \frac{i}{2}\, P_0}}
} \,.
\end{aligned}
\end{equation}
Next, the kernel  $\Sm{W}{33}{}$ was determined to be (cf.~their Eq.~(84))
\begin{equation}\label{eq:sw33}
\begin{aligned}
\Sm{W}{33}{P_sP_t}[P_0]
   & =
\frac{\rhoNS(P_t)}{4\,\SR{b}{\frac{Q}{2}+iP_0}}\frac{
    \dGNS{b}{Q\pm 2i\, P_s}\, \dGR{b}{\frac{Q}{2}-i\,P_0 \pm 2i\, P_t}}{
    \dGNS{b}{Q\pm 2i \,P_t}\, \dGR{b}{\frac{Q}{2}-i\,P_0\pm 2i\, P_s}} \\
   & \times
\int_{i\mathbb{R}}\frac{du}{i} \,
e^{-2\pi \,u\, P_s}
\bqty{
    \frac{
    \SNS{b}{u+ \frac{Q}{4} \pm i\,P_t + \frac{i}{2}\,P_0 }}{
    \SR{b}{u + \frac{3\,Q}{4} \pm i\,P_t - \frac{i}{2}\, P_0}}+
    \frac{
    \SR{b}{u + \frac{Q}{4} \pm i\,P_t + \frac{i}{2}\, P_0}}{
    \SNS{b}{u + \frac{3\,Q}{4} \pm i\,P_t - \frac{i}{2}\, P_0}}
} \,.
\end{aligned}
\end{equation}
In writing these expressions, we have stripped off a factor of $e^{-\frac{i\pi}{2}\, h_0}$ from the kernel. This is inconsequential, and can be compensated by appropriate factors in the definition of the crossing transformations.
Note that both of these expressions conform to the structural form summarized in~\cref{sec:modSsummary} and can be written straightforwardly in the uniform notation introduced therein.

Let us record some useful facts about these kernels.
\begin{itemize}[wide,left=0pt]
\item  Firstly, we may numerically check using our torus blocks that both of these expressions satisfy~\eqref{eq:SFtransform}. The kernels $\Sm{I}{33}{}$ do so for the blocks $\mathcal{F}^{\mathrm{I}}_3$, respectively (the $W$ kernel  correspondingly has $\delta = \frac{1}{2}$).
\item Second, when the Liouville momenta are real valued $P_s, \, P_t, \, P_0 \, \in \mathbb{R}$, the kernels are also real valued, $\Sm{I}{33}{P_s P_t}[P_0] \in \mathbb{R}$.
\item Both the kernels are manifestly invariant under reflections of the internal momenta $P_s$ and $P_t$, respectively. While not manifest, they are also invariant under $P_0 \to -P_0$, a reflection of the fact that the blocks are invariant under reflections of all the momenta.
\item Finally, one has to check that these $S$ modular kernels square to the identity, which in the present case implies 
\begin{equation}\label{eq:vwNSid}
\begin{aligned}
\int_{\mathbb{R}}\, dP\;
\Sm{I}{33}{P_1 P}[P_0] \,
\Sm{I}{33}{P P_2}[P_0]
   & =
\frac{1}{2} \,\bqty{\delta(P_1 - P_2) + \delta(P_1 + P_2)} \,. 
\end{aligned}
\end{equation}
\end{itemize}

There is one other useful relation that one can extract from the $S$-modular kernels. Start with the modular covariance of the torus 1-point functions~\eqref{eq:genmodcov} and express it in terms of the conformal blocks as in~\eqref{eq:Vmodchecks} or~\eqref{eq:Wmodchecks}. Then using the crossing kernel relation~\eqref{eq:SFtransform} along with the requirement of idempotency, we can distill an equation for $\mathbb{S}$ that is independent of the conformal blocks.\footnote{We also need to exploit the fact that the spectral densities, structure constants, and the $S$ modular kernels have definite properties under reflections of their arguments.}
For the NS sector, we require
\begin{equation}\label{eq:vw33rel}
\begin{aligned}
\rhoNS(P_s)\, \CV{b}(P_0,P_s,P_s) \, \Sm{V}{33}{P_s P_t}[P_0]
   & =
\rhoNS(P_t) \,\CV{b}(P_0,P_t,P_t) \, \Sm{V}{33}{P_t P_s}[P_0], \\
\rhoNS(P_s)\, \CW{b}(P_0,P_s,P_s) \, \Sm{W}{33}{P_s P_t}[P_0]
   & =
\rhoNS(P_t) \,\CW{b}(P_0,P_t,P_t) \, \Sm{W}{33}{P_t P_s}[P_0] \,.
\end{aligned}
\end{equation}
Modular covariance of $\mathcal{N} =1$ super Liouville theory in the NS sector with $c \in \mathbb{C}\setminus (-\infty, \frac{3}{2}]$ may equivalently be phrased in terms of such relations.

\subsection{Results from the bootstrap analysis}\label{sec:resultsSmod}

The NS sector $S$ modular kernels obtained by~\cite{Apresyan:2023vxl}, which we have collated in~\eqref{eq:sv33} and~\eqref{eq:sw33},  clearly satisfy all the outlined requirements of the bootstrap program. One can then attempt to fix the remaining six $S$ modular kernels systematically by examining similar ans\"atze. We outline the salient aspects of this procedure for the different sectors below.

\subsubsection{The $\widetilde{\mathrm{NS}}$-R sector crossing kernels}\label{sec:NStRcross}

The first set of $S$ modular kernels we would want to bootstrap concerns the $\widetilde{\mathrm{NS}}$ blocks, $\mathcal{F}^{\mathrm{I}}_4$. These mix with the R sector blocks $\mathcal{F}_2^{\mathrm{I}}$ under the $S$ transform, as for example can be deduced from the transformation properties of the generic character of the super Virasoro algebra
\begin{equation}
\chi_P^{\smash[b]{\widetilde{\mathrm{NS}}}}(-1/\tau)
= \int_0^\infty \, dP' \pqty{\sqrt{2}\, \cos(2\pi \, P\, P')}\,
\chi_{P'}^{\smash[b]{\mathrm{R}}}(\tau)\,.
\end{equation}
Explicitly, for the conformal blocks, we require the following relation to hold
\begin{equation}
\begin{aligned}
(-i\,\tau)^{h_0 + \delta}\,
\mathcal{F}^{\mathrm{I}}_4(P_0,P_s \mid q)
   & =
\int_{\mathbb{R}}\, dP_t\, \Sm{I}{42}{P_s P_t}[P_0]\,
\mathcal{F}^{\mathrm{I}}_2(P_0,P_t \mid \widetilde{q})  \,.
\end{aligned}
\end{equation}

\paragraph{The kernels for $\expval{V}$:}
Let us start with the kernels involving the superconformal primary $V_P$.  For the crossing from the $\widetilde{\mathrm{NS}} \to \mathrm{R}$, we deduce from our bootstrap ansatz the following answer:  
\begin{equation}\label{eq:sv42}
\begin{aligned}
\Sm{V}{42}{P_sP_t}[P_0]
   & =
\frac{\rhoR(P_t)}{2\,\SNS{b}{\frac{Q}{2}+iP_0}}\frac{
    \dGNS{b}{Q\pm 2i\, P_s}\, \dGR{b}{\frac{Q}{2}-i\,P_0 \pm 2i\, P_t}}{
    \dGR{b}{Q\pm 2i \,P_t}\, \dGNS{b}{\frac{Q}{2}-i\,P_0\pm 2i\, P_s}} \\
   & \quad  \times
\int_{i\mathbb{R}}\frac{du}{i} \,
e^{-2\pi \,u\, P_s}
\Bigg[
    \frac{
    \SNS{b}{u+ \frac{Q}{4} + i\,P_t + \frac{i}{2}\,P_0 }\,
    \SR{b}{u+ \frac{Q}{4} - i\,P_t + \frac{i}{2}\,P_0 }}{
    \SNS{b}{u + \frac{3\,Q}{4} + i\,P_t - \frac{i}{2}\, P_0}\,
    \SR{b}{u + \frac{3\,Q}{4} - i\,P_t - \frac{i}{2}\, P_0}}
    \\
    &\quad \qquad \qquad +
    \frac{
    \SR{b}{u + \frac{Q}{4} + i\,P_t + \frac{i}{2}\, P_0}\,
    \SNS{b}{u + \frac{Q}{4} - i\,P_t + \frac{i}{2}\, P_0}
    }{
    \SR{b}{u + \frac{3\,Q}{4} + i\,P_t - \frac{i}{2}\, P_0}\,
    \SNS{b}{u + \frac{3\,Q}{4} - i\,P_t - \frac{i}{2}\, P_0}
    }
    \Bigg] \,.
\end{aligned}
\end{equation}
We can make the following observations about this kernel. It is a manifestly real-valued function of the momenta, and is also even under $P_t \to -P_t$. Using reflection identities of the $\SNS{b}{z}$ and $\SR{b}{z}$ it can also be seen to be even under $P_s \to -P_s$.
Finally, it also appears to be an even function of $P_0$, which while not manifest, can be checked numerically.

In this case the idempotency requirement is non-trivial. By implementing an $S$ transform we pass from $\widetilde{\mathrm{NS}}$ to R sector, so we have to perform another $S$ transformation to get back to the starting sector. This means that we should combine $\Sm{V}{42}{}$ and $\Sm{V}{24}{}$ to test idempotency. Specifically, the constraints are
\begin{equation}\label{eq:vNStRid}
\begin{aligned}
\int_{\mathbb{R}}\, dP_{\mathrm{NS}}\,
\Sm{V}{24}{P_1 P_{\mathrm{NS}}}[P_0]\,
\Sm{V}{42}{P_{\mathrm{NS}} P_2}[P_0]
   & = \frac{1}{2} \,\bqty{\delta(P_1 - P_2) + \delta(P_1 + P_2)} \,,
\\
\int_{\mathbb{R}}\, dP_{\mathrm{R}}\,
\Sm{V}{42}{P_1 P_{\mathrm{R}}}[P_0]\,
\Sm{V}{24}{P_{\mathrm{R}} P_2}[P_0]
   & = \frac{1}{2} \,\bqty{\delta(P_1 - P_2) + \delta(P_1 + P_2)} \,.
\end{aligned}
\end{equation}
These can be manipulated to derive an expression for $\Sm{V}{24}{}$. One finds, for example,
\begin{equation}
\rhoNS(P_t)\, \CV{b}(P_0,P_t,P_t) \, \Sm{V}{42}{P_t P_s}[P_0]
=
2\,\rhoR(P_s) \,\Ce{b}(P_0;P_s,P_s) \, \Sm{V}{24}{P_s P_t}[P_0] \,.
\end{equation}
This can be immediately used to solve
\begin{equation}\label{eq:sv24}
\Sm{V}{24}{P_s P_t}[P_0]
=
\frac{\rhoNS(P_t)\, \CV{b}(P_0,P_t,P_t)}{2\,\rhoR(P_s) \,\Ce{b}(P_0;P_s,P_s)}\,
\Sm{V}{42}{P_t P_s}[P_0]\,.
\end{equation}

\paragraph{The kernels for $\expval{W}$:}  The analysis for the super descendant $W$ can be similarly carried out. For the $\widetilde{\mathrm{NS}} \to R$ direction, we find
\begin{equation}\label{eq:sw42}
\begin{aligned}
\Sm{W}{42}{P_sP_t}[P_0]
   & =
\frac{e^{\frac{i\pi}{4}}}{2\sqrt{2}}\,
\frac{\rhoR(P_t)}{\SR{b}{\frac{Q}{2}+iP_0}}\frac{
    \dGNS{b}{Q\pm 2i\, P_s}\, \dGNS{b}{\frac{Q}{2}-i\,P_0 \pm 2i\, P_t}}{
    \dGR{b}{Q\pm 2i \,P_t}\, \dGR{b}{\frac{Q}{2}-i\,P_0\pm 2i\, P_s}} \\
   & \quad  \times
\int_{i\mathbb{R}}\frac{du}{i} \,
e^{-2\pi \,u\, P_s}
\Bigg[
    \frac{
    \SNS{b}{u+ \frac{Q}{4} + i\,P_t + \frac{i}{2}\,P_0 }\,
    \SR{b}{u+ \frac{Q}{4} - i\,P_t + \frac{i}{2}\,P_0 }}{
    \SR{b}{u + \frac{3\,Q}{4} + i\,P_t - \frac{i}{2}\, P_0}\,
    \SNS{b}{u + \frac{3\,Q}{4} - i\,P_t - \frac{i}{2}\, P_0}}
    \\
    &\quad \qquad \qquad -
    \frac{
    \SR{b}{u + \frac{Q}{4} + i\,P_t + \frac{i}{2}\, P_0}\,
    \SNS{b}{u + \frac{Q}{4} - i\,P_t + \frac{i}{2}\, P_0}
    }{
    \SNS{b}{u + \frac{3\,Q}{4} + i\,P_t - \frac{i}{2}\, P_0}\,
    \SR{b}{u + \frac{3\,Q}{4} - i\,P_t - \frac{i}{2}\, P_0}
    }
    \Bigg] \,.
\end{aligned}
\end{equation}
There are now a few small differences. First, this kernel is complex. While it is even under $P_s \to -P_s$, it is actually odd under $P_t \to - P_t$. This owes to the presence of the explicit relative sign between the two terms in the integrand. It also can be checked to be  an even function of $P_0$.

To obtain the kernel for the reverse direction, we employ the analog of the idempotency condition~\eqref{eq:vNStRid}, which now reads 
\begin{equation}\label{eq:wNStRid}
\begin{aligned}
\int_{\mathbb{R}}\, dP_{\mathrm{NS}}\,
\Sm{W}{24}{P_1 P_{\mathrm{NS}}}[P_0]\,
\Sm{W}{42}{P_{\mathrm{NS}} P_2}[P_0]
   & =
\frac{P_1}{P_2}\times \frac{1}{2} \,\bqty{\delta(P_1 - P_2) + \delta(P_1 + P_2)} \,,
\\
\int_{\mathbb{R}}\, dP_{\mathrm{R}}\,
\Sm{W}{42}{P_1 P_{\mathrm{R}}}[P_0]\,
\Sm{W}{24}{P_{\mathrm{R}} P_2}[P_0]
   & = \frac{1}{2} \,\bqty{\delta(P_1 - P_2) + \delta(P_1 + P_2)} \,.
\end{aligned}
\end{equation}
We draw attention here to the extra factor of $P_1/P_2$ in the first relation --- this factor ensures that the kernels have the correct parity property. One can trace its origins to the supercharge action and its imprint on the blocks, cf.~\eqref{eq:GGhalfpolsum}. This then allows us to derive an expression for $\Sm{W}{24}{}$, obtaining
\begin{equation}\label{eq:sw24}
\Sm{W}{24}{P_s P_t}[P_0]
=
\frac{\rhoNS(P_t)\, \CW{b}(P_0,P_t,P_t)}{2\,\rhoR(P_s) \,\Co{b}(P_0;P_s,P_s)}\,
\Sm{W}{42}{P_t P_s}[P_0]\,.
\end{equation}
%

\subsubsection{The $\widetilde{\mathrm{R}}$ sector crossing kernels}\label{sec:Rtcross}

Like the NS sector, the $\widetilde{\mathrm{R}}$ sector, which is the sole odd spin structure on the torus, transforms into itself under $S$ transform. As we have noted at various points in our exposition, there are several novel features in this sector, like the vanishing of
$\expval{W}_1$, and the effective primality of the level three-half descendant $\widetilde{G}_{-\frac{3}{2}}G_{-\frac{3}{2}}  V$. We are left to determine the crossing kernel for the transformation of $\expval{V}_1$ and $\expval{\widetilde{G}_{-\frac{3}{2}} G_{-\frac{3}{2}}   V}_1$.

\paragraph{The kernel for $\expval{V}_1$:} We start with the modular kernel relating the one-point function of $\expval{V}_1$ to itself under $S$ transformation.
Following the general requirements outlined at the outset, we find
\begin{equation}\label{eq:sv11}
\begin{aligned}
\Sm{V}{11}{P_sP_t}[P_0]
   & =
\frac{1}{2\sqrt{2}}\,
\frac{\rhoR(P_t)}{\SR{b}{\frac{Q}{2}+iP_0}}\frac{
    \dGR{b}{Q\pm 2i\, P_s}\, \dGNS{b}{\frac{Q}{2}-i\,P_0 \pm 2i\, P_t}}{
    \dGR{b}{Q\pm 2i \,P_t}\, \dGNS{b}{\frac{Q}{2}-i\,P_0\pm 2i\, P_s}} \\
   & \quad  \times
\int_{i\mathbb{R}}\frac{du}{i} \,
e^{-2\pi \,u\, P_s}
\Bigg[
    \frac{
    \SNS{b}{u+ \frac{Q}{4} + i\,P_t + \frac{i}{2}\,P_0 }\,
    \SR{b}{u+ \frac{Q}{4} - i\,P_t + \frac{i}{2}\,P_0 }}{
    \SR{b}{u + \frac{3\,Q}{4} + i\,P_t - \frac{i}{2}\, P_0}\,
    \SNS{b}{u + \frac{3\,Q}{4} - i\,P_t - \frac{i}{2}\, P_0}}
    \\
    &\quad \qquad \qquad +
    \frac{
    \SR{b}{u + \frac{Q}{4} + i\,P_t + \frac{i}{2}\, P_0}\,
    \SNS{b}{u + \frac{Q}{4} - i\,P_t + \frac{i}{2}\, P_0}
    }{
    \SNS{b}{u + \frac{3\,Q}{4} + i\,P_t - \frac{i}{2}\, P_0}\,
    \SR{b}{u + \frac{3\,Q}{4} - i\,P_t - \frac{i}{2}\, P_0}
    }
    \Bigg] \,.
\end{aligned}
\end{equation}
This is an even, real-valued function of the momentum arguments. Even parity under $P_t$ reflection is manifest, while that under $P_s$ reflection requires using the reflection properties of $\SNS{b}{z}$ and $\SR{b}{z}$ under $z \to Q -z $. Working through the idempotency requirement, which reads
\begin{equation}\label{eq:sv11id}
\int_{\mathbb{R}}\, dP \, \Sm{V}{11}{P_1 P}[P_0]\,
\Sm{V}{11}{P P_2}[P_0]
= \frac{1}{2} \bqty{\delta(P_1 - P_2) + \delta(P_1 + P_2)}\,,
\end{equation}
one obtains the identity respected by the kernel
\begin{equation}\label{eq:sv11rel}
\rhoR(P_s)\, \Co{b}(P_0;P_s,P_s) \, \Sm{V}{11}{P_s P_t}[P_0]
=
\rhoR(P_t) \,\Co{b}(P_0;P_t,P_t) \, \Sm{V}{11}{P_t P_s}[P_0] \,.
\end{equation}

\paragraph{The kernel for $\expval{\widetilde{G}_{-\frac{3}{2}} G_{-\frac{3}{2}}V}_1$:}
Finally, we turn to the most exotic modular kernel, corresponding to $\widetilde{G}_{-\frac{3}{2}} G_{-\frac{3}{2}} V$. Working through the details we find
\begin{equation}\label{eq:sGv11}
\begin{aligned}
\Sm{DV}{11}{P_sP_t}[P_0]
   & =
\frac{1}{2i\,\sqrt{2}}\,
\frac{\rhoR(P_t)}{\SNS{b}{\frac{Q}{2}+iP_0}}\frac{
    \dGR{b}{Q\pm 2i\, P_s}\, \dGR{b}{\frac{Q}{2}-i\,P_0 \pm 2i\, P_t}}{
    \dGR{b}{Q\pm 2i \,P_t}\, \dGR{b}{\frac{Q}{2}-i\,P_0\pm 2i\, P_s}} \\
   & \quad  \times
\int_{i\mathbb{R}}\frac{du}{i} \,
e^{-2\pi \,u\, P_s}
\Bigg[
    \frac{
    \SNS{b}{u+ \frac{Q}{4} + i\,P_t + \frac{i}{2}\,P_0 }\,
    \SR{b}{u+ \frac{Q}{4} - i\,P_t + \frac{i}{2}\,P_0 }}{
    \SNS{b}{u + \frac{3\,Q}{4} + i\,P_t - \frac{i}{2}\, P_0}\,
    \SR{b}{u + \frac{3\,Q}{4} - i\,P_t - \frac{i}{2}\, P_0}}
    \\
    &\quad \qquad \qquad -
    \frac{
    \SR{b}{u + \frac{Q}{4} + i\,P_t + \frac{i}{2}\, P_0}\,
    \SNS{b}{u + \frac{Q}{4} - i\,P_t + \frac{i}{2}\, P_0}
    }{
    \SR{b}{u + \frac{3\,Q}{4} + i\,P_t - \frac{i}{2}\, P_0}\,
    \SNS{b}{u + \frac{3\,Q}{4} - i\,P_t - \frac{i}{2}\, P_0}
    }
    \Bigg] \,.
\end{aligned}
\end{equation}
Despite the explicit overall factor of $i$ this particular kernel is real-valued like the others discussed earlier. One can understand the reason by looking at the corresponding modular transformation of the blocks~\eqref{eq:SFtransform}: for purely imaginary $\tau$ and real momenta $P_0$, $P_s$, and $P_t$ the phase factor $e^{i\pi/4}$ of the blocks $\mathcal{F}^{\mathrm{DV}}_1$ (which are common on both sides of the equation) is irrelevant. Therefore, the kernel that relates them must be real, which can be confirmed by explicit evaluation. 

$\Sm{DV}{11}{P_sP_t}[P_0]$ has odd parity under $P_s$ and $P_t$ reflections, but is even under $P_0 \to -P_0$.  The idempotency requirement is similar structurally to~\eqref{eq:wNStRid}, again with an extra momentum factor thanks to the factor of momentum arising from supercharge action in the block. We have
\begin{equation}\label{eq:sGv11id}
\int_{\mathbb{R}}\, dP \,
\Sm{DV}{11}{P_1 P}[P_0]\,
\Sm{DV}{11}{P P_2}[P_0]
= \frac{P_1}{P_2}\times \frac{1}{2} \bqty{\delta(P_1 - P_2) + \delta(P_1 + P_2)}\,.
\end{equation}
This expression can be distilled into an identity for the kernel
\begin{equation}\label{eq:sGvvrel}
\rhoR(P_s)\, \Ce{b}(P_0;P_s,P_s) \, \Sm{DV}{11}{P_s P_t}[P_0]
=
\rhoR(P_t) \,\Ce{b}(P_0;P_t,P_t) \, \Sm{DV}{11}{P_t P_s}[P_0] \,.
\end{equation}
If we accept that this modular crossing kernel correctly implements the modular transformation of the corresponding conformal blocks for the superdescendant one-point function (\ref{eq:GGV1}) (which we check numerically below), then this relation proves that the latter is correctly modular covariant in spacelike Liouville SCFT. 

\subsection{Numerical evidence}\label{sec:numS}
\begin{figure}[htbp!]
\centering
\includegraphics[width=0.95\textwidth]{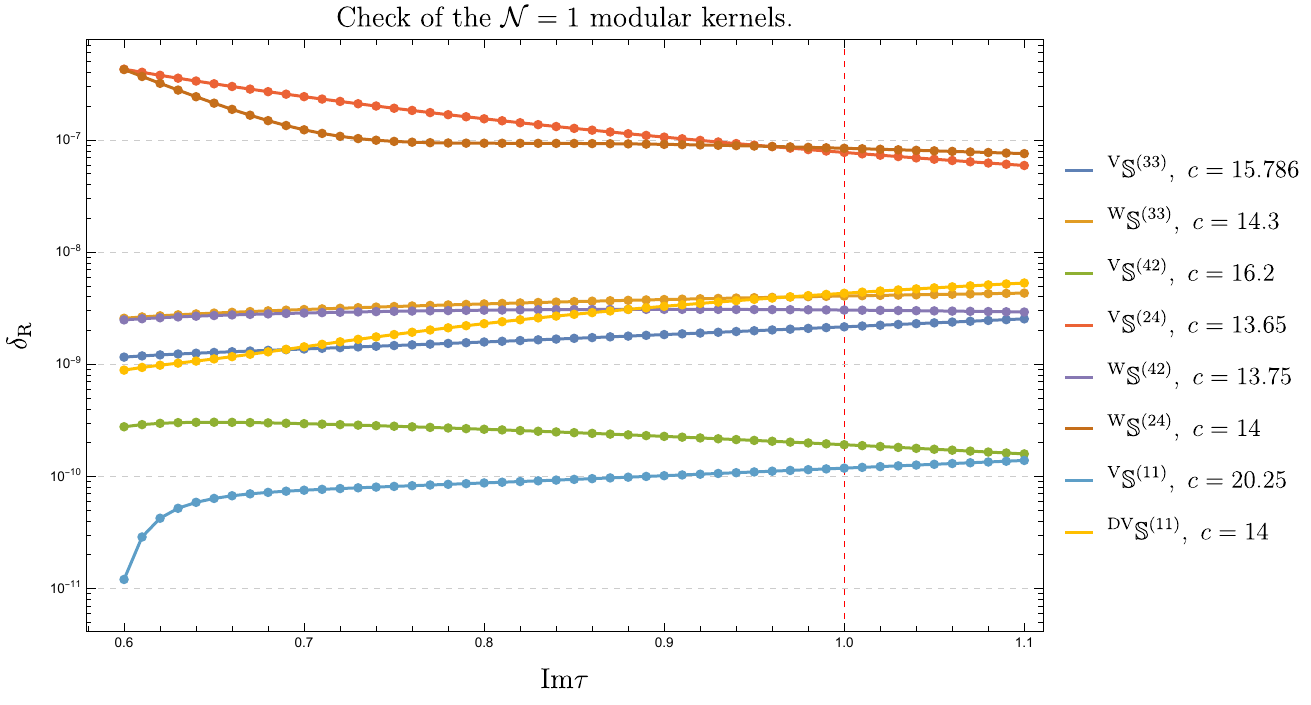}
\caption{Checks of the eight $S$ crossing kernels listed in~\eqref{eq:N1Snotation}. We check that each of them satisfies the crossing transformation~\eqref{eq:SFtransform}.  The values of the central charge used in each case are indicated in the legend.  The momenta were chosen to
be $P_0 = 0.04 + 0.03 \,i$ and $P_s = 0.3 + 0.02\,i$. The kernels were integrated against the blocks as a function of $P_t$. We have extracted data for the deviation defined in~\eqref{eq:Skerneldelta} for a range of values of $\Im(\tau)$. }
\label{fig:Skernel-check}
\end{figure}

We verify our proposal for the $S$-kernels for $\mathcal{N}=1$ super Liouville theory by an explicit crossing check. We pick values of the momenta and compute the direct and $S$-transformed values for each of the eight cases listed in~\eqref{eq:N1Snotation}. To present the results of the numerical check compactly, define
\begin{equation}\label{eq:Skerneldelta}
\delta_{\mathrm{R}}  = \abs{\frac{F(q)}{F(\widetilde{q})} -1} \,,
\end{equation}
where $F(q)$ and $F(\widetilde{q})$ are the l.h.s.~and r.h.s.~of~\eqref{eq:SFtransform}, respectively.
In practice, we fix $P_0$ and $P_s$ and examine the deviation measure $\delta_{\mathrm{R}}$  as a function of $\Im(\tau)$. The result is shown in~\cref{fig:Skernel-check}.
In performing these checks we have chosen to illustrate the behavior for the momenta $P_0$ and $P_s$ to have non-vanishing imaginary values to emphasize the meromorphic structure of the crossing kernels.

\section{Applications: the super Virasoro minimal string}\label{sec:svms}

As a concrete application of our considerations, we will analyze the worldsheet theory of the super Virasoro minimal string. This worldsheet construction is the supersymmetric generalization of the bosonic Virasoro minimal string introduced in~\cite{Collier:2023cyw}, and was briefly analyzed in~\cite{Muhlmann:2025ngz,Rangamani:2025wfa}. Recently, these theories (as we shall see there are \emph{four}) were analyzed from a chiral three-dimensional gravity perspective in~\cite{Eberhardt:2026hfh}. This perspective proves to be extremely valuable, for~\cite{Eberhardt:2026hfh} was able to derive explicit predictions for the observables.
Matrix models dual to these theories have also been analyzed in~\cite{Johnson:2024fkm,Johnson:2025vyz,Johnson:2026ugu}.

As we shall describe below, two of the theories share their observables with the bosonic string while a third is predicted to have vanishing stable NS amplitudes. Two of these theories are related to the supersymmetric JT theory in the semiclassical limit ($b \to 0$), which was the original motivation behind their study.
We will give a quick synopsis of the worldsheet theory, reviewing the basic observables (the sphere 3-point functions) that were computed in the earlier works. That will set the stage for our analysis on the torus, where, as we shall see, we will verify the predictions of~\cite{Eberhardt:2026hfh}.

\subsection{Basics of the RNS worldsheet theory}\label{sec:wsreview}

The super Virasoro minimal string comprises as matter two copies of super Liouville theory, one spacelike, and the other timelike, coupled to two-dimensional $\mathcal{N} = (1,1)$ supergravity. Specifically, the worldsheet theory's ingredients are
\begin{enumerate}
\item A spacelike super Liouville theory (denoted as $\mathrm{sL}$) with $c=\frac{3}{2}+3\,Q^2\geq \frac{27}{2}$. We set $Q=b+b^{-1}$, with $b\in \mathbb{R}_+$.
\item  A timelike super Liouville theory (denoted as $\mathrm{tL}$) with $\ch=\frac{3}{2}-3\,\Qh^2 \leq\frac{3}{2}$. We let $\Qh=\bh^{-1} - \bh$, with $\bh\in \mathbb{R}_+$.
\item A $\mathfrak{b}\mathfrak{c}-\beta\gamma$ ghost SCFT corresponding to the worldsheet superdiffeomorphisms, with central charge $c_\mathrm{gh} = -26 + 11 = -15$.
\end{enumerate}
The net central charge, $c + \ch + c_\mathrm{gh} = 0$, vanishes provided
\begin{equation}\label{eq:QQhrel}
Q^2 - \Qh^2 = 4\,, \qquad \bh = b\,.
\end{equation}

The next step is defining the worldsheet currents which have contributions from both the Liouville theories. In the process of coupling the matter theory to supergravity, we need to specify how the supercharges couple to the gravitinos and how the matter stress tensor couples to the gravitons. Energy being additive, the total matter stress-energy tensor is the sum of the spacelike and timelike ones, viz.,
\begin{equation}
\mathcal{L}_m = L_m^{\mathrm{sL}} + L_m^{\mathrm{tL}}\,.
\end{equation}
It is conventionally assumed that the total supercharge likewise is simply additive. However, this does not have to be the case.  There are two independent choices, which we shall label by $\pm$ superscripts. They are defined as\footnote{We thank Lorenz Eberhardt for helpful discussions which clarified the choices in defining consistent worldsheet string theories.
} 
\begin{equation}\label{eq:Gws}
\begin{aligned}
\mathcal{G}^-_r
   & =
G_r^\mathrm{sL} \otimes (-1)^{F_t} + \mathbb{I}^\mathrm{sL} \otimes G_r^\mathrm{tL}\,,
   & \qquad
\widetilde{\mathcal{G}}^-_r
   & =
\widetilde{G}_r^\mathrm{sL} \otimes (-1)^{F_t} +  \mathbb{I}^\mathrm{sL} \otimes \widetilde{G}_r^\mathrm{tL}\,, \\
\mathcal{G}^+_r
   & =
- G_r^\mathrm{sL} \otimes (-1)^{F_t} + \mathbb{I}^\mathrm{sL} \otimes G_r^\mathrm{tL}\,,
   & \qquad
\widetilde{\mathcal{G}}^+_r
   & =
\widetilde{G}_r^\mathrm{sL} \otimes (-1)^{F_t} +  \mathbb{I}^\mathrm{sL} \otimes \widetilde{G}_r^\mathrm{tL}\,.
\end{aligned}
\end{equation}
The cocycle factors, which involve the fermion number operator $(-1)^{F_t}$ of the timelike theory, are present to ensure that the total currents satisfy the $\mathcal{N} =1$ superconformal algebra.\footnote{The combined worldsheet superalgebra is a graded tensor algebra, where one should exercise care with fermion parity of operators in evaluating products, remembering that parity-odd operators in the two factors anticommute. Explicitly, $(A \otimes B)\, (C\otimes D) = (-1)^{\abs{B}\,\abs{C}}\, (A\,C)\otimes (B\,D)$.} We will similarly use $(-1)^{F_s}$ for the fermion number of the spacelike theory. The right moving supercharge is identical in the two cases. 

The choice of sign assigned to the worldsheet supercharges is perhaps unconventional. This assignment is predicated on ensuring that our worldsheet amplitudes agree, as far as possible, with those computed using intersection theory in~\cite{Eberhardt:2026hfh}. The names of the plus and minus theories were inspired by consideration of 3d chiral supergravity on $\mathbf{S}^1 \times \Sigma_g$. Apart from the sum over spin structures on $\Sigma_g$, we also have an option of inserting a factor of $(-1)^{F_{3d}}$, the 3d fermion parity operator along the $\mathbf{S}^1$. A natural convention is that upon dimensional reduction to $\Sigma_g$, the plus theories sum superconformal blocks with no relative sign (i.e., antiperiodic boundary conditions for fermions on the 3d temporal circle), while the minus theories compute signed sums of superconformal blocks (hence periodic boundary conditions for fermions on the 3d temporal circle). The translation to the worldsheet based on the results supports the choice~\eqref{eq:Gws}.

Let us give a quick rationale for why these choices are admissible, and how one may interpret these choices physically. Recall that the classical super Liouville theory
is characterized by a scalar $\phi$ along with its fermion superpartners $\psi$ and $\widetilde{\psi}$. The spacelike and timelike theories differ in the sign of the kinetic term (the timelike scalar has a wrong sign kinetic term). They are also characterized classically by a superpotential that includes a Yukawa coupling $\mu\, \widetilde{\psi}\, \psi\,e^{b\phi}$, where $\mu$ is the cosmological constant.
The scalar potential is insensitive to $\mathrm{sgn}(\mu)$, but clearly the Yukawa coupling depends on it. One can a priori independently choose the sign of this cosmological constant for the spacelike and timelike super Liouville theories. While the overall sign of the choice is immaterial, the relative sign does matter, leading to two different Lagrangians that can be coupled to supergravity. Flipping the relative sign of one of the Yukawa couplings (i.e., of one of the cosmological constants) in the classical action is hence equivalent to flipping the relative sign of the holomorphic supercurrent of one of the worldsheet Liouville CFTs as we have done above.

This choice was appreciated first in the context of the $\hat{c}=1$ Type 0B string~\cite{Douglas:2003up}, where it was argued that the sign of the super Liouville cosmological constant dictates the allowed physical states. The argument used the target space effective action to argue for this. One can also make the argument in the absence of a target space picture, as was done for the minimal superstrings (super Liouville coupled to $\mathcal{N} =1$ minimal models) in~\cite{Klebanov:2003wg,Seiberg:2003nm}. In fact, the features we see below for the super Virasoro minimal string can be seen in these minimal superstrings~\cite{Chakrabhavi:2024szk}. In the present case note that reversing the sign of $G_r^{\mathrm{sL}}$ is equivalent to reversing the signs of
$\CW{b}$ and $\Co{b}$, as noted in~\cref{fn:CWoddsigns}. This will be a useful mnemonic in some of the manipulations below.

The final ingredient we need is a worldsheet fermion number operator $(-1)^{F_\mathrm{ws}}$, which we require for GSO projection. We define it as  
\begin{equation}
(-1)^{F_M} =  (-1)^{F_{\mathrm{s}}}\, (-1)^{F_{\mathrm{t}}} \,, \qquad
(-1)^{F_\mathrm{ws}} = (-1)^{F_M} \, (-1)^{F_\mathrm{gh}} \,.
\end{equation}

Since we have a binary choice for our supercharge, and a binary choice for the GSO projection, viz.,
$(-1)^{F_{\mathrm{ws}}} = \pm1$ we end up with four independent theories.
We refer to these theories following~\cite{Eberhardt:2026hfh} as the Type 0A$^\pm$
and Type 0B$^\pm$ theories. The \zA{-} and \zB{-} will admit a semiclassical limit as Type 0A and 0B JT supergravity, respectively, while the \zA{+} and \zB{+} theories will share many features with the bosonic string. We define the theories more explicitly now by writing down the physical states.\footnote{As we highlight later, in the minus theories, the labels of \zA{-} and \zB{-} will end up being swapped relative to~\cite{Eberhardt:2026hfh}. Instead, our convention agrees with the definition of $\mathcal{N}=1$ JT supergravity in~\cite{Stanford:2019vob}.} We will focus here on generic states and do not dwell on exceptional states in the spectrum that arise at special Liouville momentum. Later in~\cref{sec:comparison} we point out the presence of an exceptional zero-momentum Ramond state, discussed around~\eqref{eq:Rstar}.

\paragraph{NS vertex operators:} The worldsheet superprimary operator that satisfies the super Virasoro constraints can be characterized by an unintegrated physical vertex operator (independent of our choice of supercurrents)
\begin{equation}\label{eq:NSvo11}
\wsNS{P}{\pm}{(-1)}
=
\nNS\, \tilde{\mathfrak{c}} \, \mathfrak{c}
\, e^{-\varphi-\tilde{\varphi}}
\, V_P\, \Vh_{\Ph=-i\,P}\,.
\end{equation}
Here $\nNS$ is a normalization factor for the vertex operator; we will fix this in due course.

In the $(0,0)$ picture, however, the representative depends on our choice of supercharge. In keeping with our right-left action convention for supercharges we define the matter part of the picture changing operator (PCO) as
\begin{equation}
\mathcal{X}^{\rm m}_\sigma = 
\frac{1}{4}\, e^{\tilde{\varphi}}\widetilde{\mathcal{G}}^\sigma_{-\frac{1}{2}}\, e^{\varphi}\mathcal{G}^\sigma_{-\frac{1}{2}}\,, \qquad \sigma \in \set{\pm} \,.
\end{equation}
We have included a numerical factor of a $\frac{1}{4}$ in the operator. 
In the \zA{+}/\zB{+} theories we find\footnote{The cocycle signs are important for the terms with the fermions, which, however, do not play a role in the observables we compute here.}
\begin{equation}\label{eq:NSvo00}
\begin{aligned}
\wsNS{P}{+}{(0)}
   & = 
\frac{\nNS}{4}\, \tilde{\mathfrak{c}} \, \mathfrak{c}\,
\, \widetilde{\mathcal{G}}^{+}_{-\frac{1}{2}} \mathcal{G}^{+}_{-\frac{1}{2}} 
(V_P\, \Vh_{\Ph=-i\,P})\, \\
   & =
\frac{\nNS}{4}\,  \tilde{\mathfrak{c}}\,\mathfrak{c}\,
\pqty{
+ V_P\, \Wh_{-i\,P} - W_P\, \Vh_{-i\,P}
+ \Lambda_P\, \Lambdaht_{-i\,P}
+ \Lambdat_P\, \Lambdah_{-i\,P}
} + \ldots \,,
\end{aligned}
\end{equation}
where the ellipsis refers to additional pure ghost terms that do not contribute to the torus one-point functions in the end.
On the other hand, in \zA{-}/\zB{-} theories the picture $(0,0)$ vertex operator is instead
\begin{equation}\label{eq:NSvo00p}
\begin{aligned}
\wsNS{P}{-}{(0)}
   & =
\frac{\nNS}{4}\,  \tilde{\mathfrak{c}}\,\mathfrak{c}\,
\,\widetilde{\mathcal{G}}^-_{-\frac{1}{2}}\mathcal{G}^{-}_{-\frac{1}{2}}
 (V_P\, \Vh_{\Ph=-i\,P})\,
\\
   & =
\frac{\nNS}{4}\,  \tilde{\mathfrak{c}}\,\mathfrak{c}\,
\pqty{
V_P\, \Wh_{-i\,P} + W_P\, \Vh_{-i\,P}
- \Lambda_P\, \Lambdaht_{-i\,P}
+ \Lambdat_P\, \Lambdah_{-i\,P}
} +\ldots\,.
\end{aligned}
\end{equation}
The action of the supercurrent mode was deduced from the superfield expansions~\eqref{eq:Snsop} and~\eqref{eq:Tnsop}. 

\paragraph{R vertex operators:}  In the Ramond sector, the vertex operators $\RsL{\pm}{P}$ and $\RtL{\pm}{\Ph}$ have definite fermion number. The Ramond sector ground states of the spacelike and timelike theories are
\begin{equation}
\begin{aligned}
\ket{+}_s
   & =
\lim_{z,\zb \to 0}\, e^{i\frac{\pi}{4}} \,\RsL{+}{P}(z,\zb)\,\ket{0} \,,
\qquad
\ket{-}_s =\lim_{z,\zb \to 0}\, e^{i\pi}\,\RsL{-}{P}(z,\zb)\,\ket{0}\,,
\\
\ket{+}_t
   & =
\lim_{z,\zb \to 0}\, e^{i\frac{\pi}{4}} \, \RtL{+}{\Ph}(z,\zb)\,\ket{0} \,,
\qquad
\ket{-}_t =\lim_{z,\zb \to 0}\,  e^{i\pi}\, \RtL{-}{\Ph}(z,\zb)\,\ket{0}\,.
\end{aligned}
\end{equation}
The action of the supercurrent zero mode on these states is
\begin{equation}\label{eq:spacelike and timelike supercurrent zeromode action}
G_0^{\mathrm{sL}} \ket{\pm}_s = -\frac{P}{\sqrt{2}}\, \ket{\mp}_s\,, \qquad
\widetilde{G}_0^{\mathrm{sL}} \ket{\pm}_s = \pm \frac{i\,P}{\sqrt{2}}\, \ket{\mp}_s \,,
\end{equation}
in the spacelike theory. A similar expression holds in the timelike case. Combining right and left movers we can then argue that these states satisfy
\begin{equation}
\begin{aligned}
\frac{1}{\sqrt{2}}\, (G_0^{\mathrm{sL}} + i\, \widetilde{G}^{\mathrm{sL}}_0)
\ket{+}_s
   & =
- P \,\ket{-}_s\,,
   & \qquad
\frac{1}{\sqrt{2}}\, (G^{\mathrm{sL}}_0 - i\, \widetilde{G}^{\mathrm{sL}}_0) \ket{+}_s
   &
= 0\,,
\\
\frac{1}{\sqrt{2}}\, (G^{\mathrm{sL}}_0 - i\, \widetilde{G}^{\mathrm{sL}}_0) \ket{-}_s
   & =
- P\, \ket{+}_s\,,
   & \qquad
\frac{1}{\sqrt{2}}\, (G^{\mathrm{sL}}_0 + i\, \widetilde{G}^{\mathrm{sL}}_0)\ket{-}_s
   & = 0 \,,
\\
\frac{1}{\sqrt{2}}\, (G^{\mathrm{tL}}_0 + i\, \widetilde{G}^{\mathrm{tL}}_0) \ket{+}_t
   & =
- \Ph\, \ket{-}_t\,,
   & \qquad
\frac{1}{\sqrt{2}}\, (G^{\mathrm{tL}}_0 - i\, \widetilde{G}^{\mathrm{tL}}_0) \ket{+}_t
   &
= 0\,,
\\
\frac{1}{\sqrt{2}}\, (G^{\mathrm{tL}}_0 - i\, \widetilde{G}^{\mathrm{tL}}_0) \ket{-}_t
   & =
-\Ph\, \ket{+}_t\,,
   & \qquad
\frac{1}{\sqrt{2}}\, (G^{\mathrm{tL}}_0 + i\, \widetilde{G}^{\mathrm{tL}}_0)\ket{-}_t
   &
= 0 \,.
\end{aligned}
\end{equation}

For physical states, the supercurrent zero modes
$\{\mathcal{G}^\pm_0, \widetilde{\mathcal{G}}^{\pm}_0\}$  must annihilate the state.
There is a naive Hilbert space of four states spanned by $\ket{\eta}_s\,\otimes \ket{\eta'}_t$ for $\eta,\eta' \in
\{+,-\}$. At generic nonzero momentum $P$, the physical combinations
\begin{equation}\label{eq:0BPsi}
\begin{aligned}
\ket{\Psi_+}
   & =
\frac{1}{\sqrt{2}}\, \pqty{\ket{+}_s \, \ket{-}_t
    + i\, \ket{-}_s\, \ket{+}_t}\,,
\\
\ket{\Psi_-}
   & =
\frac{1}{\sqrt{2}}\, \pqty{\ket{+}_s \, \ket{+}_t
    +  i\, \ket{-}_s\, \ket{-}_t}\,,
\end{aligned}
\end{equation}
can satisfy our requirements. With  $\Ph =-i\,P $, the states $\ket{\Psi_\pm}$ lie in the kernel of one of the supercharges:
\begin{equation}
\mathcal{G}_0^+\ket{\Psi_-} = \widetilde{\mathcal{G}}_0^+\ket{\Psi_-} =0\,, \qquad 
\mathcal{G}_0^-\ket{\Psi_+} = \widetilde{\mathcal{G}}_0^-\ket{\Psi_+} =0\,.
\end{equation}

Now let us discuss the fermion number charge of these operators. We stick to the standard convention that  $(-1)^{F_{\mathrm{gh}}} = -1$ in the Ramond sector. For both spacelike and timelike theories we will define the states $\ket{\pm}$ to be $(-1)^F$ eigenstates with eigenvalues $\pm 1$. Then, the two states above $\ket{\Psi_\pm}$ have total fermion number $(-1)^{F_{\mathrm{ws}}} = (-1)^{F_{\mathrm{s}} + F_{\mathrm{t}} + F_{\mathrm{gh}}} = \pm 1$, respectively.

\paragraph{The four string theories:} We can now put all the pieces together and define the four string theories of interest. The rationale we employ is to hew close to the standard definition of Type 0A and 0B GSO projection, with the new ingredient being the supercharge sign decoration.

\begin{itemize}[wide,left=0pt]
\item We define the Type 0A theories by requiring that the GSO projection removes the physical Ramond-sector vertex operators at generic momentum. This fixes the relative sign between the R and $\widetilde{\mathrm{R}}$ contributions in the sum over spin structures. Furthermore, the sign of the NS and $\widetilde{\mathrm{NS}}$ sectors is fixed relative to R, since the three even spin structures may be transformed into each other by $\mathrm{PSL}(2,\mathbb{Z})$ modular transformations.  
\begin{itemize}[wide,left=2pt]
\item 
For the \zA{-} theory we impose $(-1)^{F_{\rm ws}} = -1$, which removes the generic physical Ramond state $\ket{\Psi_+}$. We will see however that this does not exclude an exceptional zero-momentum Ramond insertion (\ref{eq:Rstar}). The situation is analogous to the type 0A $\hat{c}=1$ superstring \cite{Douglas:2003up, Takayanagi:2003sm}, where there was a similar distinction between propagating Ramond modes and Ramond-Ramond flux backgrounds. 
\item  
For the \zA{+} theory, the projection $(-1)^{F_\mathrm{ws}}=1$ removes the generic physical Ramond state $\ket{\Psi_-}$.
We will have to additionally account for the sign inherent in $\mathcal{G}^+$ when we assemble the observables with this projector. In this case, there is also no exceptional Ramond zero momentum state.
\end{itemize}
\item  
The Type 0B theories are defined to retain physical Ramond states at generic momentum for either definition of the total supercharge. Specifically,
\begin{itemize}[wide,left=2pt]
\item In the \zB{-} theory we retain states with $(-1)^{F_\mathrm{ws}} = 1$. This retains the  state $\ket{\Psi_+}$ defined in~\eqref{eq:0BPsi}. Its vertex operator in the $(-\frac{1}{2}, -\frac{1}{2})$ picture is (NB: $\nR$ is a vertex operator normalization that needs to be fixed)
\begin{equation}\label{eq:Rminus vertex operator}
\wsR{P}{-}{(-\frac{1}{2})}
= \nR\,  \tilde{\mathfrak{c}}\,\mathfrak{c}\,
e^{-\frac{1}{2}\,\varphi - \frac{1}{2}\, \tilde{\varphi}} \,
\pqty{\RsL{+}{P} \, \RtL{-}{\Ph = -i\,P}
   + i\,\RsL{-}{P} \, \RtL{+}{\Ph = -i\,P} }  \,.
\end{equation}
This operator vanishes in the zero momentum limit, since $\RsL{-}{P}$ and $\RtL{-}{\Ph}$ are odd under $P \to -P$, and $\Ph \to -\Ph$, respectively. 
\item The state $\ket{\Psi_-}$ defined in~\eqref{eq:0BPsi} survives the GSO projection in the Type \zB{+} theory where we project onto states with $(-1)^{F_\mathrm{ws}} = -1$. Its vertex operator in the $(-\frac{1}{2}, -\frac{1}{2})$ picture is
\begin{equation}\label{eq:Rplus vertex operator}
\wsR{P}{+}{(-\frac{1}{2})}
= \nR\, \tilde{\mathfrak{c}}\,\mathfrak{c}\, 
e^{-\frac{1}{2}\,\varphi - \frac{1}{2}\, \tilde{\varphi}} \,
\pqty{\RsL{+}{P} \, \RtL{+}{\Ph = -i\,P}
   -\,\RsL{-}{P} \, \RtL{-}{\Ph = -i\,P} }  \,.
\end{equation}
The coefficient between the operators differs from that in the state; this is again due to cocycle factors, as the fermion parity of $\RtL{-}{\Ph}$ is odd. Note that this operator survives the zero momentum limit; $\wsR{0}{+}{(-\frac{1}{2})}$ is an element of the physical BRST cohomology. The zero-momentum limit is however smooth, and thus observables can be studied at generic momenta and the limit can be smoothly taken. 
\end{itemize}
\end{itemize}
  
We thus obtain four GSO-projected super Virasoro minimal string theories, Type \zA{\pm} and \zB{\pm}. They share a common NS sector with vertex operator $\mathcal{V}_P$ (whose BRST representative depends on the picture number), but only the \zB{\pm} theories retain the generic Ramond vertex operators $\wsR{P}{\pm}{}$. In addition, the \zA{-} theory admits an exceptional zero-momentum insertion (\ref{eq:Rstar}). We now turn to observables built from the generic vertex operators, returning to the exceptional insertion in~\cref{sec:comparison}.

The justification follows from the classification of~\cite{Kaidi:2019pzj,Kaidi:2019tyf}; standard (critical) 0A theories have a contribution from the $\operatorname{\text{Arf}}$ invariant. In non-critical two-dimensional strings the contribution is $(-1)^{n\,\operatorname{\text{Arf}}}$ with $n =1$ for 0A and $n=0$ for 0B. This assumes that there are no signs lurking in the definition of worldsheet supercharge and is therefore appropriate for \zA{-} and \zB{-} theories. A chiral fermion reflection shifts $n \to n+1$. Hence, for \zA{+} and \zB{+} one imposes the projections defined above. 

Our nomenclature for the four theories, which is natural from the worldsheet GSO projection, is not fully in sync with what has been defined hitherto. Our minus theories reduce to  $\mathcal{N}=1$ JT-supergravity in the semiclassical limit, and agree with the 0A versus 0B classification used in~\cite{Stanford:2019vob}.  Indeed, 0A JT-supergravity has a non-trivial $\operatorname{\text{Arf}}$ invariant and has non-vanishing higher genus amplitudes. The matrix model duals derived in the above reference predict vanishing stable perturbative NS amplitudes for Type 0B supergravity, which is consistent with what we find. On the other hand~\cite{Eberhardt:2026hfh} assigns the theory with vanishing stable NS amplitudes to be the Type \zA{-} theory.
Our plus theories, however, agree with the 0A versus 0B classification of~\cite{Eberhardt:2026hfh}.  When we quote results below, we will translate them to our convention. The dictionary translating the conventions across the aforementioned references and ours can be found in~\cref{fn:taxonomy}.

\subsection{Worldsheet observables 1: sphere three-point function}\label{sec:3ptamp}

The basic observables we can consider in the string worldsheet are the correlation functions of the vertex operators. These, in analogy with VMS, one expects should map to the dual matrix model observables that compute volumes of supermoduli spaces.
The three-dimensional gravity description of the theory analyzed in~\cite{Eberhardt:2026hfh} predicts that the \zA{+} and \zB{+} theories share observables with the VMS, modulo an overall normalization factor in the latter case (that depends on the genus and the number of insertions). All stable NS perturbative amplitudes should vanish in the \zB{-} theory, while 
tree-level NS-sector amplitudes vanish in the \zA{-} theory.
The simplest non-vanishing NS amplitude in the \zA{-} theory is the torus 1-point function, which we will examine in~\cref{sec:torus1ptws}.

We denote the string amplitudes as
$\wsamp{g,(n_{\smash[b]{_\mathrm{NS}}}| n_{\smash[b]{_\mathrm{R}}})}{\mathrm{0A}^{\pm}}$ and $\wsamp{g,(n_{\smash[b]{_\mathrm{NS}}}| n_{\smash[b]{_\mathrm{R}}})}{\mathrm{0B}^{\pm}}$ respectively, where $n_{\smash[b]{_\mathrm{NS}}}$ is the number of NS and $n_{\smash[b]{_\mathrm{R}}} \in 2\mathbb{Z}_{\geq 0}$ is the number of R vertex operators on a genus-$g$ worldsheet. The superscript refers to the specific theory under consideration.

\paragraph{NS sector 3-point function:} The first observable we consider is the worldsheet 3-point function of NS operators. While the picture $-1$ operator is common to all four theories, its picture $0$ counterpart is sensitive to our choice of supercharge. The string amplitude is given by
\begin{equation}
\begin{aligned}
\wsamp{0,(3|0)}{\pm}(P_1,P_2,P_3)
   & =
\wsnorm\,\expval{\wsNS{P_1}{\pm}{(-1)}(0)
    \, \wsNS{P_2}{\pm}{(-1)}(1)\, \wsNS{P_3}{\pm}{(0)}(\infty)} \,.
\end{aligned}
\end{equation}

The contributions from the ghosts factor out, leaving behind the interesting piece that comes from the super Liouville theories. There is an overall normalization factor $\wsnorm$ of the worldsheet path integral (including the ghost correlators), which we need to account for.

Let us start with the \zA{-} and \zB{-} three-point amplitudes, which are identical at genus zero. We  compute the two sets of
super Liouville three-point functions and then assemble the worldsheet amplitude. First, for the minus family of theories we have 
\begin{equation}
\begin{aligned}
   &
\expval{
    V_{P_1}(0)\, \Vh_{-i\,P_1}(0)\,
    V_{P_2}(1)\, \Vh_{-i\,P_2}(1)\, \pqty{
    V_{P_3}(\infty)\, \Wh_{-i\,P_3}(\infty) +
    W_{P_3}(\infty)\, \Vh_{-i\,P_3}(\infty)}}
\\
   & =
\CV{b}(P_1,P_2,P_3)\, \CWh{b}(-i\,P_1, -i\,P_2,-i\,P_3)
+ \CW{b}(P_1,P_2,P_3)\, \CVh{b}(-i\,P_1, -i\,P_2,-i\,P_3)
\\
   & =
-2i\,  \frac{\CV{b}(P_1,P_2,P_3)}{\CV{b}(P_1,P_2,P_3)}
+ 2i\, \frac{\CW{b}(P_1,P_2,P_3)}{\CW{b}(P_1,P_2,P_3)}
\\
   & = 0 \,.
\end{aligned}
\end{equation}
The relative sign in the last line originates from the structure constant $\CWh{\bh}$ in~\eqref{eq:tLCs}.

Then for the plus family of theories we require
\begin{equation}
\begin{aligned}
   &
\expval{
    V_{P_1}(0)\, \Vh_{-i\,P_1}(0)\,
    V_{P_2}(1)\, \Vh_{-i\,P_2}(1)\, \pqty{
    V_{P_3}(\infty)\, \Wh_{-i\,P_3}(\infty) -
    W_{P_3}(\infty)\, \Vh_{-i\,P_3}(\infty)}}
\\
   & =
\CV{b}(P_1,P_2,P_3)\, \CWh{b}(-i\,P_1, -i\,P_2,-i\,P_3)
- \CW{b}(P_1,P_2,P_3)\, \CVh{b}(-i\,P_1, -i\,P_2,-i\,P_3)
\\
   & =
- 2i\, \frac{\CV{b}(P_1,P_2,P_3)}{\CV{b}(P_1,P_2,P_3)}
- 2i\, \frac{\CW{b}(P_1,P_2,P_3)}{\CW{b}(P_1,P_2,P_3)}
\\
   & = -4i  \,.
\end{aligned}
\end{equation}

Thus, modulo an overall numerical factor (which we can normalize to unity), the \zA{+} and \zB{+} theories have a non-vanishing NS 3-point function. On the other hand, the 3-point function vanishes in both \zA{-} and \zB{-} theories; to wit,
\begin{equation}\label{eq:sph3ptNS-all}
\begin{aligned}
\wsamp{0,(3|0)}{\mathrm{0A}^+}(P_1,P_2,P_3)
&= 
\wsamp{0,(3|0)}{\mathrm{0B}^+}(P_1,P_2,P_3)
=
-i\, \wsnorm\, \nNS^3 = 1\,, \\
\wsamp{0,(3|0)}{\mathrm{0A}^-}(P_1,P_2,P_3)
&=
\wsamp{0,(3|0)}{\mathrm{0B}^-}(P_1,P_2,P_3)
=
0\,.
\end{aligned}
\end{equation}
We have presaged in the first equality that the amplitude in \zA{+} and \zB{+} can be set to unity by fixing $\wsnorm$ in terms of $\nNS$, cf.~\eqref{eq:wsspherenorm}.

\paragraph{Mixed NS and R sector 3-point functions:}
Let us next turn to correlators involving the generic Ramond vertex operators of the \zB{\pm} theories. When the number of Ramond insertions is odd, the correlators vanish. So there is only one 3-point function  to consider in each of the \zB{\pm} theories, viz.,
\begin{equation}
\begin{aligned}
\wsamp{0,(1|2)}{\mathrm{0B}^{\pm}}(P_1;P_2,P_3)
   & = \wsnorm\,\expval{
    \wsNS{P_1}{\pm}{(-1)}(0)\,
    \wsR{P_2}{\pm}{(-\frac{1}{2})}(1)\,
    \wsR{P_3}{\pm}{(-\frac{1}{2})}(\infty)
    }\,.
\end{aligned}
\end{equation}
Note that the total picture number being $-2$ we do not need additional PCO insertions to evaluate the worldsheet correlator.

First consider the  \zB{-} theory, where we have to compute the super Liouville correlator 
\begin{equation}
\begin{aligned}
 &
\expval{
V_{P_1}(0) \Vh_{-i P_1}(0)\,
\pqty{\RsL{+}{P_2} \,\RtL{-}{-i P_2} + i\,
    \RsL{-}{P_2} \,\RtL{+}{-iP_2}}(1)\,
\pqty{\RsL{+}{P_3}\, \RtL{-}{-iP_3} + i\,
    \RsL{-}{P_3} \,\RtL{+}{-i P_3}}(\infty)}
\\
 & =
\expval{
V_{P_1}(0)\,\RsL{+}{P_2}(1)\, \RsL{+}{P_3}(\infty)}\,
\expval{\Vh_{-i\,P_1}(0)\,\RtL{-}{-i\,P_2}(1)\, \RtL{-}{-i\,P_3}(\infty)}
\\
 & \qquad \qquad
-\expval{
   V_{P_1}(0)\,\RsL{-}{P_2}(1)\, \RsL{-}{P_3}(\infty)}\,
\expval{\Vh_{-i\,P_1}(0)\,\RtL{+}{-i\,P_2}(1)\, \RtL{+}{-i\,P_3}(\infty)}
\\
 & =
\frac{1}{2} \pqty{- \Ce{b}(P_1;P_2,P_3) \,\Coh{b}(-iP_1;-iP_2,-iP_3)
    + \Co{b}(P_1;P_2,P_3) \, \Ceh{b}(-iP_1;-iP_2,-iP_3)}
\\
 & =
0\,.
\end{aligned}
\end{equation}
Next, for the 0B$^{+}$ theory we require
\begin{equation}
\begin{aligned}
 &
\expval{
V_{P_1}(0) \Vh_{-i P_1}(0)\,
\pqty{\RsL{+}{P_2} \,\RtL{+}{-i P_2}
    -\,\RsL{-}{P_2} \,\RtL{-}{-iP_2}}(1)\,
\pqty{\RsL{+}{P_3}\, \RtL{+}{-iP_3}
    -\,\RsL{-}{P_3} \,\RtL{-}{-i P_3}}(\infty)}
\\
 & =
\expval{
V_{P_1}(0)\,\RsL{+}{P_2}(1)\, \RsL{+}{P_3}(\infty)}\,
\expval{\Vh_{-i\,P_1}(0)\,\RtL{+}{-i\,P_2}(1)\, \RtL{+}{-i\,P_3}(\infty)}
\\
 & \qquad \qquad
- \expval{ V_{P_1}(0)\,\RsL{-}{P_2}(1)\, \RsL{-}{P_3}(\infty)}\,
\expval{\Vh_{-i\,P_1}(0)\,\RtL{-}{-i\,P_2}(1)\, \RtL{-}{-i\,P_3}(\infty)}
\\
 & =
\frac{1}{2} \pqty{\Ce{b}(P_1;P_2,P_3) \,\Coh{b}(-iP_1;-iP_2,-iP_3)
    + \Co{b}(P_1;P_2,P_3) \, \Ceh{b}(-iP_1;-iP_2,-iP_3)}
\\
 & =
1\,.
\end{aligned}
\end{equation}
The relative sign between the two terms on the right-hand side of the first equality requires an explanation. It arises from a cocycle factor. Recall that for the supercharges to satisfy the worldsheet superconformal algebra we had to introduce a factor of $(-1)^{F_t}$ in the tensor product~\eqref{eq:Gws}. This factor grades the odd fermion parity of the timelike super Liouville sector with an extra sign. In particular, the computation $\pqty{\RsL{-}{P_2} \otimes \RtL{-}{-iP_2}}(1) \times \pqty{\RsL{-}{P_3} \otimes \RtL{-}{-i P_3}}(\infty)$ requires one to commute the timelike Ramond operator past the spacelike one to obtain the expression in terms of the product of spacelike and timelike correlators; a cocycle factor contributes a sign in this process.\footnote{ One can equivalently motivate the sign by viewing the mixed three-point function as a matrix element $\mel{\Psi_\pm}{\mathcal{V}}{\Psi_\pm}$. The conjugate state
$\bra{\Psi_\pm}$, which is annihilated by $G_0^\dagger = G_0$ and $\widetilde{G}_0^\dagger = \widetilde{G}_0$ ends up being
$\bra{\Psi_\pm} = \frac{1}{\sqrt{2}} \pqty{\bra{+}_s \bra{\mp}_t + i \bra{-}_s\bra{\pm}_t}$. Computing the matrix element one immediately recovers the sign as written.  } The end result is consistent with the intuition that the signs of $\CW{b}$ and $\Co{b}$ for the spacelike structure constants are effectively reversed in the plus theories (cf.~\cref{fn:CWoddsigns}).

The physical string amplitudes are then
\begin{equation}\label{eq:sph3ptmix-all}
\begin{aligned}
\wsamp{0,(1|2)}{\mathrm{0B}^+}(P_1;P_2,P_3)
 & =
\, \wsnorm\, \nNS\, \nR^2\ = 1\,,
\\
\wsamp{0,(1|2)}{\mathrm{0B}^-}(P_1;P_2,P_3)
 & =
0\,.
\end{aligned}
\end{equation}
In the case of \zB{+} we have previewed that we will be able to choose the Ramond-sector vertex operator normalization such that the mixed amplitude is set to unity. The mixed amplitude involving the zero-momentum Ramond insertion will be discussed separately around (\ref{eq:mixed-three-point-amplitude-with-Ramond-punctures}). 
 
\subsection{Worldsheet observables 2: torus one-point function}\label{sec:torus1ptws}

To compute the torus one-point function of an NS sector operator, we require one PCO insertion, to account for the moduli space which is of super dimension $1 | 1$. So the NS sector operator is in picture $0$ (we will see that in the odd spin structure, it will be convenient to keep the PCO and the vertex operator well-separated). In string theory, we are required to sum over the four spin structures. The weighted sum is dictated by the GSO projection. For diagonal GSO projection of the Type 0 theories we only have a relative sign between the even spin structure part and  the odd spin structure part. Specifically, the torus one-point amplitude is
\begin{equation}\label{eq:0A-and-0B-spin-structure-sums}
\begin{aligned}
\wsamp{1,(1|0)}{\mathrm{0A}^\sigma}(P)
 & =
\tnorm\left(\expval{\wsNS{P}{\sigma}{(0)}(0)}_{\mathrm{NS}} +
\expval{\wsNS{P}{\sigma}{(0)}(0)}_{\widetilde{\mathrm{NS}}} +
\expval{\wsNS{P}{\sigma}{(0)}(0)}_\mathrm{R} -\sigma
\expval{\wsNS{P}{\sigma}{(0)}(0)}_{\widetilde{\mathrm{R}}} \right) ,\\
\wsamp{1,(1|0)}{\mathrm{0B}^\sigma}(P)
 & =
\tnorm\left(\expval{\wsNS{P}{\sigma}{(0)}(0)}_{\mathrm{NS}} +
\expval{\wsNS{P}{\sigma}{(0)}(0)}_{\widetilde{\mathrm{NS}}} +
\expval{\wsNS{P}{\sigma}{(0)}(0)}_\mathrm{R} +\sigma
\expval{\wsNS{P}{\sigma}{(0)}(0)}_{\widetilde{\mathrm{R}}} \right)\,.
\end{aligned}
\end{equation}
Note that we have used translational invariance to fix the location of the vertex operator on the torus.  
The relative signs appearing in the odd spin structure implement GSO projections that remove physical Ramond sector vertex operators at generic momentum in the 0A theories and retain them in the 0B theories.

Let us compile the worldsheet integrals we need to compute. Focusing on the even spin structures,
$a = 2,3,4$ we have\footnote{Strictly speaking, it only makes sense to consider the worldsheet moduli space integral of the \emph{sum} of the $a=2,3,4$ sectors over the fundamental domain, since the torus one-point functions in each sector mix under modular transformations.}

\begin{equation}\label{eq:evenss1pt}
\begin{aligned}
\expval{\wsNS{P_{\mathrm{ext}}}{\pm}{(0)}}_a
 & =  \frac{\nNS}{4}\,
\int_{\mathcal{F}_0}\, d^2\tau\, \mathcal{Z}^\mathrm{gh}_a\, \bigg(
\expval{V_{P_\mathrm{ext}}}_a\,  \expval{\Wh_{-iP_\mathrm{ext}}}_a \mp
\expval{W_{P_\mathrm{ext}}}_a\,  \expval{\Vh_{-iP_\mathrm{ext}}}_a\bigg)
\\
 & = \frac{\nNS}{4}\,
\int_{\mathcal{F}_0}\, d^2\tau\, \frac{\abs{\eta(\tau)}^6}{\abs{\vartheta_a(\tau)}^2}
\bigg(
\expval{V_{P_\mathrm{ext}}}_a\,  \expval{\Wh_{-iP_\mathrm{ext}}}_a \mp
\expval{W_{P_\mathrm{ext}}}_a\,  \expval{\Vh_{-iP_\mathrm{ext}}}_a\bigg)\,.
\end{aligned}
\end{equation}
In the second line we include the trace over the ghost Hilbert space explicitly. This factor will cancel against the elliptic factors in the conformal blocks.

The story for $a=1$ is different as explained earlier. One needs to account for the fact that the superconformal ghosts have zero modes in this sector. Therefore, a naive picture-$0$ vertex operator correlator is ill-defined and needs to be regulated. This can be seen as follows: should we use the standard story and work with the picture zero operator, the analog of~\eqref{eq:evenss1pt} would instruct us to compute the torus 1-point function of the picture raised $W$ operator in the spacelike and timelike theories. However, $\expval{W}_1 = \expval{\Wh}_1 = 0$ from~\eqref{eq:FblockR}, since the corresponding block is identically zero. The $\beta\gamma$ ghost determinant in sector $a$ is $\abs{\eta(\tau)/ \vartheta_a(0\mid\tau)}^2$ which diverges, thanks to the aforementioned zero modes.
Clearly, this ambiguous $0 \times \infty$ product needs to be evaluated correctly to obtain the physical answer.

\paragraph{The odd spin structure:} To resolve this ambiguity, we retain the PCO at a fixed point separated from the vertex operator and evaluate the ghost correlator before taking any coincidence limit. Doing so leads to the $\widetilde{\mathrm{R}}$ sector correlator being given by the following~\cite{Yin:2026str} 
    \begin{equation}\label{eq:oddss1pt}
    \begin{aligned}
    \expval{\wsNS{P_\mathrm{ext}}{\pm}{(0)}(0)}_{\widetilde{\mathrm{R}}}
     & =
      \frac{\nNS}{4}\,
    \int_{\mathcal{F}_0}\, d^2\tau\,
    \pqty{
    \expval{V_{P_{\mathrm{ext}}}}_1 \,
    \expval{
        \widetilde{G}_{-\frac{3}{2}}
        G_{-\frac{3}{2}}
        \Vh_{-iP_{\mathrm{ext}}}}_1 
    \mp \expval{
        \widetilde{G}_{-\frac{3}{2}}\,
        G_{-\frac{3}{2}}
        V_{P_{\mathrm{ext}}}}_1\,
    \expval{\Vh_{-iP_{\mathrm{ext}}}}_1}\,.
    \end{aligned}
    \end{equation}
To clarify the notation: the  l.h.s.~is a shorthand for the full ghost decorated operator, whose correlator is defined by the pure matter correlator on the r.h.s.

Let us explain how this result is derived, starting from the torus one-point function in the odd spin structure. We retain the external vertex operator in the $(-1,-1)$ picture at the origin, with representative $ \wsNS{P}{\pm}{(-1)}$ given in~\eqref{eq:NSvo11}, and keep the PCO insertion at a fixed point $(z,\bar z)$ away from the origin.\footnote{The bosonic modulus of the torus requires insertion of the operator $\mathfrak{B}_\tau\, \mathfrak{B}_{\overline{\tau}}$. This part is common to all spin sectors and produces the familiar $\abs{\eta(\tau)}^4$ measure factor from the $\mathfrak{b}\mathfrak{c}$ ghost system.} For the purposes of this computation, the relevant part of the picture changing operator is $ e^{\tilde{\varphi}} \widetilde{\mathcal{G}}(\zb)\,e^{\varphi} \mathcal{G}(z)$.
We then compute the resulting worldsheet correlator. The remaining observable is then\footnote{We remind the reader that we will often use $a=1$ interchangeably with $\widetilde{\mathrm{R}}$ to denote the odd spin structure sector.}
\begin{equation}\label{eq:GGVderive1}
\expval{e^{\tilde{\varphi}}(\zb) \, e^{\varphi}(z)\,e^{-\varphi-\tilde{\varphi}}(0)}_{\beta\gamma,\widetilde{\mathrm{R}}}\times \expval{\widetilde{\mathcal{G}}(\zb)\, \mathcal{G}(z)\, V_{P_{\mathrm{ext}}} (0)\, \Vh_{-iP_{\mathrm{ext}}}(0)}_{\widetilde{\rm R}} \,.
\end{equation}
Let's first discuss the ghost correlator. It can be computed exactly as follows. 
In any spin-structure sector the ghost correlator on the torus is given by~\cite{Verlinde:1987sd}
\begin{equation}
    \expval{e^{\varphi}(z)e^{-\varphi}(0)}_{\beta\gamma,\,a}= \eta(\tau)\,\frac{2\pi\vartheta_1(z|\tau)}{\vartheta_1'(0|\tau)\vartheta_a(z|\tau)}.
\end{equation}
In the $a=1$ case relevant for the odd spin structure, we thus have~\cite{Yin:2026str}
\begin{equation}
    \expval{e^{\tilde{\varphi}}(\zb) \, e^{\varphi}(z)\,e^{-\varphi-\tilde{\varphi}}(0)}_{\beta\gamma,\widetilde{\mathrm{R}}} = |\eta(\tau)|^{-4}\, .
\end{equation}
This is in particular independent of $z$ and moreover exactly cancels the contribution of the $\mathfrak{b}\mathfrak{c}$ ghost correlator. 

To extract the remaining matter correlator, write $U = V_{P_{\rm ext}}\mathsf{V}_{-iP_{\rm ext}}$, and define the reduced correlator
\begin{equation}
    F(z,\bar z) \equiv \left\langle \widetilde{\mathcal G}(\bar z)\mathcal G(z)U(0)\right\rangle_1\, .
\end{equation}
In this computation we should regard $\bar z$ and $z$ as independent since the left- and right-moving PCOs could be inserted at separate positions. In the odd spin structure the supercurrents are periodic around both cycles, and thus $F$ is an elliptic function in $z$ (with $\bar z$ fixed). The function $F$ has at most a simple pole at the origin. However such a doubly-periodic function is necessarily constant in $z$; an elliptic function cannot have a single simple pole with a non-vanishing residue. A similar argument applies for $\bar z$. The constant term can be extracted by the replacement $\mathcal{G}(z)U(0)\to \oint\frac{dz}{2\pi i z}\mathcal{G}(z)U(0)= \mathcal{G}_{-\frac{3}{2}}U(0)$. Thus, we have
\begin{equation}
    F(z,\bar z) = \left\langle \widetilde{\mathcal{G}}_{-\frac{3}{2}}\mathcal{G}_{-\frac{3}{2}}U(0)\right\rangle_1\, ,
\end{equation}
 and so the torus one-point function of the external vertex operator in the odd spin structure is given by the level-$\frac{3}{2}$ superdescendant correlator constructed in (\ref{eq:oddss1pt}).

\paragraph{Normalization of the torus amplitude:} The main
part of the worldsheet torus path integral we need to fix is the  ghost contribution to the amplitude before assembling the final result. We do this in a couple of steps.

First, the bosonic modulus is soaked up by the insertion of $(2\pi)^2 \, \mathfrak{b}\tilde{\mathfrak{b}}$, cf.~\cite[Equation (7.3.2)]{Polchinski:1998rq}. The $\mathbb{Z}_2$ reflection symmetry gives a factor of $1/2$, so there is an overall $\frac{1}{2}(2\pi)^2$ from the diffeomorphism ghosts. In addition, the $\mathfrak{bc}$ correlator gives an overall factor of $\abs{\eta(\tau)}^4$.

The canonical normalization of the PCO has a factor of $1/4$, which has been explicitly included in the picture raised operator. The normalization of the spin structure sum contributes a factor of $1/2$. The $\beta\gamma$ contribution in the even spin structures produces a factor of $\abs{\frac{\eta(\tau)}{\vartheta_a(0\mid \tau)}}^2$. Combined with the diffeomorphism ghosts this produces the modular prefactor written out in~\eqref{eq:evenss1pt}. Therefore, in the even spin sector the overall normalization factor  is
\begin{equation}\label{eq:torusNfactor}
\tnorm = \frac{1}{2} \times \frac{(2\pi)^2}{2} = \pi^2\,.
\end{equation}

\paragraph{The string amplitudes:}
All told, the torus $1$-point amplitude we need to compute is assembled from the following super Liouville theory ingredients.   
First introduce the building blocks for the correlators in each of the spin structure sectors 
\begin{equation}\label{eq:Ipm-evenodd}
\begin{aligned}
\mathcal{I}_a^\mp 
&= 
 \int_{\mathcal{F}_0} d^2\tau\; 
\abs{\frac{\eta^3(\tau)}{\vartheta_a(\tau)}}^2 \;
\big[
\expval{V}_a \, \expval{\Wh}_a  
\pm 
\expval{W}_a \, \expval{\Vh}_a\big] \,,  
\qquad 
a = 2, 3, 4\,,
\\
\mathcal{I}_1^\mp  
&= 
\int_{\mathcal{F}_0} d^2\tau\;\big[
\expval{V}_1 \, \expval{\mathsf{DV}}_1
\pm 
\expval{DV}_1 \, \expval{\Vh}_1
\big]\,.
\end{aligned}
\end{equation}
The explicit expressions for each of the ingredients are given in~\cref{sec:wsbblocks}.
The torus one-point amplitudes $\wsamp{1,(1|0)}{}$ for the four worldsheet string theories are given in terms of the following combinations: 
\begin{equation}\label{eq:wsamp-ABpm}
\begin{aligned}
\wsamp{1,(1|0)}{{\mathrm{0A}^+}}
 & =
\frac{\nNS}{4}\, \tnorm\,
\pqty{\mathcal{I}^+_3 +  \mathcal{I}^+_4 + \mathcal{I}^+_2 - \mathcal{I}^+_1}\,,
\\
\wsamp{1,(1|0)}{{\mathrm{0B}^+}}
 & =
\frac{\nNS}{4}\,\tnorm\,
\pqty{\mathcal{I}^+_3 +  \mathcal{I}^+_4 + \mathcal{I}^+_2 + \mathcal{I}^+_1}\,,
\\
\wsamp{1,(1|0)}{{\mathrm{0A}^-}}
 & =
\frac{\nNS}{4}\, \tnorm\,
\pqty{\mathcal{I}^-_3 +  \mathcal{I}^-_4 + \mathcal{I}^-_2 + \mathcal{I}^-_1}\,, \\
\wsamp{1,(1|0)}{{\mathrm{0B}^-}}
 & =
\frac{\nNS}{4}\, \tnorm\,
\pqty{\mathcal{I}^-_3 +  \mathcal{I}^-_4 + \mathcal{I}^-_2 - \mathcal{I}^-_1}\,.
\end{aligned}
\end{equation}
Since $\tnorm$ is fixed, given a prediction for the amplitudes from intersection theory, we will be able to fix the normalization of the amplitude, cf.~\eqref{eq:NSvnorm}.

\paragraph{A check on GSO signs:}
A further check that the GSO sign choices correctly project out Ramond states with generic momentum in the 0A theories is the following. Consider the combination of the $a=2$ and $a=1$ integrands entering $\mathcal I_2^\sigma-\sigma\mathcal I_1^\sigma$ that describes the Ramond degeneration channel of the 0A$^\sigma$ amplitude. At fixed internal momenta $(P,\widehat P)$, suppressing common overall factors and the arguments of the structure constants, the leading behavior in the degenerating limit is given by
\begin{equation}
\begin{aligned}
    2i\rho^{(b)}_{\rm R}(P)\varrho^{(b)}_{\rm R}(\widehat{P})(q\, \bar q)^{\frac{P^2+\widehat{P}^2}{2}}\left[(P^2+\widehat P^2)\left(
    C^{(b)}_{\mathrm{even}}\mathsf{C}^{(b)}_{\mathrm{odd}}-\sigma C^{(b)}_{\mathrm{odd}}\mathsf{C}_{\mathrm{even}}^{(b)}\right)+\mathcal{O}(q,\bar q)\right]\, .
\end{aligned}
\end{equation}
At fixed internal momenta, integrating over the modulus in a neighborhood of the degeneration limit $q\to 0$ produces a potential pole $1/(P^2+\widehat{P}^2)$ corresponding to an on-shell Ramond state. However, in the 0A theories, the specific combination that appears in the worldsheet integrand conspires to produce the factor $(P^2+\widehat{P}^2)$ seen above that exactly cancels the putative on-shell pole at $P^2 + \widehat{P}^2 = 0$. 
We take this as a further signal that the generic Ramond sector states have been correctly projected out in the 0A theories. It does not preclude the existence of an exceptional zero-momentum insertion of \zA{-}, which must be analyzed separately as we do shortly.

\paragraph{Comments on numerical implementation:} The building blocks $\mathcal{I}^\pm$  can be evaluated numerically. 
First we evaluate the one-point functions of the individual operators in the spacelike and timelike theories separately at a fixed point in the torus moduli space. For this we use the explicit representations~\eqref{eq:1ptNS},~\eqref{eq:1ptR}, and~\eqref{eq:GGV1}. The blocks are evaluated using the recursion relation to $8^{\mathrm{th}}$ order for fixed $P_{\mathrm{ext}}$. The internal momentum is constrained to lie in $[0,P_{\mathrm{max}}]$ or $[-P_{\mathrm{max}},P_{\mathrm{max}}]$ for the spacelike and timelike theories, respectively. We find $P_{\mathrm{max}} =6.0$ suffices for clean convergence of the momentum integrals (the high momentum region is suppressed by the conformal blocks). For the timelike super Liouville theory we perform the momentum integral along a contour shifted up by a small imaginary part in the $P$ plane, viz., $P\to P+ i\,\epsilon$. As in our modular covariance checks, a shift by $\epsilon =0.08$ seems optimal for numerical integration.

\begin{figure}[hb!]
\begin{subfigure}[b]{0.5\textwidth}
    \centering
    \includegraphics[width=0.95\textwidth]{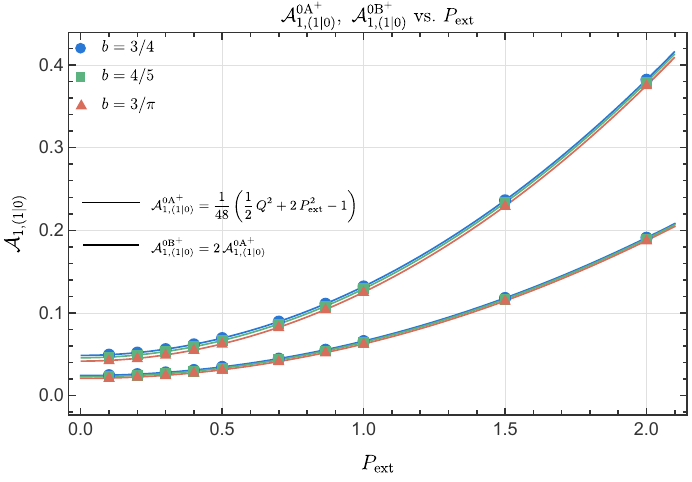}
\end{subfigure}
\begin{subfigure}[b]{0.5\textwidth}
    \centering  \includegraphics[width=0.95\textwidth]{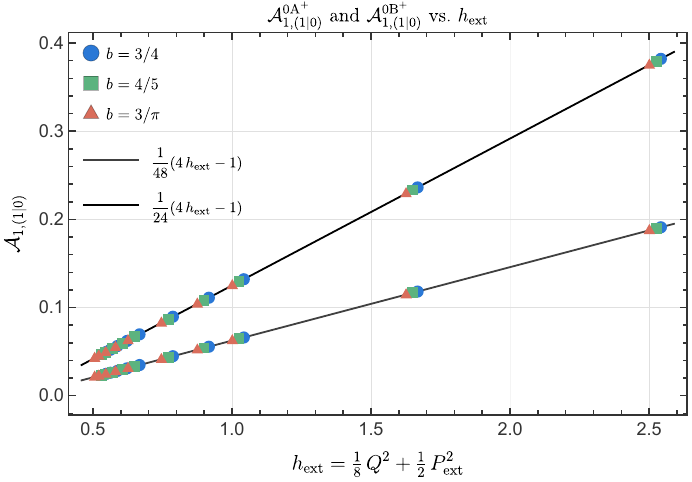}
\end{subfigure}
\\[1ex]
\begin{center}
    \begin{subfigure}[b]{0.5\textwidth}
    \centering
    \includegraphics[width=0.95\textwidth]{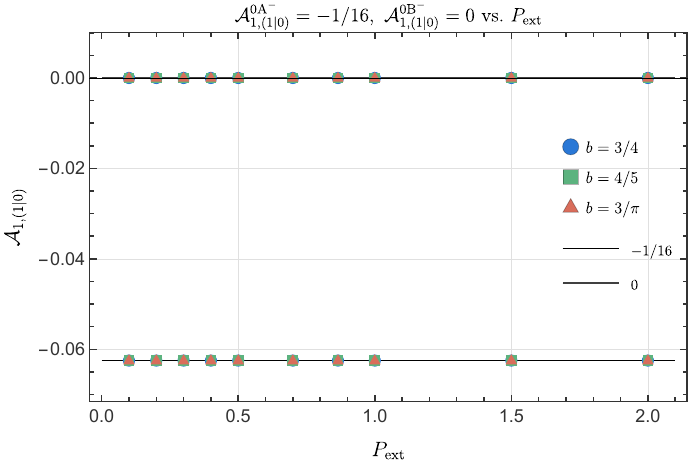}
    \end{subfigure}
\end{center}
\caption{String amplitudes for \zA{\pm} and \zB{\pm} theories. We plot the raw amplitudes computed at discrete values of the external momenta and compare them against the intersection theory and matrix model predictions. The last step involves fixing an overall normalization, which we explain later in the text. }
\label{fig:string-amps}
\end{figure}

We then scan over a range of $\tau \in \mathcal{F}_0$ and assemble the integrands of $\mathcal{I}^\pm$. The zero mode contributions ensure that the cusp is well-behaved; the integrands decay rapidly for large $\Im(\tau) = \tau_2$ as $\tau_2^{-2}$ in all four  sectors. We therefore split the integration over $\mathcal{F}_0$ into an `ear region' with $\tau_2 < \tau_{2,\mathrm{mid}} \sim 3$, an intermediate rectangular domain $\tau_{2,\mathrm{mid}} < \tau_2 < \tau_{2,\mathrm{max}} \sim 100$ and a `cusp region' $\tau_2 > \tau_{2,\mathrm{max}}$. Note that in the rectangular and cusp regions, we can approximate the blocks by unity, since the leading contribution from $q^{\frac{1}{2}}$ terms is already suppressed (for $\tau_{2,\mathrm{mid}} = 3$ the corrections are already of order $10^{-4}$). In these domains the $\tau_1 = \Re(\tau)$ integral enforces level matching (with a regulator contribution from the timelike theory owing to the $P\to P+ i\,\epsilon$ shift), and the $\tau_2$ integral is a simple exponential.

The integral in the intermediate rectangle is computed numerically by discretization.  The integral in the cusp region can be performed analytically. The asymptotics of the spacelike and timelike amplitudes give factors of $\tau_2^{-3/2}$ and $\tau_2^{-1/2}$, respectively, and so the integrand decays as $\tau_2^{-2}$. Using the integrand data from the upper half of the rectangular domain, a polynomial fit (in $1/\tau_2$) provides a good approximation to the integrand, which is then analytically evaluated.  In the ear region we employ a Chebyshev-Lobatto grid. We pick $32$ grid points for $\tau_1 \in [-\frac{1}{2},\frac{1}{2}]$ and a similar number of nodes for $\tau_2 \in [\sqrt{1-\tau_1^2},\tau_{2,\mathrm{mid}}]$. The Chebyshev grid does better sampling at the corners near $\tau_1 = \pm \frac{1}{2}$ and $\abs{\tau} =1$, allowing for smaller errors. The results of the direct numerical evaluation are displayed in~\cref{fig:string-amps}.

\paragraph{The string amplitudes:} Evaluating the amplitudes numerically, and upon fixing the overall normalization we find that they are very simple functions of the external momenta. We claim
that the correctly normalized string amplitudes are given by 
\begin{equation}
\begin{aligned}
\wsamp{1,(1|0)}{{\mathrm{0A}^+}}(P_{\rm ext})
 & = \frac{4\,h_{\mathrm{ext}} - 1}{48}\,,
\\
\wsamp{1,(1|0)}{{\mathrm{0B}^+}}(P_{\rm ext})
 & =
\frac{4\,h_{\mathrm{ext}} - 1}{24}
= 2\,\wsamp{1,(1|0)}{{\mathrm{0A}^+}}\,,
\\
\wsamp{1,(1|0)}{{\mathrm{0A}^-}}(P_{\rm ext})
 & =
-\frac{1}{16}\,,
\\
\wsamp{1,(1|0)}{{\mathrm{0B}^-}}(P_{\rm ext})
 & =
0\,.
\end{aligned}
\label{eq:torus1ptSVMS}
\end{equation}
with
\begin{equation}
h_{\mathrm{ext}} = \frac{1}{8}\, Q^2 + \frac{1}{2}\, P_{\mathrm{ext}}^2\,.
\end{equation}
In determining these we have fixed the vertex operator normalization $\nNS$ in terms of the torus ghost correlation factor $\tnorm$ up to an overall phase, cf.~\eqref{eq:NSvnorm}. This choice is conditioned on matching the worldsheet calculation against the intersection theory result.  

\begin{figure}[hpbt!]
\centering
\includegraphics[width=0.95\textwidth]{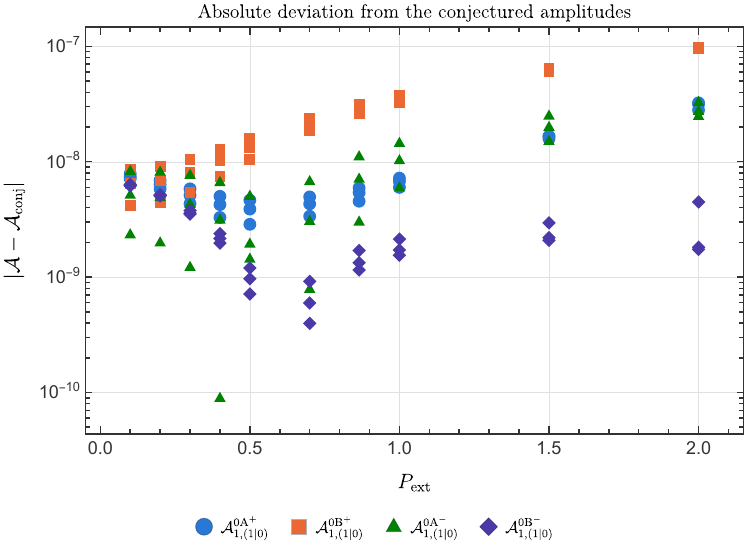}
\caption{Plot of the deviation of string amplitude from the expected matrix model result. }
\label{fig:string-amps-deviation}
\end{figure}

Agreement with this behavior is clearly visible in the plots of the amplitudes in~\cref{fig:string-amps}. We also show the deviation from the expected result
\begin{equation}
\delta \wsamp{1,(1|0)}{} = \abs{\wsamp{1,(1|0)}{\mathrm{string}} -  \wsamp{1,(1|0)}{\mathrm{matrix}}} \,,
\end{equation}
in~\cref{fig:string-amps-deviation}. The deviation is at most of order $10^{-7}$.  

It is instructive to dissect the amplitude and ask how the building blocks from the worldsheet behave. The correct combinations to consider are the modular invariant spin structure pieces. For the even spin structure numerical data supports the following 
\begin{equation}
\begin{aligned}
\frac{\nNS}{4}\, \tnorm\, \pqty{ 
\mathcal{I}_3^+ + \mathcal{I}_4^+ + \mathcal{I}_2^+ }
& = \frac{4\,h_{\mathrm{ext}} -1}{32}\,,
\\
\frac{\nNS}{4}\, \tnorm\, \pqty{ 
\mathcal{I}_3^- + \mathcal{I}_4^- + \mathcal{I}_2^- }
&= -\frac{1}{32} \,.
\end{aligned}
\end{equation}
The odd spin structure combinations also simplify, and turn out to be related to the even spin structure result. Explicitly,  
\begin{equation}
\begin{aligned}
\mathcal{I}^{+}_1 
& = 
\frac{1}{3}\, \pqty{ 
\mathcal{I}_3^+ + \mathcal{I}_4^+ + \mathcal{I}_2^+ }\,,
\\
\frac{\nNS}{4}\, \tnorm\,
\mathcal{I}^-_1 
& = 
 -\frac{1}{32}\,.
\end{aligned}
\end{equation}
At this point it is clear that the string amplitudes follow by basic algebra. A concise summary of our results is collated in~\cref{tab:summary}.

\begin{table}[h]
\begin{center}
\renewcommand{\arraystretch}{1.45}
\setlength{\tabcolsep}{5pt}
\small
\begin{tabular}{l||c|c|c|c|c|c|c|c}
Theory & supercurrents & GSO & R & $(-1)^{F_{\mathrm{ws}}}$
& sign of $\mathcal{I}_1$ & $\wsamp{0,(3|0)}{}$ & $\wsamp{0,(1|2)}{}$
& $\wsamp{1,(1|0)}{}$ \\
\hline\hline
\zA{+} & $(\mathcal{G}^+,\widetilde{\mathcal{G}}^+)$ & $+1$
  & $|\Psi_-\rangle$ ($\cross$) & $-1$ & $-$ & $1$ & ---
  & $\tfrac{4h_{\mathrm{ext}}-1}{48}$ \\
\zB{+} & $(\mathcal{G}^+,\widetilde{\mathcal{G}}^+)$ & $-1$
  & $|\Psi_-\rangle$ ($\checkmark$) & $-1$ & $+$ & $1$ & $1$
  & $\tfrac{4h_{\mathrm{ext}}-1}{24}$ \\
\zA{-} & $(\mathcal{G}^-,\widetilde{\mathcal{G}}^-)$ & $-1$
  & $|\Psi_+\rangle$ ($\cross$) & $+1$ & $+$ & $0$ & ---
  & $-\tfrac{1}{16}$ \\
\zB{-} & $(\mathcal{G}^-,\widetilde{\mathcal{G}}^-)$ & $+1$
  & $|\Psi_+\rangle$ ($\checkmark$) & $+1$ & $-$ & $0$ & $0$ & $0$ \\
\end{tabular}
\end{center}
\caption{The four super Virasoro minimal strings: the choice of total supercurrents~\eqref{eq:Gws}, the GSO projection, the generic physical Ramond ground state, its worldsheet fermion parity, the sign with which $\mathcal{I}_1$ enters (\ref{eq:wsamp-ABpm}) and the sphere three-point~\eqref{eq:sph3ptNS-all} and (\ref{eq:sph3ptmix-all}) and torus one-point~\eqref{eq:torus1ptSVMS} amplitudes. The entry for the mixed amplitude refers to generic (nonzero momentum) Ramond states; in \zA{-} we recall that the exceptional zero-momentum Ramond state has the mixed amplitude $\mathcal{A}^{\mathrm{0A}^-}_{0,(1|2)}(P_1;*,*) = 1$.}
\label{tab:summary}
\end{table}
 %

\subsection{Comparison of string amplitudes with intersection theory and matrix models}\label{sec:comparison}
 
With the worldsheet computations in place, we finally turn to the last remaining ingredient, viz., fixing the vertex operator normalization and the overall normalization of the worldsheet path integral (which involves fixing factors of the ghost correlators). The benchmark is the intersection theory prediction of~\cite{Eberhardt:2026hfh}. Perhaps it is worth noting at the outset that consistency of our calculation requires that the non-vanishing amplitudes agree with these predictions up to an overall scale. So we only have one factor to fix across all four of the torus amplitudes, which we shall now proceed to do.

In comparing to the matrix integrals, we find it convenient to mostly adopt the conventions of the bosonic Virasoro minimal string~\cite{Collier:2023cyw}. In particular we will take $E = -z^2$ to be the energy coordinate in the matrix integral, with $E = P^2$ on the physical sheet. Notice that this differs by a factor of two from the definition of the conformal weights in the NS sector, $h - \tfrac{Q^2}{8} = \tfrac{P^2}{2}$. Let $R_{g,n}(E_1,\ldots,E_n)$ be the connected resolvent of the matrix integral and denote by $\omega_{g,n}$ its corresponding differential
\begin{equation}
    \omega_{g,n}(z_1,\ldots,z_n) = \prod_{i=1}^n(-2z_i\,\d z_i)R_{g,n}(-z_1^2,\ldots,-z_n^2)\, .
\end{equation}
In each sector the quantum volumes are extracted from the associated resolvents by inverse Laplace transform exactly as in the bosonic theory
\begin{equation}\label{eq:bosonic dictionary}
    \begin{aligned}
        \mathsf{V}^{(b)}_{g,n}(P_1,\ldots,P_n) &= \int_\Gamma\prod_{i=1}^n\left[\frac{\d z_i}{2\pi i}\frac{\sqrt{2}z_i}{P_i}\mathrm{e}^{4\pi z_i P_i}\right]R_{g,n}(-z_1^2,\ldots,-z_n^2)\\
        &=\prod_{i=1}^n \mathop{\mathrm{Res}}_{z_i=0}\left[-\frac{\mathrm{e}^{4\pi z_i P_i}}{\sqrt{2}P_i}\right]\omega_{g,n}(z_1,\ldots z_n)\, .
    \end{aligned}
\end{equation}
Here $\Gamma$ is a collection of vertical contours in the complex $z_i$ planes situated to the right of the poles at $z_i=0$. The second equality applies to the stable differentials, whose only poles occur at $z_i = 0$. The inverse relation is given by
\begin{equation}
    \omega_{g,n}(z_1,\ldots,z_n) = \int_0^\infty \left[\prod_{i=1}^n-4\sqrt{2}\pi P_i\d P_i\,\mathrm{e}^{-4\pi z_i P_i}\right]\mathsf{V}_{g,n}^{(b)}(P_1,\ldots,P_n)\left(\prod_{i=1}^n\d z_i\right)\, .
\end{equation}

Going forward we will use the dictionary (\ref{eq:bosonic dictionary}) in all the sectors for which the amplitude is non-vanishing. Topological recursion determines the higher resolvent differentials (and thus the higher string amplitudes) in terms of the initial data $\omega_{0,1}$, $\omega_{0,2}$ as follows~\cite{Eynard:2007kz}
\begin{equation}
    \begin{aligned}\label{eq:omega TR}
      \omega_{g,n}(z_1,\boldsymbol{z}) 
      &= 
      \mathop{\mathrm{Res}}_{z=0}K(z_1,z)\Big(\omega_{g-1,n+1}(z,-z,\boldsymbol{z})\\
      &\quad\quad+ 
   \sum_{h=0}^g\sideset{}{'} \sum_{\mathcal{I}\subseteq\{z_2,\ldots,z_n\}}\omega_{h,1+|\mathcal{I}|}(z,\boldsymbol{z}_{\mathcal{I}})\omega_{g-h,1+|\mathcal{I}^c|}(-z,\boldsymbol{z}_{\mathcal{I}^c})\Big)\,,
    \end{aligned}
\end{equation}
where the prime means that we omit $(h,\mathcal{I}) = (0,\emptyset),(g,\boldsymbol{z})$, with $\boldsymbol{z} = \{z_2,\ldots,z_n\}$, and the recursion kernel is given by
\begin{equation}\label{eq:recursion kernel}
    K(z_1,z) = \frac{\int_{-z}^z\omega_{0,2}(z_1,\cdot)}{4\omega_{0,1}(z)}\, .
\end{equation}
In practice, in applying (\ref{eq:omega TR}) one should notice that there is a sign that enters due to the $-z$ argument in one of the resolvent differentials on the right-hand side. So for example with
\begin{equation}
    \omega_{0,2}(z_1,z_2) = \frac{\d z_1\, \d z_2}{(z_1-z_2)^2}
\end{equation}
we have
\begin{equation}
    \omega_{0,2}(z,-z) = -\frac{\d z^2}{4 z^2}\, .
\end{equation}

\paragraph{1. Intersection theory prediction for \zA{+}/\zB{+}:}
The amplitudes of these two theories were argued in~\cite{Eberhardt:2026hfh} to be proportional to the ordinary bosonic quantum volumes. The precise claim is that 
\begin{equation}\label{eq:vol-zABplus}
\begin{aligned}
\wsamp{g,(n_{\smash[b]{_\mathrm{NS}}}| n_{\smash[b]{_\mathrm{R}}})}{X^+}(P_1,\ldots,P_n)
=
\mathsf{V}^{(b)}_{g,n}(P_1,\ldots,P_n)\,
\begin{dcases}
1\,,
 & \qquad X = \mathrm{0A}
\\
2^g\,,
 & \qquad X = \mathrm{0B}\,,
\end{dcases}
\end{aligned}
\end{equation}
where $n =n_{\smash[b]{_\mathrm{NS}}} + n_{\smash[b]{_\mathrm{R}}}$, with $n_R = 0$ in \zA{+}. Here the volumes $\mathsf{V}^{(b)}_{g,n}{}$ are undecorated, as they correspond to those associated with the spectral curve of bosonic VMS. The corresponding spectral curve is given by~\cite{Collier:2023cyw}\footnote{Strictly speaking, $y_{\rm bos}(z)$ differs from what was reported in~\cite{Collier:2023cyw} by a sign. The reason for this is that there the resolvents were computed ignoring the sign mentioned above; equivalently one can assign a sign to the definition of the recursion kernel (\ref{eq:recursion kernel}).}
\begin{equation}
    \mathsf{x}_{\rm bos}(z) = -z^2, \quad \mathsf{y}_{\rm bos}(z) = \frac{2\sqrt{2}\pi\sin(2\pi b z)\sin(2\pi b^{-1} z)}{z},\quad\omega_{0,2}(z_1,z_2) = \frac{\d z_1\, \d z_2}{(z_1-z_2)^2}\, ,
\end{equation}
and thus
\begin{equation}
    \omega_{0,1}(z) = -\mathsf{y}_{\rm bos}(z)\,\d\mathsf{x}_{\rm bos}(z) =  4\sqrt{2}\pi \sin(2\pi b z)\sin(2\pi b^{-1} z)\, \d z\, .
\end{equation}
Applying topological recursion (\ref{eq:omega TR}) to this curve we find~\cite{Collier:2023cyw}
\begin{equation}
    \omega_{1,1}(z) = -\frac{1}{24\sqrt{2}\pi}\left(\frac{b^2+b^{-2}}{8z^2}+\frac{3}{16\pi^2 z^4}\right)\,dz\,,
\end{equation}
and so the worldsheet amplitude is computed by applying the dictionary (\ref{eq:bosonic dictionary})
\begin{equation}
    \wsamp{1,(1|0)}{\mathrm{0A}^+}(P_{\rm ext}) = \mathsf{V}_{1,1}^{(b)}(P_{\rm ext}) = \frac{1}{24}\left(\frac{b^2+b^{-2}}{4}+P_{\rm ext}^2\right) = \frac{4h_{\rm ext}-1}{48}\, .
\end{equation}
The factor of $2$ relation between the \zA{+} and \zB{+} amplitudes that we observed numerically in (\ref{eq:torus1ptSVMS}) follows immediately from~\eqref{eq:vol-zABplus}.

\paragraph{2. Intersection theory prediction for \zB{-}:}  
Our worldsheet results for the \zB{-} theory are consistent with the vanishing of all stable perturbative amplitudes (with NS-sector insertions) computed from the intersection numbers weighted by the Arf theta class in~\cite{Eberhardt:2026hfh}. In particular this predicts a vanishing torus one-point amplitude
\begin{equation}\label{eq:mm-0Aminus}
\wsamp{1,(1| 0)}{\mathrm{0B}^-}(P_{\mathrm{ext}})  = 0 \,.
\end{equation}
This is numerically consistent (within the precision of the computation) with the amplitude we computed directly from the worldsheet in~\eqref{eq:torus1ptSVMS}. However, there appears to be a small subtlety in the identification of the minus theories. The corresponding theory was labeled \zA{-} in the intersection theory construction of~\cite{Eberhardt:2026hfh}. Here we have chosen to refer to it as \zB{-}, since it retains physical Ramond vertex operators at generic momentum. Below we will explain in more detail the origin of this discrepancy when we discuss the role of Ramond punctures in the \zA{-} theory.

\paragraph{3. Intersection theory prediction for \zA{-}:} 
Complementarily to the discussion above, here the relevant comparison is between our \zA{-} theory and the intersection numbers weighted by the theta class, which were assigned to the \zB{-} theory in~\cite{Eberhardt:2026hfh}. With the same qualification given above regarding the apparent mismatch in the labeling of the minus theories, the proposal for the amplitudes involving NS-sector vertex operators in the \zA{-} theory is thus~\cite{Eberhardt:2026hfh}
\begin{equation}
    \begin{aligned}
        \mathsf{V}_{g,n}^{(b),\mathrm{0A}^-}(P_1,\ldots,P_n) &= (-1)^n 2^{1-g-n}\int_{\overline{\mathcal{M}}_{g,n}} \Theta_{g,n}\exp\bigg(\frac{b^2+b^{-2}}{8}\kappa_1 + \frac{1}{2}\sum_{i=1}^n P_i^2 \psi_i\\
        &\quad\quad\quad\quad\quad\quad\quad\quad\quad\quad\quad\quad\quad\quad- \sum_{k\geq 1}\frac{B_{2k}(1-2^{-2k})\kappa_{2k}}{2k(2k)!}\bigg)\, .
    \end{aligned}
\end{equation}
The corresponding spectral curve can be derived directly from this expression. In the conventions of this paper, the spectral curve is given by~\cite{Eberhardt:2026hfh}\footnote{Compared to $\mathsf{y}_{\mathrm{0B}^-}$ in~\cite{Eberhardt:2026hfh}, $\mathsf{y}_{\mathrm{0A}^-}$ here differs by a sign due to a relative sign difference in the definition of the recursion kernel (\ref{eq:recursion kernel}).}
\begin{equation}\label{eq:0A-minus-spectral-curve}
   \mathsf{x}_{\mathrm{0A}^-}(z) = -z^2,\quad 
   \mathsf{y}_{\mathrm{0A}^-}(z) = -\frac{2\sqrt{2}\pi \cos(2\pi b z)\cos(2\pi b^{-1} z)}{z},\quad\omega_{0,2}(z_1,z_2) = \frac{\d z_1 \, \d z_2}{(z_1-z_2)^2}\, ,
\end{equation}
and so
\begin{equation}
    \omega_{0,1}(z) = -\mathsf{y}_{\mathrm{0A}^-}(z)\,\d \mathsf{x}_{\mathrm{0A}^-}(z) = -4\sqrt{2}\pi \cos(2\pi b z)\cos(2\pi b^{-1} z)\,dz\,.
\end{equation}
That $\omega_{0,1}$ has a nonzero constant leading coefficient at the branch point $z=0$ is the signature of a spectral curve with a ``hard edge''~\cite{norbury2024superweilpeterssonmeasuresmoduli,Chidambaram:2022cqc,Norbury:2020vyi,Do:2016odu,Chekhov:2017aot,Eberhardt:2026hfh}. Carrying out topological recursion on this spectral curve, we find
\begin{equation}
    \omega_{1,1}(z) = \frac{\d z}{32\sqrt{2}\pi z^2}\, .
\end{equation}
Applying the dictionary (\ref{eq:bosonic dictionary}), this leads to the amplitude
\begin{equation}
    \wsamp{1,(1|0)}{\mathrm{0A}^-}(P_\mathrm{ext}) = \mathsf{V}_{1,1}^{(b),\mathrm{0A}^-}(P_{\mathrm{ext}}) = -\frac{1}{16}\,  .
\end{equation}
Once again this is consistent with the amplitude that we computed numerically directly from the worldsheet in~\eqref{eq:torus1ptSVMS}.

A priori, it is perhaps curious that this hard edge spectral curve, whose leading density of eigenvalues is related to the Ramond Cardy formula~\eqref{eq:rhons}, governs the NS-sector amplitudes in a theory without Ramond-sector vertex operators at generic momentum. There is, however, no contradiction: the hard edge is not intrinsically tied to the presence of physical Ramond-sector states. Indeed, as discussed above, the GSO projection in \zA{-} removes the on-shell factorization pole corresponding to generic physical Ramond states. In addition, the observables one computes in the NS sector are signed traces with the insertion of the non-anomalous fermion parity operator~\cite{Stanford:2019vob}. The fact that the graded NS vacuum character ($\widetilde{\mathrm{NS}}$ sector) branches into the ungraded Ramond characters (R sector) with the Ramond density of states may be viewed as evidence that we should have anticipated the nature of the spectral curve. Moreover, as we will now see, the \zA{-} theory admits an exceptional zero-momentum Ramond insertion. The hard edge spectral curve written above governs the NS-sector amplitudes without such insertions. It would be interesting to understand more directly how the spectral curve~\eqref{eq:0A-minus-spectral-curve} emerges from the worldsheet theory.
 
\paragraph{Ramond punctures in \zA{-}:}
Although we have defined the \zA{-} theory via a GSO projection that removes the physical Ramond vertex operators at generic momenta (\ref{eq:Rminus vertex operator}) in the minus theory, there is a small but important exception. It turns out that this projection actually \emph{does not} preclude the existence of exceptional zero-momentum Ramond vertex operators. We will associate these vertex operators with Ramond ``punctures,'' for reasons that will become clear shortly. Indeed, at $P = \mathsf{P}=0$ the zeromodes of the timelike and spacelike supercurrents annihilate the Ramond ground states (cf.~(\ref{eq:spacelike and timelike supercurrent zeromode action})), so the physical state conditions no longer single out the states $\ket{\Psi_\pm}$ of (\ref{eq:0BPsi}). The GSO projection of the \zA{-} theory allows the states created by the Ramond-sector operators $R_0^+\mathsf{R}^+_0$ and $R^-_0 \mathsf{R}^-_0$. However oddness under reflections $P\to -P$ implies that both $R^-_{0}$ and $\mathsf{R}^-_{0}$ vanish in the operator normalizations used here, leaving the non-vanishing insertion $R^+_0\mathsf{R}^+_0$. See \cite{Klebanov:2003wg} for a similar discussion in the context of type 0A minimal superstrings. 
This operator is exceptional in the physical spectrum of the \zA{-} theory in that it does not extend to a family of physical Ramond vertex operators for any $P\ne 0$. It nevertheless coincides with the regular zero-momentum limit of the \zB{+} Ramond vertex operator (\ref{eq:Rplus vertex operator}).

We can define an on-shell vertex operator representing the Ramond-sector puncture in the $(-\frac{1}{2},-\frac{1}{2})$ picture as usual\footnote{In what follows we will denote the exceptional zero-momentum insertion with a $*$. There are subtleties involving different picture representatives of such exceptional Ramond vertex operators, which were explored for the critical superstrings in~\cite{Berkovits:1997mc}. For the observables we study here, the picture $(-\tfrac{1}{2}, -\tfrac{1}{2} )$ representative suffices.}
\begin{equation}\label{eq:Rstar}
    \mathcal{R}_*^{(-\frac{1}{2})} = \nR\, \tilde{\mathfrak{c}}\mathfrak{c}\,e^{-\frac{1}{2}\varphi-\frac{1}{2}\tilde\varphi}R^+_0\mathsf{R}^+_0\, .
\end{equation}
The coupling of this exceptional zero-momentum Ramond puncture to the NS-sector can be computed by the mixed three-point amplitude given by
\begin{equation}\label{eq:mixed-three-point-amplitude-with-Ramond-punctures}
    \mathcal{A}_{0,(1|2)}^{\mathrm{0A}^-}(P_1;*,*) = \wsnorm\expval{\wsNS{P_1}{-}{(-1)}(0)\,\mathcal{R}_*^{(-\frac{1}{2})}(1)\,\mathcal{R}_*^{(-\frac{1}{2})}(\infty)}\, .
\end{equation}
Notice that the relevant structure constants simplify when we set the Ramond momenta to zero
\begin{equation}
\begin{aligned}
    C_{\mathrm{even}}^{(b)}(P_1;0,0) &= C_{\mathrm{odd}}^{(b)}(P_1;0,0) \equiv C_*^{(b)}(P_1)\ne 0\,,\\
    \mathsf{C}_{\mathrm{even}}^{(\mathsf{b})}(\mathsf{P}_1;0,0) &= \mathsf{C}_{\mathrm{odd}}^{(\mathsf{b})}(\mathsf{P}_1;0,0)  = \frac{1}{C_*^{(\mathsf{b})}(i\mathsf{P}_1)}\, .
\end{aligned}
\end{equation}
So the sphere three-point function that enters the mixed amplitude with Ramond punctures (\ref{eq:mixed-three-point-amplitude-with-Ramond-punctures}) is straightforwardly computed to be
\begin{equation}
\begin{aligned}
 &
\expval{
V_{P_1}(0) \Vh_{-i P_1}(0)\,
\left(\RsL{+}{0} \,\RtL{+}{0}\right)(1)\,
\left(\RsL{+}{0}\, \RtL{+}{0}\right)(\infty)}
\\
 & =
\expval{
V_{P_1}(0)\,\RsL{+}{0}(1)\, \RsL{+}{0}(\infty)}\,
\expval{\Vh_{-i\,P_1}(0)\,\RtL{+}{0}(1)\, \RtL{+}{0}(\infty)}
\\
 & =
\frac{C_{\mathrm{even}}^{(b)}(P_1;0,0) + C_{\mathrm{odd}}^{(b)}(P_1;0,0)}{2}\frac{\mathsf{C}_{\mathrm{even}}^{(b)}(-iP_1;0,0) + \mathsf{C}_{\mathrm{odd}}^{(b)}(-iP_1;0,0)}{2}
\\
 & =
1\,.
\end{aligned}
\end{equation}
Thus the mixed amplitude involving Ramond punctures in \zA{-} coincides with the zero-momentum limit of the \zB{+} mixed three-point amplitude
\begin{equation}\label{eq:punctures mixed amplitude}
    \mathcal{A}_{0,(1|2)}^{\mathrm{0A}^-}(P_1;*,*) = \wsnorm\, \nNS\, \nR^2 = 1\, .
\end{equation}
In the last equality we have used the normalization of $\nR$ fixed below by requiring that the \zB{+} mixed three-point amplitude be unity. The mixed amplitude involving the $\mathcal{R}_*$ punctures inherits this normalization. 
With the corresponding normalization of the puncture insertions, this agrees with the result expected from intersection theory dictated by the $0|0$-dimensional moduli space of a sphere with one NS and two Ramond punctures~\cite{Stanford:2019vob,Eberhardt:2026hfh}.

We interpret this exceptional zero-momentum vertex operator as a natural candidate for the Ramond punctures of~\cite{Stanford:2019vob}. There, it was emphasized that a Ramond puncture is not simply the zero-length limit of a Ramond boundary.\footnote{The former is obtained from the latter by further removing an odd modulus.} This parallels the observation above that the operator $\mathcal{R}_*$ is not the $P\to 0$ limit of a physical Ramond vertex operator in the minus theory. Moreover, in~\cite{Stanford:2019vob} it was proposed that including a gas of Ramond punctures in the type 0A $\mathcal{N}=1$ JT supergravity theory leads to a simple deformation of the dual matrix integral. It is thus natural to expect that including a gas of the zero-momentum Ramond punctures (\ref{eq:Rstar}) in \zA{-} $\sVMS$ leads to a similar deformation of the hard-edge spectral curve (\ref{eq:0A-minus-spectral-curve}). Summing over these punctures may be interpreted as turning on a background Ramond-Ramond flux. A similar distinction occurs in the type 0A $\hat{c}=1$ superstring, which has no propagating Ramond-Ramond modes but nevertheless admits Ramond-Ramond flux backgrounds~\cite{Douglas:2003up,Takayanagi:2003sm}. In the $\hat{c}<1$ type 0A minimal superstring, the analogous Ramond ground state operators were similarly identified with Ramond-Ramond fluxes \cite{Klebanov:2003wg}.

The distinction between Ramond punctures and Ramond-sector vertex operators with nonzero momentum also clarifies the relationship between our labels for the minus theories and those of~\cite{Eberhardt:2026hfh}. The theory governed by the hard-edge spectral curve admits Ramond punctures but not Ramond vertex operators at generic momenta. On the other hand the theory with the vanishing stable NS amplitudes has the opposite property: it admits the Ramond vertex operator $\wsR{P}{-}{}$ for any $P\ne 0$, but the $P\to 0$ limit of that operator vanishes. In~\cite{Eberhardt:2026hfh} the 0A theories are defined by weighting the sum over spin structures with $(-1)^{\operatorname{\text{Arf}}}$, and consequently exclude Ramond punctures. We instead assign the 0A/0B labels based on the presence of generic (non-zero momentum) Ramond states, adopting the convention of the literature on the $\hat{c}=1$ superstring~\cite{Douglas:2003up, Takayanagi:2003sm}, which does not exclude the exceptional Ramond puncture in the \zA{-} theory. For the plus theories, these two conventions coincide, but for the minus theories, they lead to the opposite assignments.

There is no analogous zero-momentum insertion representing a Ramond puncture in the \zA{+} theory because the candidates allowed by the GSO projection involve either $R^-_0$ or $\mathsf{R}^-_0$, which vanish as operators.

\paragraph{Fixing the worldsheet normalizations:} With the intersection theory results at hand we can fix the normalization of the worldsheet factors. We have the vertex operator normalizations $\nNS$ and $\nR$, and the overall normalization of the worldsheet path integral $\wsnorm$.

Comparing the \zA{+} amplitude from the worldsheet with the corresponding answer computed from the bosonic VMS spectral curve, we find agreement provided  
\begin{equation}\label{eq:NSvnorm}
\frac{\nNS}{4}\, \tnorm\,e^{-i\frac{\pi}{2}} = \frac{1}{2}
\;\; \Longrightarrow \;\;
\nNS =  \frac{2\, e^{i\frac{\pi}{2}}\,}{\tnorm}\,.
\end{equation}
The phase factor arises because each of the building blocks of the string amplitude in~\eqref{eq:wsamp-ABpm} ends up being purely imaginary. This is obvious for the NS sector amplitudes owing to an explicit factor of $i$ in the structure constants $\CW{b}$ and $\CWh{\bh}$, cf.~\eqref{eq:sLCVandW} and~\eqref{eq:tLCs}. In  the Ramond sector the factor arises from the blocks~\eqref{eq:1ptR} and~\eqref{eq:GGV1}.

Having fixed the normalization of the vertex operator, we can also fix the worldsheet sphere path integral normalization. Demanding that the non-vanishing NS sector three-point function at genus-zero~\eqref{eq:sph3ptNS-all} in the \zA{+}/\zB{+} theories be unity determines
\begin{equation}\label{eq:wsspherenorm}
-i\, \wsnorm \, \nNS^3 = 1 \;\; \Longrightarrow\;\;
\wsnorm = -\frac{1}{8}\, \tnorm^3\,.
\end{equation}
Finally, demanding that the mixed NS-R three-point amplitude of \zB{+} be unity fixes $\nR$ in~\eqref{eq:sph3ptmix-all}
\begin{equation}
\wsnorm\, \nNS \, \nR^2 = 1 \;\; \Longrightarrow\;\;
\nR = e^{-i\,\frac{\pi}{4}}\, \nNS = \frac{2\,e^{i\,\frac{\pi}{4}}}{\tnorm}\,.
\end{equation}
%

\section{Discussion}\label{sec:discuss}

The main thrust of the discussion was to bring $\mathcal{N}=1$ super Liouville CFT on par with its bosonic counterpart. We examined the theory on the torus, and amended existing results in the literature to provide a complete formulation of torus superconformal blocks. These were checked to satisfy modular crossing numerically. Moreover, to facilitate analytic checks, we also bootstrapped the modular $S$-kernels, and verified them.

The primary motivation for our study was to build a solid foundation of the super Liouville data to facilitate applications in worldsheet string theory. To this end, we applied the technology to the study of the $\sVMS$ theory, which was proposed in~\cite{Muhlmann:2025ngz,Rangamani:2025wfa}. While the original construction made some preliminary observations, thanks to the development relating these theories to 3d chiral supergravity~\cite{Eberhardt:2026hfh} various aspects became much more transparent. 

For instance, the fact that there are four distinct worldsheet theories characterized by the choice of (diagonal) GSO projection, and by the relative combination of spacelike and timelike supercharges, was made manifest. This observation, in particular, makes clear that the sign ambiguity noticed in the bootstrap construction of the timelike super Liouville structure constants, whilst innocuous for the superconformal theory itself,\footnote{Although the relative signs of the timelike structure constants must be chosen compatibly with modular covariance.} has a non-trivial imprint in the worldsheet construction. We chose to include this sign choice as a relative sign for the total left moving supercharge and used the (generic) physical Ramond spectrum to fix the 0A/0B labels. The resulting comparisons of the NS amplitudes with the intersection theory predictions of~\cite{Eberhardt:2026hfh} agree for the plus labels and exchange the minus labels.

As an application, we computed the NS sphere three-point and torus one-point amplitudes and found agreement with the dual matrix model predictions, with the theory labels specified above. We also found that the \zA{-} theory, despite having no physical Ramond vertex operators at generic momentum, admits an exceptional zero-momentum Ramond operator with precisely the three-point coupling expected of a Ramond puncture. While the final answers are very simple, it bears emphasizing that the worldsheet results in individual spin-structure sectors are anything but.  It is only after a non-trivial spin-structure sum and integration over moduli that one arrives at the simple polynomial form of the amplitudes. One may wonder if the standard RNS perturbation theory that we used is to blame, and whether there is a more straightforward way to arrive at the results (see below). Nevertheless, the string worldsheet provides a strong consistency check of our results, for had any sector been computed incorrectly, we would not have landed on the matrix model answer. 

We highlight a few open questions that our analysis throws up for further contemplation. 

\paragraph{Further $\sVMS$ amplitudes:} A natural next step in the program is to compute the genus-$0$ four-point amplitudes.
For the pure NS four-point amplitude, confirming that the worldsheet reproduces the matrix model predictions given in~\cite{Eberhardt:2026hfh} would be a highly nontrivial computation due to the need for multiple picture-changing operators, potentially leading to spurious singularities which may require vertical integration~\cite{Sen:2015hia}.\footnote{We have preliminary evidence~\cite{Collier:2026wip} that this nevertheless works out leading to non-vanishing amplitudes proportional to the bosonic VMS volume in the plus theories, and vanishing amplitudes in the minus theories.} 

Amplitudes involving Ramond-sector vertex operators provide further tests. For generic momenta these are present in the \zB{\pm} theories, while the \zA{-} theory supports the exceptional zero-momentum Ramond insertion $\mathcal{R}_*$. Within the \zB{+} theory the Ramond insertions were argued to have the same correlators as NS vertex operators from the intersection theory perspective in~\cite{Eberhardt:2026hfh}. In the \zB{-} theory, the primary question is whether the perturbative vanishing extends to amplitudes involving Ramond-sector operators. Within RNS perturbation theory, R and NS vertex operators enter on a very different footing. It would be very interesting to understand the structural mechanism underlying the agreement of the corresponding amplitudes.

On the matrix model side, super Weil-Petersson volumes for surfaces with Ramond punctures and Neveu-Schwarz boundaries have recently been obtained from topological recursion~\cite{Stanford:2019vob,norbury2024superweilpeterssonmeasuresmoduli,Alexandrov:2024kuj} (see also~\cite{Johnson:2026jls,Johnson:2026jgs,Johnson:2026ugu}). Since these volumes carry no momentum labels for the Ramond punctures, the natural worldsheet counterparts of the latter are the zero-momentum insertions (\ref{eq:Rstar}) of the \zA{-} theory. The exceptional Ramond insertion has a non-vanishing mixed three-point amplitude (\ref{eq:punctures mixed amplitude}) that supports this interpretation. It would be interesting to subject this identification to further tests via sphere four-point amplitudes involving two or four zero-momentum Ramond insertions. One could also understand how summing over these insertions, which may be interpreted as turning on Ramond-Ramond flux, deforms the hard-edge spectral curve (\ref{eq:0A-minus-spectral-curve}).

\paragraph{A free string theory (Type \zB{-}):} From the worldsheet construction, perhaps the \zB{-} theory is the most bizarre. As we have seen, it is built from standard ingredients like its other $\sVMS$ cousins. Nevertheless, its stable perturbative NS amplitudes are all argued to vanish~\cite{Stanford:2019vob,Eberhardt:2026hfh}. 
As far as we can tell there is no hint at the level of worldsheet building blocks of the mechanism for this remarkable cancellation, and the torus one-point amplitude did not shed any light on this front. Understanding how the RNS formalism manages to obscure this surprising fact would be instructive.

\paragraph{Supermoduli space formulation of string perturbation theory:}
Perhaps one lesson suggested by the aforementioned observation is that we should exploit the full power of the superconformal symmetry, and work with the definition of the string amplitude as the integral of an appropriate superform on supermoduli space~\cite{Witten:2012bh} (see also the comments in~\cite{Knighton:2026wva} where this technology was applied to \AdS{3} string theory). Ideally, the reframing will allow one to see how the worldsheet amplitudes satisfy topological recursion and/or explain the vanishing of the \zB{-} amplitudes simply from the presence of unsaturated fermionic moduli in the integrand. The former, however, seems ambitious, considering that we do not at present understand how the VMS amplitudes themselves satisfy topological recursion.

\paragraph{ZZ-instantons and non-perturbative effects:}
The dual matrix integrals also encode effects that are non-perturbative in the string coupling. Aspects of these eigenvalue instanton effects were reproduced from the worldsheet in bosonic VMS and CLS via ZZ-instanton amplitudes in~\cite{Collier:2023cyw,Collier:2024mlg}. Similar technology has been developed for the minimal superstring in~\cite{Chakrabhavi:2024szk,Barman:2026jjm} (see also the pioneering work on ZZ-instantons~\cite{Balthazar:2019rnh,Balthazar:2019ypi} in the $c=1$ string and~\cite{Balthazar:2022apu} in the type 0B $\hat{c}=1$ string). To investigate non-perturbative effects in $\sVMS$ from the worldsheet, the principal necessary ingredient is the set of conformal boundary conditions of $\mathcal{N}=1$ Liouville SCFT. For the spacelike theory these are well-understood (see~\cite{Fukuda:2002bv,Ahn:2002ev}), but the conformal boundary conditions for the timelike SCFT remain to be explored. We expect these investigations to be interesting for two reasons. Firstly, some of the matrix models are governed by spectral curves characterized by a hard edge where the eigenvalue density diverges, and thus the non-perturbative effects should qualitatively differ from the usual eigenvalue tunneling in theories with a soft spectral edge~\cite{Johnson:2021owr,Stanford:2019vob}. Secondly, 
since all stable perturbative NS amplitudes are argued to vanish in the \zB{-} theory, any nontrivial non-perturbative effect would constitute the leading contribution to the corresponding process. Understanding these effects would shed light on the extent to which this perturbatively free theory is really trivial.

\paragraph{The super complex Liouville string $\widehat{\mathrm{CLS}}$:} Another obvious application of the formalism we developed is to combine two super Liouville theories with $c_\pm =\frac{15}{2} \pm i\, \lambda$. As noted at the outset the worldsheet theory has been analyzed at the level of low-lying genus-zero amplitudes in~\cite{Du:2025lya}, leading to a proposal of two spectral curves for dual two-matrix integrals (one of which is equivalent to that of bosonic CLS).  
However, just as in $\sVMS$, one expects four distinct string theories distinguished by the GSO projection together with the choice of supercharge, and the former is invisible to the pure NS low-lying sphere amplitudes. 
Investigations of the torus amplitudes in super CLS indeed point towards a modification of this picture, and hint at a richer set of dual matrix integrals~\cite{Collier:2026wip}.  

\paragraph{The $\mathcal{N}=1$ fusion kernel bootstrap:} It is natural to consider extending the S-kernel bootstrap program developed here to the case of $\mathcal{N}=1$ fusion kernels, which implement crossing transformations of sphere four-point conformal blocks~\cite{Hadasz:2007wi, Chorazkiewicz:2008es,Chorazkiewicz:2011zd, Poghosyan:2016kvd}. The power of this approach lies in exploiting the analytic properties of the kernels, as dictated by those of the corresponding conformal blocks, thereby circumventing the need to solve the Moore–Seiberg consistency relations directly. These relations are arguably more intricate in the $\mathcal{N}=1$ case, where different crossing kernels arise depending on whether the internal and external operators belong to the NS or R sectors. It is worth mentioning that a significant body of work already exists on the construction of the \emph{6j symbols} of the quantum group $U_q\left(\mathfrak{osp}(1|2)\right)$ for all possible cases, including both the NS and R sectors~\cite{Hadasz:2013bwa,Pawelkiewicz:2013wga,Apresyan:2022erh, Apresyan:2023twk}. These objects are expected to be related to the various $\mathcal{N}=1$ fusion kernels through an appropriate choice of normalization. To the best of our knowledge, this map has not yet been made precise and requires a detailed understanding of the corresponding sphere four-point conformal blocks and their recursion relations.

\paragraph{The gravitational path integral:} 
The worldsheet string theory in the $b\to 0$ limit should reduce to semiclassical gravity. Two of the string theories (the minus theories) become super-JT gravity analyzed in~\cite{Stanford:2019vob}, while the other two (the plus theories) have not been discussed hitherto. The string construction and the three-dimensional perspective~\cite{Eberhardt:2026hfh} make these natural. As discussed in~\cref{sec:wsreview} the distinction between the plus and minus at the level of semiclassical description lies in the relative sign of the superpotential term (equivalently, the cosmological constant) of the two Liouville theories. In addition, one has to make the choice of summing over spin structures (akin to GSO projection). However, before one does so, one has to better understand the relation between timelike Liouville CFT and the timelike Liouville path integral.
The conformal bootstrap perspective, which we have chosen, picks a specific spectrum and integration contour, and assumes reflection symmetry. Neither is derived in its present form from the path integral, cf.~\cite{Anninos:2024iwf,Usciati:2025cdn, Chatterjee:2025yzo} for recent developments. Nor has there been any attempt to understand timelike super Liouville theory directly. Progress in these directions could help understand the semiclassical worldsheet observables directly from the gravitational path integral.

\paragraph{More supersymmetry:} 
A natural extension is to incorporate $\mathcal{N}=2$ supersymmetry. The superconformal algebra  comprises a $\mathrm{U}(1)$ R-symmetry, which through spectral flow relates the NS and R sectors. One reason to focus on this example is that  $\mathcal{N}=2$ JT-supergravity arises naturally as the near-horizon theory of supersymmetric black holes~\cite{Heydeman:2020hhw} (dual matrix models have been analyzed in~\cite{Turiaci:2023jfa}). $\mathcal{N}=2$ super Liouville theory, which is a unitary two-dimensional CFT with central charge $c \geq 3$ is a canonical starting point for a string theoretic embedding. Thanks to extended supersymmetry, the spacelike version is mirror dual to $\mathrm{SL}(2,\mathbb{R})_k/U(1)$ Kazama--Suzuki supercoset~\cite{Hori:2001ax}. This duality was recently exploited in~\cite{Muhlmann:2026pdc} to extract the structure constants. Understanding the timelike analog (cf.~\cite{Anninos:2023exn} for preliminary semiclassical analysis), and constructing a bona-fide string by coupling to 2d worldsheet supergravity remains open.

\section*{Acknowledgments} 

It is a great pleasure to acknowledge Lorenz Eberhardt, Sylvain Ribault,
Victor Rodriguez, Ashoke Sen, Douglas Stanford, and Xi Yin for illuminating 
discussions. We thank Lorenz Eberhardt and Victor Rodriguez for helpful comments on a draft of the manuscript. We acknowledge the use of GPT Codex 5.4/5.5/6.0 (OpenAI) and Claude Opus
4.7/4.8/5.0 (Anthropic) for assistance with numerical computations and
checks.

S.~C.~was supported by U.S.~Department of Energy grant DE-SC0026551, and thanks the organizers of the Amsterdam String Summer Workshop for hospitality during the course of this work.
B.~M.~gratefully acknowledges funding provided by the Leinweber Foundation at
the Institute for Advanced Study and the National Science Foundation with grant
number PHY-2514611. M.~R.~was supported by U.S.~Department of Energy grant
DE-SC0009999 and by funds from the University of California. M.R.~would also like to acknowledge the Galileo Galilei Institute, the Benasque Center for Science, IHES, CERN, and the Aspen Center for Physics (which is supported by National Science Foundation grant PHY-2210452), for providing a stimulating environment during this project. I.~T.~is supported by the
European Union (ERC, FUNBOOTS, project number 101043588). Views and opinions
expressed are however those of the author(s) only and do not necessarily reflect those of the European Union or the European Research Council. Neither the European Union nor the granting authority can be held responsible for them.
J.~Z.~acknowledges support from a Graduate
Program Fellowship from the Department of Physics at the University of California, Davis. J.~Z.~would also like to acknowledge Simons Center for Geometry and Physics, Stony Brook University, and Institute for Advanced Study,  Tsinghua University for hospitality during the course of this work.

\appendix

\section{Translation of conventions}\label[appendix]{sec:conventions}

We compile a quick reference for the translation between our conventions and those used in~\cite{Hadasz:2012im}. 
\begin{itemize}[wide,left=0pt]
\item  The external dimensions are parameterized by $\lambda$ which is related to $P$ via
\begin{equation}
\lambda^2 = -4\, P^2 \,.
\end{equation}
\item The Ramond states are labeled by 
\begin{equation}
\beta^2 = - \frac{P^2}{2} \,.
\end{equation}
\item Insofar as the structure constants are concerned, the dictionary reads
\begin{equation}
C^{\lambda,\overline{\lambda}}_{\Delta,\overline{\Delta}}
= \CV{b}(P_1(\lambda),P(\Delta),P(\Delta))
\,, \qquad
\widetilde{C}^{\lambda,\overline{\lambda}}_{\Delta,\overline{\Delta}}
= \CW{b}(P_1(\lambda),P(\Delta),P(\Delta))\, ,
\end{equation}
where we have indicated the map from the $\lambda,\Delta$ notation to our momenta.
\item In the Ramond sector we used definite parity combinations.
\begin{equation}
C^\pm = \frac{\Ce{b} \pm \Co{b}}{2}\,, \qquad
C^{(\pm)} = \frac{C^+ \pm C^-}{2} \,.
\end{equation}
These relate to the combinations defined in~\cite{Hadasz:2012im} up to a factor of two, i.e., 
\begin{equation}
C^{(+)} = \frac{1}{2}\, \Ce{b}\,, \qquad C^{(-)} = \frac{1}{2}\, \Co{b}\,.
\end{equation}
More explicitly, our translation should read
\begin{equation}
C^{P(+)}_{P'} = \frac{1}{2}\, \Ce{b}(P;P',P') \,, \qquad
C^{P(-)}_{P'} = \frac{1}{2}\, \Co{b}(P;P',P') \,.
\end{equation}
\item We also redefine the recursion data for the superconformal blocks as follows  
\begin{equation}
\begin{aligned}
& \mathcal{H}^{\lambda}_{\Delta} \longrightarrow \mathcal{H}^{\mathrm{V}}_4(q)\,,
\qquad 
\widetilde{\mathcal H}^{\lambda}_{\Delta}
\longrightarrow\mathcal H_3^{\mathrm{V}},
\qquad 
\mathcal H_{\beta,\textrm{e}}^{\lambda(+)}
\longrightarrow\mathcal H_{\rm even}^{\mathrm{V}},
\qquad 
\mathcal H_{\beta,\textrm{e}}^{\lambda(-)}
\longrightarrow\mathcal H_{\rm odd}^{\mathrm{V}}\,,
\\
& \mathcal{H}^{*\lambda}_{\Delta} \longrightarrow \mathcal{H}^{\mathrm{W}}_4(q)\,,
\qquad 
\widetilde{\mathcal H}^{*\lambda}_{\Delta}
\longrightarrow\mathcal H_3^{\mathrm{W}},
\qquad 
\mathcal H_{\beta,\textrm{e}}^{*\lambda(-)}
\longrightarrow\mathcal H_{\rm odd}^{\mathrm{W}}\,.
\end{aligned}
\end{equation}
Note that the NS and $\widetilde{\mathrm{NS}}$ labels of the explicit coefficients in~\cite{Hadasz:2012im} 
(either from their recursion or from low-order coefficients given in their Appendix C) are interchanged. This should clarify the swap of labels in our identification above. In addition,~\cite{Hadasz:2012im} claim that $\mathcal H^{*\lambda(-)}_{\beta,\mathrm e}=\mathcal H^{\lambda(-)}_{\beta,\mathrm e}$. However, this assertion cannot be true --- $\mathcal H^{\rm W}_{\rm odd}\neq\mathcal H^{\rm V}_{\rm odd}$  --- as we explicitly demonstrated in~\cref{sec:torusblocks}.
\item Finally, the conformal blocks of~\cite{Hadasz:2012im} map onto ours in the following manner (modulo the corrections discussed in~\cref{sec:torusblocks}):
\begin{equation}
\begin{aligned}
\mathcal{F}_3^{\mathrm{V}}(P_1,P,P)
& =
\mathcal{F}_{\Delta_P}^{P_1} \,,
& \qquad
\mathcal{F}_4^{\mathrm{V}}(P_1,P,P)
& =
\widetilde{\mathcal{F}}_{\Delta_P}^{P_1} \,,
\\
\mathcal{F}_3^{\mathrm{W}}(P_1,P,P)
 & =
\mathcal{F}_{\Delta_P}^{*P_1} \,,
 & \qquad
\mathcal{F}_4^{\mathrm{W}}(P_1,P,P)
 & =
\widetilde{\mathcal{F}}_{\Delta_P}^{*P_1} \,,
\\
\mathcal{F}_2^{\mathrm{V}}(P_1,P,P)
 & =
\mathcal{F}_{\Delta_P, \mathrm{e}}^{P_1(+)} \,,
 & \qquad
\mathcal{F}_1^{\mathrm{V}}(P_1,P,P)
 & =
\mathcal{F}_{\Delta_P,\mathrm{e}}^{P_1(-)} \,,
\\
\mathcal{F}_2^{\mathrm{W}}(P_1,P,P)
 & =
\mathcal{F}_{\Delta_P,\mathrm{e}}^{*P_1(-)} \,,
 & \qquad
\mathcal{F}_1^{\mathrm{W}}(P_1,P,P)
 & =
\mathcal{F}_{\Delta_P,\mathrm{e}}^{*P_1(+)} \,.
\end{aligned}
\end{equation}
Note that here $\Delta_P$ should be appropriately picked for internal NS versus R operators. Also note that these identifications are defined by the Hilbert space trace. The elliptic factors entering the decomposition should be corrected based on the previous point. 
\end{itemize}

\section{Building blocks of string amplitudes}\label[appendix]{sec:wsbblocks}
To keep all the conventions explicit, we record in one place all the torus one-point functions of interest in the spacelike and timelike theories. These enter as the building blocks of the string worldsheet amplitudes. In writing these, we make explicit the phases and the complex conjugation rules inherent in assembling the left and right moving contributions. We also indicate the dependence on the Liouville parameter $b$ in the blocks, which we had refrained from doing in the text.

\paragraph{Spectral densities:}
\begin{equation}
\rhoNS(P) = 4\sinh(\pi b P)\sinh(\pi P/b)\,, \qquad
\rhoR(P) = 2\sqrt{2}\cosh(\pi b P)\cosh(\pi P/b)\,.
\end{equation}
Timelike spectral densities:
\begin{equation}
\varrho_{\mathrm{NS}}^{(\mathsf{b})}(\mathsf{P}) = \frac{(i\mathsf{P})^2}{2\rho_{\mathrm{NS}}^{(\mathsf{b})}(i\mathsf{P})}, \qquad \varrho_{\mathrm{R}}^{(\mathsf{b})}(\mathsf{P}) = \frac{1}{2\rho_{\mathrm{R}}^{(\mathsf{b})}(i\mathsf{P})}
\end{equation}
$\Ph$ is integrated on the shifted contour $\Ph \in \mathbb{R} 
+ i\,\epsilon$.

\paragraph{Spacelike super Liouville amplitudes:}
\begin{equation}\label{eq:genus1-space}
\begin{aligned}
\expval{V_{P_\mathrm{ext}}}_a^\mathrm{sL}
&= 
\abs{\frac{\vartheta_a(q)}{\eta(q)^3}}\,
\int_0^\infty dP\;\rhoNS(P)\;\CV{b}(P_\mathrm{ext},P,P)\;
\norm{q^{P^2/2}\,\mathcal{H}_a^{\mathrm{V}}(b,q,P_\mathrm{ext},P)}^2\,,
\\
\expval{W_{P_\mathrm{ext}}}_a^\mathrm{sL}
&= 
\abs{\frac{\vartheta_a(q)}{\eta(q)^3}}\, 
\int_0^\infty dP\;\rhoNS(P)\;\CW{b}(P_\mathrm{ext},P,P)\;
\norm{q^{P^2/2}\,\mathcal{H}_a^{\mathrm{W}}(b,q,P_\mathrm{ext},P)}^2\,,
\\
\expval{V_{P_\mathrm{ext}}}_2^\mathrm{sL}
&= 
\abs{\frac{\vartheta_2(q)}{\eta(q)^3}}\, 
\int_0^\infty dP\;\rhoR(P)\;\Ce{b}(P_\mathrm{ext};P,P)\;
\norm{q^{P^2/2}\,\mathcal{H}_{\mathrm{even}}^{\mathrm{V}}(b,q,P_\mathrm{ext},P)}^2\,,
\\
\expval{W_{P_\mathrm{ext}}}_2^\mathrm{sL}
&= 
2i\,
\abs{\frac{\vartheta_2(q)}{\eta(q)^3}}\,
\int_0^\infty dP\;\rhoR(P)\;\Co{b}(P_\mathrm{ext};P,P)\;P^2\;
\norm{q^{P^2/2}\,\mathcal{H}_{\mathrm{odd}}^{\mathrm{W}}(b,q,P_\mathrm{ext},P)}^2\,,
\\
\expval{V_{P_\mathrm{ext}}}_1^\mathrm{sL}
&= 
2\,\int_0^\infty dP\;\rhoR(P)\;\Co{b}(P_\mathrm{ext};P,P)\;
\norm{q^{P^2/2}\,\mathcal{H}_{\mathrm{odd}}^{\mathrm{V}}(b,q,P_\mathrm{ext},P)}^2\,,
\\
\expval{\widetilde{G}_{-3/2} G_{-3/2} V_{P_\mathrm{ext}}}_1^\mathrm{sL}
&= 
i\int_0^\infty dP\;\rhoR(P)\;\Ce{b}(P_\mathrm{ext};P,P)\;P^2\;
\norm{q^{P^2/2}\,\mathcal{H}_{\mathrm{even}}^{\mathrm{W}}(b,q,P_\mathrm{ext},P)}^2\,.
\end{aligned}
\end{equation}
Here $a=3,4$ in the first two lines.

\paragraph{Timelike super Liouville amplitudes:}
\begin{equation}\label{eq:genus1-time}
\begin{aligned}
\expval{\Vh_{\Ph_\mathrm{ext}}}_a^\mathrm{tL}
= 
\abs{\frac{\vartheta_a(q)}{\eta(q)^3}}\,
\int_{\mathbb{R} 
+ i\,\epsilon }d\Ph\;
& 
\rhoNStL(\Ph)\;
\CVh{\bh}(\Ph_\mathrm{ext},\Ph,\Ph)\;
q^{\Ph^2/2}\,\bar{q}^{\Ph^2/2}\;
\\
&\times 
\mathcal{H}_a^{\mathrm{V}}(i\bh,q,\Ph_\mathrm{ext},\Ph)\;
\mathcal{H}_a^{\mathrm{V}}(i\bh,\bar{q},\Ph_\mathrm{ext},\Ph)\,,
\\
\expval{\Wh_{\Ph_\mathrm{ext}}}_a^\mathrm{tL}
= 
\abs{\frac{\vartheta_a(q)}{\eta(q)^3}}\,
\int_{\mathbb{R} 
+ i\,\epsilon} d\Ph\;
& 
\rhoNStL(\Ph)\;
\CWh{\bh}(\Ph_\mathrm{ext},\Ph,\Ph)\;
q^{\Ph^2/2}\,\bar{q}^{\Ph^2/2}\;
\\
&\times  
\mathcal{H}_a^{\mathrm{W}}(i\bh,q,\Ph_\mathrm{ext},\Ph)\;
\mathcal{H}_a^{\mathrm{W}}(i\bh,\bar{q},\Ph_\mathrm{ext},\Ph)\,,
\\
\expval{\Vh_{\Ph_\mathrm{ext}}}_2^\mathrm{tL}
= 
\abs{\frac{\vartheta_2(q)}{\eta(q)^3}}\,
\int_{\mathbb{R} 
+ i\,\epsilon } d\Ph\;
&
\rhoRtL(\Ph)\;
\Ceh{\bh}(\Ph_\mathrm{ext};\Ph,\Ph)\;
q^{\Ph^2/2}\,\bar{q}^{\Ph^2/2}\;
\\
&\times
\mathcal{H}_{\mathrm{even}}^{\mathrm{V}}(i\bh,q,\Ph_\mathrm{ext},\Ph)\;
\mathcal{H}_{\mathrm{even}}^{\mathrm{V}}(i\bh,\bar{q},\Ph_\mathrm{ext},\Ph)\,,
\\
\expval{\Wh_{\Ph_\mathrm{ext}}}_2^\mathrm{tL}
= 
2i\,
\abs{\frac{\vartheta_2(q)}{\eta(q)^3}}\,
\int_{\mathbb{R} 
+ i\,\epsilon } d\Ph\;
&
\rhoRtL(\Ph)\;
\Coh{\bh}(\Ph_\mathrm{ext};\Ph,\Ph)\; \Ph^2\;
q^{\Ph^2/2}\,\bar{q}^{\Ph^2/2}\;
\\
&\times 
\mathcal{H}_{\mathrm{odd}}^{\mathrm{W}}(i\bh,q,\Ph_\mathrm{ext},\Ph)\;
\mathcal{H}_{\mathrm{odd}}^{\mathrm{W}}(i\bh,\bar{q},\Ph_\mathrm{ext},\Ph)\,,
\\
\expval{\Vh_{\Ph_\mathrm{ext}}}_1^\mathrm{tL}
= 
2\int_{\mathbb{R} 
+ i\,\epsilon } d\Ph\;
&
\rhoRtL(\Ph)\;
\Coh{\bh}(\Ph_\mathrm{ext};\Ph,\Ph)\;
q^{\Ph^2/2}\,\bar{q}^{\Ph^2/2}\;
\\
&\times 
\mathcal{H}_{\mathrm{odd}}^{\mathrm{V}}(i\bh,q,\Ph_\mathrm{ext},\Ph)\;
\mathcal{H}_{\mathrm{odd}}^{\mathrm{V}}(i\bh,\bar{q},\Ph_\mathrm{ext},\Ph)\,,
\\
\expval{\widetilde{G}_{-3/2} G_{-3/2}\Vh_{\Ph_\mathrm{ext}}}_1^\mathrm{tL}
= 
i\int_{\mathbb{R} 
+ i\,\epsilon } d\Ph\;
&\rhoRtL(\Ph)\;
\Ceh{\bh}(\Ph_\mathrm{ext};\Ph,\Ph)\; \Ph^2\;
q^{\Ph^2/2}\,\bar{q}^{\Ph^2/2}\;
\\
&\times
\mathcal{H}_{\mathrm{even}}^{\mathrm{W}}(i\bh,q,\Ph_\mathrm{ext},\Ph)\;
\mathcal{H}_{\mathrm{even}}^{\mathrm{W}}(i\bh,\bar{q},\Ph_\mathrm{ext},\Ph)\,.
\end{aligned}
\end{equation}
As above, here $a=3,4$ in the first two lines. In the worldsheet amplitude evaluation we use $\bh = b$ and $\Ph_{\mathrm{ext}} = - i\,P_{\mathrm{ext}}$.

  
\input{N1VMS-refs.tex}
\end{document}

%% file: svms-macros.tex
\newcommand{\sVMS}{\widehat{\text{VMS}}}
\newcommand{\SF}[1]{\mathring{\mathscr{#1}}}

\newcommand{\Prod}[2]{\prod\limits_{#1}^{#2}}

\newcommand{\zb}{\overline{z}}
\newcommand{\thetab}{\bar{\theta}}

\newcommand{\Lambdat}{\widetilde{\Lambda}}
\newcommand{\hP}{h_{_P}}
\newcommand{\hPt}{\widetilde{h}_{_P}}

\newcommand{\RsL}[2]{R^{#1}_{#2}}

\newcommand{\dGb}[1]{\Gamma_b\pqty{#1}}
\newcommand{\dGbX}[2]{\Gamma^{(b)}_{_\mathrm{#1}}\pqty{#2}}
\newcommand{\dGNS}[2]{\Gamma^{(#1)}_{_\mathrm{NS}}\pqty{#2}}
\newcommand{\dGR}[2]{\Gamma^{(#1)}_{_\mathrm{R}}\pqty{#2}}
\newcommand{\Sb}[1]{S_b\pqty{#1}}
\newcommand{\SbX}[2]{S^{(b)}_{_\mathrm{#1}}\pqty{#2}}
\newcommand{\SNS}[2]{S^{(#1)}_{_\mathrm{NS}}\pqty{#2}}
\newcommand{\SR}[2]{S^{(#1)}_{_\mathrm{R}}\pqty{#2}}

\newcommand{\CV}[1]{C^{(#1)}_{_\mathrm{V}}}
\newcommand{\CW}[1]{C^{(#1)}_{_\mathrm{W}}}
\newcommand{\Co}[1]{C^{(#1)}_\mathrm{odd}}
\newcommand{\Ce}[1]{C^{(#1)}_\mathrm{even}}

\newcommand{\rhoNS}{\rho_{_\mathrm{NS}}^{(b)}}
\newcommand{\rhoR}{\rho_{_\mathrm{R}}^{(b)}}
\newcommand{\rhoNStL}{\varrho_{_\mathrm{NS}}^{(b)}}
\newcommand{\rhoRtL}{\varrho_{_\mathrm{R}}^{(b)}}

\newcommand{\bh}{\mathsf{b}}
\newcommand{\ch}{\mathsf{c}}
\newcommand{\Qh}{\mathsf{Q}}
\newcommand{\Ph}{\mathsf{P}}
\newcommand{\Vh}{\mathsf{V}}
\newcommand{\Wh}{\mathsf{W}}
\newcommand{\Lambdah}{\mathsf{\Lambda}}
\newcommand{\Lambdaht}{\widetilde{\mathsf\Lambda}}

\newcommand{\hhP}{\mathsf{h}_{_\mathsf{P}}}
\newcommand{\hhPt}{\widetilde{\mathsf{h}}_{_\mathsf{P}}}
\newcommand{\RtL}[2]{\mathsf{R}^{#1}_{#2}}

\newcommand{\CVh}[1]{\mathsf{C}^{(#1)}_{_\mathsf{V}}}
\newcommand{\CWh}[1]{\mathsf{C}^{(#1)}_{_\mathsf{W}}}
\newcommand{\Coh}[1]{\mathsf{C}^{(#1)}_\mathsf{odd}}
\newcommand{\Ceh}[1]{\mathsf{C}^{(#1)}_\mathsf{even}}

\newcommand{\Sm}[3]{{}^{\smash[b]{\scriptstyle{\mathrm{#1}}}}\mathbb{S}^{(#2)}_{#3}}

\newcommand{\nNS}{\mathsf{N}_{_\mathrm{NS}}}
\newcommand{\nR}{\mathsf{N}_{_\mathrm{R}}}
\newcommand{\wsnorm}{\mathsf{C}_{\mathbf{S}^2}}
\newcommand{\tnorm}{\mathsf{C}_{\mathbf{T}^2}}

\newcommand{\zA}[1]{0A$^{#1}$}
\newcommand{\zB}[1]{0B$^{#1}$}
\newcommand{\wsNS}[3]{\tensor*[_{#2}]{\mathcal{V}}{^{\smash[b]{^{#3}}}_{#1}}}
\newcommand{\wsR}[3]{\tensor*[_{#2}]{\mathcal{R}}{^{\smash[b]{^{#3}}}_{#1}}}
\newcommand{\wsamp}[2]{\mathcal{A}_{#1}^{#2}}

%% file: N1VMS-refs.tex
\providecommand{\href}[2]{#2}\begingroup\raggedright\endgroup